\documentclass[a4paper,11pt]{article}
\pdfoutput=1 

\usepackage{jheppub} 
                     
\usepackage[T1]{fontenc} 

\usepackage{amsmath}
\usepackage{amssymb}
\usepackage{latexsym}
\usepackage{graphicx}
 
\usepackage{tikz, pgf}
\usetikzlibrary{shapes.misc}

\usetikzlibrary{decorations.pathmorphing}	
\usetikzlibrary{decorations.markings}
\tikzset{
    sugra/.style={decorate, decoration={snake}, draw=black},
    suym/.style={draw=black, postaction={decorate},
        },
    hwbou/.style={draw=black, postaction={decorate}, ultra thick
        },
    vector/.style={decorate, decoration={snake}, draw},
	provector/.style={decorate, decoration={snake,amplitude=2.5pt}, draw},
	antivector/.style={decorate, decoration={snake,amplitude=-2.5pt}, draw},
    fermion/.style={draw=black, postaction={decorate},
        decoration={markings,mark=at position .55 with {\arrow[draw=black]{>}}}},
    fermionbar/.style={draw=black, postaction={decorate},
        decoration={markings,mark=at position .55 with {\arrow[draw=black]{<}}}},
    fermionnoarrow/.style={draw=black},
    gluon/.style={decorate, draw=black,
        decoration={coil,amplitude=4pt, segment length=5pt}},
    scalar/.style={dashed,draw=black, postaction={decorate},
        decoration={markings,mark=at position .55 with {\arrow[draw=black]{>}}}},
    scalarbar/.style={dashed,draw=black, postaction={decorate},
        decoration={markings,mark=at position .55 with {\arrow[draw=black]{<}}}},
    scalarnoarrow/.style={dashed,draw=black},
    electron/.style={draw=black, postaction={decorate},
        decoration={markings,mark=at position .55 with {\arrow[draw=black]{>}}}},
	bigvector/.style={decorate, decoration={snake,amplitude=4pt}, draw},
}
\definecolor{light-gray}{gray}{0.80}

\usetikzlibrary{shapes}
\usetikzlibrary{fit}

\allowdisplaybreaks[1]

\numberwithin{equation}{section}

\newcommand{\be}{\begin{equation}}
\newcommand{\ee}{\end{equation}}
\newcommand{\bea}{\begin{eqnarray}}
\newcommand{\eea}{\end{eqnarray}}
\def\nn{\nonumber}

\newcommand{\gcou}{\textrm{g}}

\def\hwgymsq{ \lambda^2 }

\newcommand{\gii}{g_{II} }
\newcommand{\giia}{g_{IIA} }
\newcommand{\giib} {g_{IIB}}
\newcommand{\ghet}{g_{het}}
\newcommand{\ghe} {g_{he}}
\newcommand{\gho} {g_{ho}}
\newcommand{\gia}{g_{IA}}
\newcommand{\gib}{g_I}

\newcommand{\raiia}{r_{IIA}}
\newcommand{\raiib}{r_{IIB}}

\newcommand{\rahe}{r_{he}}
\newcommand{\raho}{r_{ho}}
\newcommand{\raia}{r_{IA}}
\newcommand{\raib}{r_{I}}
\newcommand{\raten}{R_{10}}
\newcommand{\raele}{R_{11}}
\newcommand{\ciele}{{L}}

\newcommand{\lepl}{\ell_{11}}

\newcommand\lesth{\ell_H}
\newcommand\lesti{\ell_I}
\newcommand\lestii{\ell_{II}}

\newcommand{\Tr}{\textrm{Tr}}
\newcommand{\tr}{\textrm{tr}}
\newcommand{\trs}{\textrm{tr}_{\mathcal{S}}}
\newcommand{\curv}{{R}}

\def \xgs{X_8^{(\textrm{gs})}} 
\def \xvw{X_8^{(\textrm{vw})}} 
\def \ygs{Y_8^{(\textrm{gs})}} 
\def \yvw{Y_8^{(\textrm{vw})}}

\def\half{ \frac{1}{2}}  
\def\threeh{\frac{3}{2}}

\def\quart{\frac{1}{4}}

\newcommand{\coc}{\mathcal{C}}
\def\sm{\smallskip}

\newcommand{\IZ}{\mathbb{Z}}

\def\ba{\begin{align}}
\def\ea{\end{align}}
\def\bse{\begin{subequations}}
\def\ese{\end{subequations}}

\def\cN{{\cal N}}

\def\ba{{\bf a}}

\def\half{{1\over 2}}

\title{\boldmath Type I/heterotic duality and M-theory amplitudes}
\author[a]{Michael B. Green }
\author[a,b]{, Arnab Rudra}
\affiliation[a]{Department of Applied Mathematics and
Theoretical Physics\\
Wilberforce Road, Cambridge CB3 0WA, UK}
\affiliation[b]{Center for Quantum Mathematics and Physics (QMAP)\\
Department of Physics, University of California, Davis, CA 95616 USA}
\emailAdd{M.B.Green@damtp.cam.ac.uk} 
\emailAdd{rudra@ucdavis.edu}

\abstract{This paper investigates relationships between low-energy  four-particle scattering amplitudes with external gauge particles and gravitons in the $E_8\times E_8$ and $SO(32)$ heterotic string theories and the type I and type IA  superstring theories by considering a variety of tree level and one-loop Feynman diagrams describing such amplitudes in eleven-dimensional supergravity in a Ho\u rava--Witten background compactified on a circle.   This accounts for a number of perturbative and non-perturbative aspects of low order  higher derivative terms in the low-energy expansion of string theory amplitudes, which are expected to be protected by half maximal supersymmetry from receiving corrections beyond one or two loops.  It also suggests the manner in which type I/heterotic duality may be realised for certain higher derivative interactions that are not so obviously protected.  For example, our considerations suggest that $\curv^4$ interactions (where $\curv$ is the Riemann curvature)  might receive no perturbative corrections beyond one loop by virtue of a conspiracy involving contributions from  (non-BPS) $\mathbb{Z}_2$ D-instantons in the type I  and heterotic $SO(32)$ theories. }

\begin{document}
\maketitle
\flushbottom


\section{Overview}
\label{sec:overview} 

The dualities of M-theory that relate string theories in different  regions of moduli space have been well-studied over the past 20 years.    In particular, the interrelationships between various theories with ten-dimensional $\mathcal{N}=1$ supersymmetry follow from the considerations of Ho\u rava and Witten  (HW) \cite{Horava:1995qa,Horava:1996ma}).\footnote{We will refer to theories with half-maximal supersymmetry as $\mathcal{N}=1$ theories, which reflects their ten-dimensional supersymmetry.}  
They   understood that the $E_8\times E_8$ heterotic string  (referred to as the HE theory in the following) is equivalent to eleven-dimensional M-theory on an interval in the $x^{11}$ direction\footnote{The coordinates of eleven-dimensional Minkowski space will be dlabelled $x^\mu$ with $\mu =1,2,\dots,x^{11}$ and with $x^1$ chosen to be the time coordinate.}   of length $\ciele=\pi \lepl \raele\, $ ($\lepl $ is the eleven-dimensional Planck scale) -- in other words in a background space-time with geometry $\mathbb{M}^{10} \times S^1/\mathbb{Z}_2 $, where $\mathbb{M}^{10}$ is ten-dimensional Minkowski space.   This is equivalent to the compactification of the eleven-dimensional theory on an orbifold of a circle of radius $\lepl\raele$ so that the eleventh dimension is an interval that terminates on ten-dimensional boundaries.  Consistency of eleven-dimensional supergravity in the presence of  these boundaries requires boundary degrees of freedom that correspond to an  independent $\mathcal{N}=1$ supersymmetric  $E_8$ gauge theory restricted to each ten-dimensional  boundary.   The  $E_8\times E_8$ heterotic string coupling constant is   $\ghe=\raele^{3/2}$, and so the limit $\raele\to 0$ is the  weak coupling limit of the HE theory. 

\sm

\begin{figure}[h]
\begin{center}
\begin{tikzpicture}[scale=.40]
\draw [thick]  [>=latex,->]  (-3.4,3.6) -- (3.4,3.6);
\draw [thick]  [>=latex,->]   (-3.4,-3.6) -- (3.4,-3.6);
\draw [thick]  [>=latex,->] (-3.6,-3.6) -- (-3.6,3.6);
\draw [thick] [>=latex,->]  (3.6,-3.6) -- (3.6,3.6);
\draw (6.,6.) node{\text{ M theory on $\mathbb{M}^{9}\times   S^1/\mathbb{Z}_2\times S^1 $}};
\draw (6.,-6.) node { \text{Type IA}};
\draw (-6,-6.) node {\text{SO(32)  Heterotic/Type I}};
\draw (-6,6) node {\text{E$_8$ $\times$ E$_8$ Heterotic}};
\draw (-4.8,0) node { $\raten$ };
\draw (4.8,0) node {$\raten$};
\draw (0,4.5) node {$\raele$};
\draw (0,-4.5) node {$\raele$};
\draw (4.35,-4.35) node {$(0,\infty)$};
\draw (4.35,4.35) node {$(\infty,\infty)$};
\draw (-4.35,4.35) node {$(\infty,0)$};
\draw (-4.35,-4.35) node {$(0,0)$};
\end{tikzpicture}
\end{center}
\caption{ Dualities relating M-theory on  $\mathbb{M}^9 \times S^1/\mathbb{Z}_2\times S^1$ to $\mathcal{N}=1$ string theories}
\label{fig:Imth1}   
\end{figure}
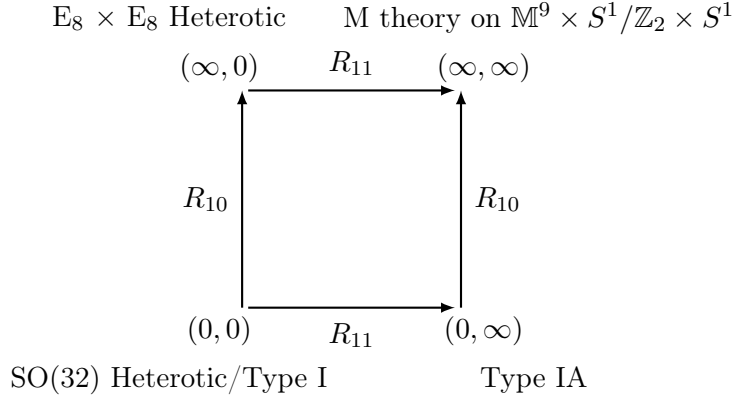

When compactifying on an additional spatial circle in $x^{10}$  of radius $\lepl\raten$ so that the background is  $\mathbb{M}^9 \times S^1 \times S^1/\mathbb{Z}_2 $  Wilson lines may be added, which break the gauge symmetry of the HE theory.   Choosing the Wilson lines so that  $E_8$ is broken to $SO(16)$ on each boundary (following \cite{Horava:1995qa,Horava:1996ma}) the HE theory is related by T-duality to a compactification of the heterotic $Spin(32)/\mathbb{Z}_2$ (referred to as the HO theory in the following) to 9 dimensions in the presence of an HO Wilson line that breaks $SO(32)$ to $SO(16)\times SO(16)$. The HO theory has a coupling constant $\gho =\raele/\raten$.   This, in turn is related by a weak/strong duality transformation (which will be referred to as the ``$S$ transformation'' in the following) to the type I theory compactified on a circle, where the circle direction is $x^{11}$ and an appropriate Wilson line again breaks $SO(32)$ to  $SO(16)\times SO(16)$.  The type I coupling constant is $\gib=\raten/\raele$.  Finally, T-duality of the type I theory along the $x^{11}$ direction relates it to the type IA theory with coupling constant $\gia=\raten^{3/2}$ in a configruration where there are eight D8-branes and their mirror images coincident with each of the two orientifold 8-planes.  We see from this circle of dualities, which is illustrated in  figure \ref{fig:Imth1}, that the type IA and HE theories are related by interchanging $\raten$ and $\raele$.  
The precise correspondence between the parameters of the various $\mathcal{N}=1$ superstring theories and the parameters of the HW theory is reviewed in appendix~\ref{sec:dictionary}.

\sm

 In this paper we will  investigate these duality relationships further by considering certain scattering amplitudes in M-theory determined by tree-level  and one-loop Feynman diagrams of eleven-dimensional supergravity in the Ho\u rava --Witten background.  We will consider amplitudes involving the scattering of four gauge particles  as well as  those involving four gravitons (for economy of space we will not consider mixed gauge particle/graviton amplitudes or amplitudes involving dilaton fluctuations).   

\sm

Of course, we do not expect the Feynman diagram approximation  to capture the detailed behaviour of  M-theory, but it should provide some information about the low-lying terms in its low-energy approximation.   Since we will discuss scattering amplitudes of massless states in $D=9$ dimensions, much of the rich structure of the theory, such as that associated with the effects of M2-branes and M5-branes will not enter in this perturbative approximation.  Furthermore, we will treat the  Ho\u rava--Witten background geometry as rigid, ignoring in particular the fluctuations of the boundaries, which are associated with the dynamics of the HE dilaton.     

\sm

 The perturbative approximation to supergravity should be valid at momentum scales   $k << \lepl^{-1}$, where $k$ is a characteristic momentum in the scattering amplitude and $\lepl$ is the eleven-dimensional Planck scale. In this low-energy regime we would expect supergravity to reproduce terms in the low-energy expansions of scattering amplitudes, which are insensitive to a Planck-scale cutoff.  However, we will be considering the transformation of frames involved in discussing the low-energy approximations to the dual heterotic and type I string theories (the relationships between these frames are reviewed in  appendix~\ref{sec:dictionary}).  These dualities involve compactification on length scales $<< \lepl$, where perturbation theory cannot in general be justified.  However, just as in the analogous discussion of four-graviton scattering in the type II superstring  \cite{Green:1997as},  we expect that the BPS properties of the low order terms in the low-energy expansion should justify these  approximations.
 
\sm 

The precise pattern of non-renormalisation conditions in theories with half-maximal supersymmetry (ten-dimensional $\mathcal{N}=1$ supersymmetry) is not completely understood.  Following \cite{Berkovits:2009aw} and references therein, single-trace contributions to the low-energy expansion of the  four-gluon scattering amplitude, which contribute terms of order $s^n\,t_8 \tr F^4$ to the effective action,\footnote{Here, the symbol $\tr$ denotes the trace in the fundamental representation of $SO(16)$.} are expected to get contributions from all orders in perturbation theory, apart from the cases $n =-2$ (which is the Yang--Mills pole term)  and $n=0$, which should receive no corrections beyond one loop.  Double-trace  contributions of order $s^n\, t_8(\tr F^2)^2$ are expected to get contributions from all orders in perturbation theory apart from the cases $n=-1$ (which is the gravitational pole term), $n=0$, which should receive no contributions beyond one loop, and $n=1$, which should receive no contributions beyond two loops.\footnote{More precisely, these statements apply to the heterotic theories, while the rules in the type I and type IA theories are slightly modified from the ones described in \cite{Berkovits:2009aw} .}  The non-renormalisation conditions on terms in the low-energy expansion of the four-graviton amplitude are naively expected to parallel those of the double-trace terms of the same dimension in the Yang--Mills amplitudes.   However, this would suggest that $t_8t_8\, \curv^4$  (where $R$ is the Riemann curvature and $t_8$ is an eighth-rank tensor that will be discussed later)  gets contributions from all loops, whereas explicit multi-loop calculations in supergravity \cite{Bern:2012cd} suggest that $t_8t_8\, \curv^4$ gets no contributions beyond one loop  (see also \cite{Tourkine:2012ip}).  
Additional non-renormalisation conditions apply to those parity-conserving interactions that are related by supersymmetry to the parity-violating anomaly-cancelling terms, which only get one-loop contributions  \cite{Tseytlin:1995fy,Tseytlin:2000sf}.  
 
 \subsection*{Outline}
 In section~\ref{sec:feyn} we will discuss the Feynman rules of relevance to the calculations that follow.  These differ from conventional bulk Feynman rules by virtue of the presence of space-like boundaries, which correspond to the fixed points of the orbifold of $S^1$.\footnote{For the most part we will use the language of the ``downstairs'' formalism in this paper.  This is the description in which $x^{11}$ is restricted to the interval with two boundaries, $0\le x^{11} \le \pi \lepl  \raele$.  In the ``upstairs'' formalism $x^{11}$ spans the circle, $0\le x^{11} \le 2\pi \lepl  \raele$ with fixed points at $x^{11}=0$ and $x^{11}=\pi \lepl R_{11}$ implied by the orbifold condition.}   The discussion of the  propagators will follow that given in \cite{Dasgupta:2000xu}, while the vertices for bulk fields coupling to the boundary fields was given in  \cite{Horava:1995qa,Horava:1996ma}.  In order to clarify the discussion, in appendix~\ref{sec:A} we will give a brief review of the action and Feynman rules in the HW background.  We will also introduce a streamlined notation for terms quartic in field strengths and curvatures that arise in the low-energy ten-dimensional $\mathcal{N}=1$ superstring actions.  
 Terms in the action that are quadratic in gravitational field strengths, would vanish on shell in ordinary gravity theories, but in the context of the Ho\u rava--Witten background such interactions are localised on the boundaries and they give rise to two-point and three-point amplitudes involving Kaluza--Klein modes of the bulk fields,  that will be discussed in section \ref{sec:boundint}
 
 \sm 
 
In order to set the scene for the discussion of four-particle Yang--Mills amplitudes in supergravity in the Ho\u rava--Witten background, in section  \ref{sec:YMamps} we summarise a number of detailed properties of these amplitudes that emerge in the various $\mathcal{N} = 1$ perturbative string theories.  Many of these features are well known from earlier work, but some details of the more subtle properties will be presented elsewhere \cite{Rudra:2016xx}.

\sm

 Section \ref{sec:YMtrees} is concerned with  the low-energy expansion of four-particle supergravity tree amplitudes with external gauge particles in the $\mathbb{M}^{9}\times  S^1 \times S^1/\mathbb{Z}_2 $ background.  The simplest example is the standard Yang--Mills tree amplitude localised in either boundary, which simply  translates into the leading low-energy contribution (the massless Yang--Mills pole contribution) in the various $\mathcal{N}=1$ string theories.  

\sm
 
The contribution of the Yang--Mills tree with a gravitational propagator will then be considered.  This is a generalisation of the gauge boson amplitude considered in \cite{Dasgupta:2000xu} (which also considered the scattering of massive $SO(16)$ spinors, which are not considered here). The low-energy expansion of this term leads to double-trace contributions of the schematic form $s^n\,t_8\, (\tr_i F_i^2)\, (\tr_j F_j^2)$, where $i,j=1,2$ and  $t_8$ is a standard eighth rank tensor, and the subscripts $1$ and $2$ label the different $SO(16)$ subgroups.   These effective interactions arise in very different ways in the  various string theories.  For example, the $n=0$ term arises  as a purely tree-level effect  in the heterotic theories  but arises at two loops (spherical world-sheets with three boundaries) in the type I and IA theories.   This term is expected to be protected from higher order renormalisation.  More generally, such terms  are tree-level contributions in the HE and HO theories but arise from world-sheet with $n+2$ boundaries (and no handles) in the type I theories.  However, only the interactions with $n=-1,0, 1$ (where $n=-1$ denotes the term with the gravitational pole) are protected against renormalisation by higher loop effects.

\sm
 
In section \ref{sec:loopamp} we will consider the Yang--Mills loop amplitude localised on either boundary and compactified on a circle of radius $\lepl\raten $ and with a Wilson line breaking the symmetry, which generates Kaluza--Klein towers of massless and massive $SO(16)$ adjoint states and massive $SO(16)$ spinor  states circulating in the loop.  After expressing this as a sum over windings of the loop we will argue that the ultraviolet divergent zero winding term must vanish after renormalisation.  The sum of non-zero windings gives a finite coefficient for the $t_8\,(\tr_1 F_1^4+ \tr_2 F_2^4)$ interaction that contributes at one loop to the decompactified HO theory  and at disk level to the type I and IA theories, but does not contribute to the decompactified HE theory.  Here the trace is in the {\it fundamental} representation for each $SO(16)$ subgroup, whereas the naive expectation based on conventional gauge theory would be for the trace to be in the adjoint representation. 
  These observations are in accord with string theory expectations.
We will also comment on the relation of this interaction to the chiral gauge anomaly cancelling terms of the HO and type I theories.

\sm
 
  In section~\ref{sec:graviamps} we will summarise some detailed properties four-graviton amplitudes in $\mathcal{N} = 1$ perturbative string theories.  This will include a review of the relationships between the different kinds of $\curv^4$ terms that arise as higher derivative interactions.  Those that are related by $\mathcal{N}=1$ supersymmetry to the anomaly cancelling terms are again not expected to be renormalised.  However, the interaction of the form $t_8t_8 \curv^4$ is known to get contributions at tree level and one loop in all the theories. Based on naive dimensional analysis this is not a BPS interaction, but there is a possibility that it has special features since the expected  three-loop divergence in $D=4$, $\mathcal{N}=4$ supergravity is known to be absent  \cite{Bern:2012cd}  (see also \cite{Tourkine:2012ip}).  It is of interest to understand how this can be consistent with $S$-duality between the HO and type I theories since the individual perturbative contributions do not transform into each other.   
   
   \sm

In section~\ref{sec:gravitrees} we will consider tree-level graviton scattering amplitudes in supergravity in the Ho\u rava--Witten orbifold background.  Apart from the bulk tree-level supergravity amplitude, there are tree amplitudes with one of the vertices localised on one or on both  boundaries.   These generate terms of the form $t_8\, \tr \curv^4$ and $t_8 \, (\tr \curv^2)^2$ (where the trace is in the fundamental representation of the tangent-space group.), which are the gravitational analogues of the gauge interactions of section  \ref{sec:YMtrees}.  These parity-conserving interactions combine with the familiar parity-violating terms needed for chiral anomaly cancellation in the HE theory (as discussed in \cite{Horava:1996ma}) to form a sum of ${\cal N}=1$ 
superinvariants.

\sm

One-loop contributions to the  four-graviton amplitude that generate terms in the effective action of order $\curv^4$ will be considered in section~\ref{sec:gravloop}.    In section~\ref{subsec:gravloop2} we will consider the one-loop amplitude   in which the external gravitons interact with a supermultiplet of gauge  particles localised in either boundary and compactified on $S^1$.  This is the supergravity analogue of the four gauge particle loop considered earlier.  It gives a contribution to the $\curv^4$ interaction  that is a linear combination of $t_8\, \tr \curv^4$  and $t_8\, (\tr \curv^2)^2$. 
 In order to evaluate the amplitude  
in which the external gravitons couple to bulk supergravity states circulating in the loop we will introduce an adaptation of the eleven-dimensional  world-line superparticle formalism that was used to describe one loop in eleven-dimensional supergravity compactified on $S^1$ in  \cite{Green:1997as, Green:1999by}. This  formalism will be reviewed briefly in section~\ref{subsec:susyloops} and extended to implement the $\mathbb{Z}_2$ orbifold of the Hor\u ava--Witten background.  We will see that the component of the loop with a circulating superparticle carrying zero Kaluza--Klein mode ($m=0$)  in the $x^{11}$ interval gives the  one-loop supergravity contribution to the ${\cal N}=1$ theory that combines with the gauge loops of section~\ref{subsec:gravloop2} to complete the parity conserving part of the superinvariants that also contain the parity-violating anomaly cancelling interactions.  It is expected that this is not renormalised by higher supergravity or string theory loop contributions. 
 
\sm
 
 The piece of the supergravity loop in which the circulating particles carry non-zero Kaluza--Klein charges in the interval gives rise to an effective interaction of the form $t_8 t_8 \curv^4$, much as in the type II theories.   This has a coefficient that is a function of the ratio $\raele/\raten= \gho =\gib^{-1}$.  This coefficient, which is the discussed in section~\ref{subsec:gravloop1}, is simply the non-holomorphic Eisenstein series $E_{3/2}(i /\gho)$ of the type that arises in the nine-dimensional type II theories, with the important distinction that the pseudsocalar field (the real part of the type IIB coupling constant) is set to zero since it is projected out by the orientifold that takes type II to type I.   As remarked above, even though naive dimensional analysis suggests that this interaction is not protected by supersymmetry, it is known not to have the three-loop $\curv^4$ divergence in four-dimensional $\cN=4$ supergravity that would have been expected for an unprotected interaction..  This function, which is invariant under S-duality, has some interesting features.  Firstly, it correctly reproduces the tree-level and one-loop coefficients in the heterotic and type I/IA theories and has no further perturbative corrections.  Furthermore, it possesses an infinite set of D-instanton contributions in both the HO and type I theories but not in the HE and type IA theories.  These non-BPS objects are identified with the effects of wrapped euclidean world-lines of $D0$-branes.  One might be skeptical that these non-BPS effects can be predicted accurately, although it has been forcibly argued \cite{Witten:1998cd} that the presence of $\mathbb{Z}_2$ instantons associated with the homotopy relation $\pi_9(SO(32))= \mathbb{Z}_2$ is an essential  feature in the type I theory. This homotopy condition would also allow instantons in the HO theory, but their origin is much more questionable.  The fact that $\pi_9(E_8)= 0$ means such instantons  must be absent in the HE theory, which is indeed a property of the amplitude presented in section~\ref{subsec:gravloop1}.

\sm

Higher order terms in the low-energy expansion can arise from many sources.  Firstly, there are higher order contributions from the low-energy expansion of the tree and loop terms described above.  In addition there are many further loop diagrams that are very complicated to analyse.   Since these correspond to non-BPS protected interactions it is unclear to what extent they illustrate genuine features of the string theories.   Nevertheless, in section~\ref{sec:higherloop}  we will consider the most intriguing example of such a diagram, that contributes to the four gauge particle amplitude at order $s\, t_8\tr_i F_i^4$.  This is a loop of gauge particles localised in a boundary, but with one gauge propagator replaced by a gravitational propagator. It is associated with a function that transforms in a non-trivial manner under S-duality.  Although we do not expect that it is the complete story,  since we are ignoring  intrinsically ``stringy'' effects that we do not have control over,  it is of interest that this diagram contains perturbative and instanton contributions in the HO and type I theories with sensible powers of the coupling constants. 

\sm
 
A summary and discussion of these results are given in section \ref{sec:result}.
 
 \section{Feynman diagrams in the  Ho\u rava--Witten geometry}
\label{sec:feyn}

In this section we will review the Feynman rules in the Ho\u rava--Witten geometry that enter into the calculations in the subsequent sections.  The action described in \cite{Horava:1995qa,Horava:1996ma} takes the form 
\begin{eqnarray} 
S = S_{sugra} + S_{YM} + S_{boundary}\,,
\label{totact}
\end{eqnarray} 
which is the sum of the bulk eleven-dimensional supergravity action, the ten-dimensional super Yang--Mills action and a boundary contribution that includes the gauge and gravitational Chern--Simons interactions that are required in order to ensure the absence of chiral anomalies, as discussed in \cite{Horava:1996ma}).  Whereas $S_{sugra}$ and  $S_{YM} $ are second order in derivatives,  $S_{boundary}$ includes gravitational interactions of fourth order in derivatives, namely the gravitational Chern--Simons term, a $\curv^2$ term (where $R$ is the riemann curvature) and a $(\partial H)^2$ term (where $H$is the field strength of the two-form potential).  The action $S$ includes all the interactions that contribute to three-point functions.  Ten-dimensional $\mathcal{N}=1$ supersymmetry guarantees that these interactions are not renormalised by loop corrections. 

\sm

 In discussing the gauge particle and graviton tree diagrams it will be sufficient to use the bosonic components of the Feynman rules obtained from the  action  \eqref{totact}.   However, the discussion of loop amplitudes  necessarily involves supermultiplets of circulating particles.  For the purpose of evaluating one-loop amplitudes in this paper it will prove efficient  to make use of  an extension of the supersymmetric first-quantised light-cone gauge description of   eleven-dimensional supergravity  compactified on a circle in \cite{Green:1999by}.  This involves vertices describing the emission of a supergraviton from a superparticle world-line.    The extension to include vertices localised in the Ho\u rava--Witten boundaries will be briefly described in section \ref{sec:boundint} (its application to a loop amplitude in the bulk will be described in sections~\ref{subsec:susyloops} and \ref{subsec:gravloop1}).

\sm

\subsection{The action in the HW background}
\label{subsec:compvert}
 
 In writing the various terms in the action below we will only be explicit about the interactions involving bosonic fields - the fermionic terms required by supersymmetry are explicit in many references to earlier papers.   
The bosonic terms that enter in the total action $S$ in \eqref{totact} are the following.
 The purely eleven-dimensional supergravity action is given by  \cite{Cremmer:1978km} 
\begin{eqnarray} 
	S_{sugra}&=&\frac{1}{2\kappa_{11}^2}	 \Bigg[\int_{\mathcal{M}_{11}} d^{11}x\ \sqrt{-G^{(11)}} \left(\curv-\frac{1}{2}|G_4|^2 \right)
	-\frac{1}{6}\int_{\mathcal{M}_{11}} C_3 \wedge G_4 \wedge G_4 \Bigg]\,,
\label{sugra} 
\end{eqnarray} 
where $G_4 = d C_3$ is the four-form field strength associated with the three-form potential $C_3$ and $G^{(D)}$ is the determinant of the $D$-dimensional space-time metric (with $D=11$ in the above case).  In our later applications the integration domain $\mathcal{M}_{11}$, will be the space $\mathbb{M}^{9} \times S^1\times S^1/\mathbb{Z}_2$ of the Ho\u rava--Witten geometry compatified on $S^1$.

\sm

The ten-dimensional supersymmetric Yang--Mills action \cite{Brink:1976bc}  is given in its generally coordinate invariant form by the sum of terms on each boundary
\begin{eqnarray}
\label{sym}
	S_{\textrm{YM} }=-\frac{1}{4\hwgymsq}\int_{\mathcal{M}_{11}} d^{11}x\ \sqrt{-G^{(11)}} \,  \frac{1}{30}\left[\Tr_1 \left(
F_{\mu \nu}F^{\mu \nu}\right)\,  \delta(x^{11}) + \Tr_2 \left(
F_{\mu \nu}F^{\mu \nu}\right)\,\delta(x^{11} - \ciele) \,\right]\,,
\end{eqnarray}
where the symbol $\Tr_i$ indicates a trace in the adjoint representation of $(E_8)_i$ (which will be broken to its $SO(16)$ subgroup upon compactification on $S^1$) where the subscript $i$ labels the boundary.  The Yang--Mills fields depend only on the ten dimensional space-time of the boundaries, but the metric degrees of freedom also depend on $x^{11}$ so this interaction includes the vertex coupling bulk gravity to Yang--Mills fields with arbitrary $p_{11}$ momentum. 
 
 \sm
 
The interactions that we are including in the boundary term in \eqref{totact} comprise the Yang-Mills and Lorentz Chern--Simons interactions, as discussed in \cite{Horava:1995qa,Horava:1996ma}, together with $\curv^2$ and $(\partial \, H)^2$ interactions, where $H= dC$, and the only non-zero boundary components of $C$ are  $C_{\mu\nu 11}$.  So we will write 
\begin{eqnarray} 
S_{boundary} = S_{CS} + S_{\curv^2}+ S_{(\partial \, H)^2}\,.
\label{sbound}
\end{eqnarray} 
 The Chern--Simons terms in $S_{CS}$  are obtained by modifying the supergravity action \eqref{sugra} by replacing $G_4$ by  
\begin{eqnarray}
G_4 = dC^{(3)} + \Omega_3\,,
\label{newg4}
\end{eqnarray} 
where $\Omega_3$ is  defined by
\begin{eqnarray} 
\Omega_3 = 
-\delta(x^{11})\left(\Omega_{3\textrm{YM}}-\frac{1}{2}\Omega_{3\textrm{L}} \right) 
- \delta(x^{11}-\ciele)\left(\Omega_{3\textrm{YM}}-\frac{1}{2}\Omega_{3\textrm{L}} \right)\,,
\label{deltadef}
\end{eqnarray} 
and we need to impose  the condition that only the components $C^{(3)}_{\mu\nu 11}$ of the bulk three-form are non-zero at the boundary (and we are again  ignoring fermionic contributions). 
The quantities $\Omega_{3\textrm{YM}}$ and $\Omega_{3\textrm{L}}$ are the standard Yang-Mills and Lorentz Chern--Simons forms.  
The action now includes the boundary terms 
\begin{eqnarray}
S_{CS} = \left(\frac{\kappa_{11}^2}{\lambda^2}\right)\int_{\mathcal{M}_{11}} d^{11} x  \sqrt{-G^{(11)}}\, \left(dC^{(3)} \wedge * \Omega_3+ \half \Omega_3 \wedge * \Omega_3\right)\,,
\label{boundint} 
\end{eqnarray}
where the $*$ operation is with respect to the ten-dimensional boundary space-time.
The first term determines the Chern--Simons contribution to on-shell three-point functions (to be discussed in the following section)  in which  two states may be Yang--Mills gauge bosons or  bulk gravitons and the third state  is a Kaluza--Klein mode of the bulk $C_{\mu\nu 11}$ field.  The second term in \eqref{boundint} is a contact term that plays an important r\^ole in the context of the four-particle  amplitudes  to be considered later.  This term is singular since it involves the  integral of the product of two $\delta(x^{11})$ (or $\delta(x^{11} - L)$) factors,  arising from the product of two $\Omega_3$ factors localised on the same boundary, resulting in a term proportional to $\delta(0)$  in $S_{CS}$.   Similar divergent terms  in the low energy action and their regularisation were briefly discussed in  section 4 of \cite{Horava:1996ma}.  Such divergences cancel in physical amplitudes.   For example, in four-particle tree amplitudes beginning and ending on the same boundary the contact term divergence cancels  with a similarly divergent contribution from the Feynman diagram in which a propagator joins two three-point vertices located in the same boundary.

Note also  that the presence of $S_{CS}$ leads to a modification of the Bianchi identity for the four-form field strength, which has an anomalous boundary contribution of the form
\begin{eqnarray}
(d G)_{11 \mu\nu\rho\sigma}  = -6\left(\frac{\kappa_{11}^2}{\lambda^2}\right) \, &&\left[ \delta(x^{11})\left(\frac{1}{30} \Tr_1 F_{[ \mu\nu}F_{\rho\sigma]}- \frac{1}{2}\, \tr\, R_{[ \mu\nu}R_{\rho\sigma]}\right)\,\right.
\nonumber\\
&&\left.
 + \delta(x^{11} -\ciele)\,\left(\frac{1}{30} \Tr_2 F_{[ \mu\nu}F_{\rho\sigma]}- \frac{1}{2}\, \tr\, R_{[ \mu\nu}R_{\rho\sigma]}\right)\right]\,,    
\label{boundG}
\end{eqnarray}
which is necessary in order to ensure the cancellation of chiral gauge and gravitational anomalies.  

\sm

The second higher derivative gravitational term in $S_{boundary}$ in \eqref{sbound} is the four-derivative gravitational term that is quadratic in curvature tensors, and gives  higher derivative contributions to the two-graviton and three-graviton vertices. 
This $\curv^2$ term has an action of the form
\begin{eqnarray}
S_{\curv^2}=\frac{1}{8\lambda^2}\int_{\mathcal{M}_{11}} d^{11} x \sqrt{-G^{(11)}} \Big(\delta(x^{11})+\delta(x^{11}-\ciele)\Bigr) \, \curv_{\mu \nu \rho \sigma}\curv^{\mu \nu \rho \sigma} 
\label{grav3pt1}
\end{eqnarray}
(where $\curv $ is the Riemann curvature scalar and $\curv_{\mu\nu}$ is the Ricci tensor). In writing this expression we have set to zero the coefficients of  $\curv^2$ and $\curv_{\mu\nu}\, \curv^{\mu\nu} $ interactions, which vanish on shell and  can be removed by field redefinitions.  
The $\curv^2$ interaction  in \eqref{grav3pt1} does not vanish on shell -- it contributes to on-shell graviton two-point functions localised on the boundaries.   The gravitons carry arbitrary Kaluza--Klein momentum, $p_{11}$ while the tangential momentum $k_\mu$ is conserved.    It also contributes an on-shell three-graviton vertex localised on the boundary.
These contributions to the amplitude will be discussed in the next section.

\sm

The final boundary interaction between bulk bosons that contributes to two-particle and three-particle amplitudes has the form (see \cite{Gross:1986mw}) 
\begin{eqnarray} 
S_{(\partial\,H)^2} =  \frac{1}{8\lambda^2}\int_{\mathcal{M}_{11}} d^{11} x \sqrt{-G^{(11)} } \Big(\delta(x^{11})+\delta(x^{11}-\ciele)\Bigr) \,  \left(\partial^\mu H^{\nu\rho\sigma}\, \partial_\rho H_{\sigma\mu\nu} + \partial_\mu H^{\mu\rho\sigma}\, \partial^\nu H_{\nu\rho \sigma}\right)  \,,
\nonumber\\ 
\end{eqnarray} 
where $H_{\mu\nu\rho} = \partial_{[\mu} C_{\nu\rho] 11}$.  This interaction again contributes non-vanishing on-shell two-point and three-point functions with a pair of external on-shell bulk $C$-fields, as will also be discussed in the next section.

\sm

We note, for future reference that according to Ho\u rava and Witten the gravitational coupling, $\kappa_{11}$ and the gauge coupling, $\lambda$ are related  to the eleven-dimensional Planck length by
\begin{eqnarray}
2\kappa_{11}^2 =(2\pi)^8 \, \lepl^9\, \qquad \lambda^2 =2\pi (4\pi \kappa_{11}^2)^{2/3} = (2\pi )^7\, \lepl^6\,.
\label{gravigauge}
\end{eqnarray}

\subsection{Bulk and boundary to boundary gravity propagators}
\label{subsec:prop}
 
Before considering the scalar propagator on an orbifold we recall the form of the momentum space propagator on a circle of radius $\lepl\raele$, which is given by
\begin{eqnarray}
D(p^2, m) =\frac{1}{p^2+p_{11}^2}\,,
\label{propdef}
\end{eqnarray}
where the eleven-dimensional momentum $(p^\mu, p^{11})$ has components  $p^\mu$ in the ten Minkowski space dimensions and the momentum in the $x^{11}$ direction is quantised in units of $\lepl\raele$
\begin{equation}
\label{momeleven} 
p_{11} = \frac{m}{\raele\, \lepl} = \frac{m\, \pi}{\ciele} \,.
\end{equation} 
 The norm $p^2= p_\mu p^\mu$   ($\mu=1,2,\dots,10$) is with respect to the 
Minkowski metric ${\rm diag}(-, + ,\dots, +)$.     
The propagator  between points $x^{11}$ and $y^{11}$  on $S^1$  is  given by the Fourier sum
\begin{eqnarray}
	G(p; x^{11}-y^{11})= \frac{1}{2 \ciele}\sum_{m=-\infty}^{\infty}e^{ i\, p_{11}\,(x^{11}-y^{11}) }\,D(p^2, m) \,,
\label{Iprop5}
\end{eqnarray}
where the Minkowski space directions $1,2, \dots, 10$ have been left in momentum space.

\sm

 We are interested in formulating Feynman rules in the presence of the $\mathbb{Z}_2$ orbifold boundary conditions at $x^{11} = 0$ and $x^{11}+L$, as was considered in \cite{Dasgupta:2000xu}, in which case the momentum conjugate to $x^{11}$ is quantised as given in \eqref{momeleven}. The  scalar propagator between two points in the orbifold direction is obtained by 
imposing the additional orbifold boundary conditions, which require the propagator to be invariant under $x^{11} \to - x^{11}$.  This is achieved by identifying the propagator on the orbifold as the combination 
\begin{eqnarray}
	\mathcal{G}(p; x^{11},y^{11}) &=&	G(p; x^{11}-y^{11})+	G(p; x^{11}+y^{11})
\nonumber\\
&=& \frac{1}{2 \ciele}\sum_{m=-\infty}^{\infty}  \,\frac{1}{p^2+p_{11}^2} \left( e^{i p_{11} (x^{11}-y^{11}) } + e^{i p_{11}(x^{11}+y^{11}) } \right) 
\nonumber\\
	&=& \frac{1}{\ciele}\sum_{m=-\infty}^{\infty}  \,\frac{1}{p^2+p_{11}^2} \cos\left(p_{11} x^{11}\right) \,\cos\left(p_{11}y^{11}\right)  \,.
\label{Iprop6}
\end{eqnarray} 

In considering loop amplitudes later it will be useful to note the following product relations
\begin{eqnarray}
	\int_0^{\ciele } dy^{11} \mathcal{G}(p_1; x^{11},y^{11})\, 	\mathcal{G}(p_2; y^{11},z^{11})	= \frac{1}{\ciele}  \sum_{m=-\infty}^{\infty}  \left(\prod_{r=1}^2\frac{1}{p_r^2+p_{11}^2}\right)  \,   \cos\left(p_{11} x^{11}\right) \,\cos\left(p_{11}z^{11}\right) \,,\nonumber\\ 
\label{productg}
\end{eqnarray}
and 
\begin{eqnarray}
	&&\int_0^L dx^{11}  \int_0^Ldy^{11}  \int_0^Ldz^{11} \int_0^L dw^{11} \,	\mathcal{G}(p_1; x^{11},y^{11})	\,
	\mathcal{G}(p_2;y^{11},z^{11}) \,	\mathcal{G}(p_3; z^{11},w^{11})	\, \mathcal{G}(p_4; w^{11},x^{11})	
\nonumber\\	
&&=\frac{1}{2}\left[\prod_{r=1}^4
\left(\frac{1}{p_r^2}\right)+	\sum_{m=-\infty}^{\infty}
\prod_{r=1}^4\left(\frac{1}{p_r^2+p_{11}^2}\right)
\right]
\label{productfour} 
\end{eqnarray}
We will later be interested in supergraviton propagators that are constrained to begin and end on either boundary, which involve the following two slightly different expressions for the propagator.

\subsection*{Endpoints on  the same boundary}
Setting
$x^{11}=0=y^{11}$ in \eqref{Iprop6} gives
\begin{eqnarray}
\mathcal{G}(p^M; 0,0)&=&\frac{1}{ \ciele } \sum_{m=-\infty}^{\infty}\frac{1}{p^2 + p_{11}^2}=\frac{1}{ \pi \lepl \raele }\sum_{m=-\infty}^{\infty}\int_0^\infty  d\sigma e^{-\sigma(p^2 +p_{11}^2)}
\nonumber\\
&=&\frac{1}{\sqrt \pi }\sum_{n=-\infty }^{\infty}\int d\sigma \sigma^{-\frac{1}{2}}e^{-\sigma p^2-\frac{\ciele^2}{\sigma}n^2 }
\label{Iprop8}
\end{eqnarray}
where the last step involves a Poisson summation.  This transforms the sum over Kaluza--Klein modes of charge $m$  into the sum of windings of the propagator around $x^{11}$ with winding number $n$.  The integral can be performed explicitly, giving
\begin{eqnarray}
\mathcal{G}(p^M; 0,0)&=&\frac{1}{\sqrt{p^2}}\sum_{n=-\infty}^{\infty}e^{-|2n|\sqrt{p^2}\ciele }
=\frac{1}{\sqrt{p^2}}\left(1+ 2\sum_{n=1}^{\infty}e^{-|2n|\sqrt{p^2}\ciele }\right)
\nonumber\\
&=&\frac{\cosh(\sqrt{p^2}\ciele)}{\sqrt{p^2}\ \sinh(\sqrt{p^2}\ciele) }=\frac{1}{\sqrt{p^2}\ \tanh(\sqrt{p^2}\ciele) }\,.
	\label{Iprop2}
\end{eqnarray}
The propagator $\mathcal{G}(p^M; \ciele,\ciele)$ is given by the same expression.

\subsection*{Endpoints on different boundaries}

Setting $x^{11}=0$ and $y^{11}=\ciele$ in \eqref{Iprop6} gives  
\begin{eqnarray}
\mathcal{G}(p^M; 0, \ciele)&=&\frac{1}{\ciele} \sum_{m=-\infty}^{\infty}\frac{(-1)^m}{p^2 +p_{11}^2}= \frac{1}{\ciele} \sum_{m=-\infty}^{\infty}(-1)^m\int_0^\infty  d\sigma e^{-\sigma\left( p^2 +p_{11}^2\right)}
\nonumber\\
&=&\frac{1}{\sqrt \pi }\sum_{n=-\infty }^{\infty}\int d\sigma \sigma^{-\frac{1}{2}}e^{-\sigma p^2-\frac{\ciele^2}{\sigma}(n+\frac{1}{2})^2 }\,,
\label{Iprop7}
\end{eqnarray}
where the last step again follows by Poisson summation.   Once again, performing the integral gives the expression
\begin{eqnarray}
\mathcal{G}(p^M; 0,\ciele)&=&\frac{1}{\sqrt{p^2}}\sum_{n=-\infty}^{\infty}e^{-|2n+1|\sqrt{p^2} \ciele  }
\nonumber\\
&=& \frac{1}{\sqrt{p^2}\ \sinh(\sqrt{p^2} \ciele) }\,.
\label{Iprop1}
\end{eqnarray}
The  propagators of the bosonic supergravity fields, the metric and three-form field,  involve extra numerator factors arising from their spin.  These were presented in detail in  \cite{Dasgupta:2000xu} and were important in determining the tree diagram with a propagator joining the boundaries.  In subsequent sections we will make use of the expressions in \cite{Dasgupta:2000xu} in  order to avoid repeating those details here. 

\sm

It will prove convenient to introduce a matrix notation for the various propagators terminating on the Hora\u va--Witten walls by defining the components of a matrix $D_{ij}$ ($i,j=1,2$) by
\begin{subequations}
\begin{eqnarray}
D_{11}(p^2, m) &=& D_{22}(p^2, m)  = \mathcal{G}(p^M; 0,0)= \mathcal{G}(p^M; \ciele,\ciele) 
  \label{dmatrixdefa}
\\
\nonumber\\
  D_{12}(p^2, m)  &=& D_{21}(p^2, m) = \mathcal{G}(p^M; 0,\ciele)\,.
  \label{dmatrixdefb}
\end{eqnarray}
\end{subequations}

\sm

We may now begin to consider the Feynman diagrams that describe gauge and graviton scattering amplitudes in the 
Ho\u rava--Witten  supergravity background.  We will begin with the simplest cases of two-point and three-point functions (which only receive contributions from the Chern--Simons and $\curv^2$ interactions). 

\subsection{On-shell boundary two-point and three-point functions}
\label{sec:boundint}

\subsubsection*{Yang--Mills/graviton three-point function}
The Yang--Mills action, $S_{YM}$ \eqref{sym} that is localised on the boundaries contributes to the usual ten-dimensional interactions between gauge bosons and gravitons, together with the three-point interaction between Yang-Mills states and  a graviton carrying non-zero  $p_{11}$.  The on-shell three-point function between Yang--Mills states and a bulk graviton polarised in directions parallel to the boundaries is given by 
 \begin{eqnarray} 
A^{\textrm{YM grav}}= \epsilon^{(1)\mu_1}\zeta^{(2)\mu_2\nu_2}\epsilon^{(3)\mu_3}\
\left(\eta_{\mu_3\mu_1}k^{(3)}_{\mu_2}+\eta_{\mu_2\mu_3}k^{(2)}_{\mu_1}+\eta_{\mu_1\mu_2}k^{(1)}_{\mu_3}\right)k^{(1)}_{\nu_2}
\textrm{tr}_{\textrm{v}}\left(T^{a_1}T^{a_3}\right)\,.
\label{ymgravvert}
\end{eqnarray} 
In this expression the  external gauge bosons have null Minkowski momenta, $k^{(r)}_\mu$ ($r=1,3$), with $|k^{(r)}|=0$ and  polarisation vectors, $\epsilon^{(r)}_{\mu}$ satisfy $k^{(r)}_\mu \, \epsilon^{(r)\mu}=0$.   The external graviton has eleven-dimensional  momentum $p^{(2)} = (-(k^{(1)} + k^{(3)})_\mu, p^{(2)}_{11})$ and symmetric polarisation tensor $\zeta^{(2)[\mu\nu]}$, which satisfies $(k^{(1)} + k^{(3)})_\mu \zeta^{\mu \nu}=0$.  When viewed from ten dimensions this bulk state has Kaluza--Klein mass given by (mass$)^2 = (p^{(2)}_{11})^2$, which follows from the  eleven-dimensional  massless condition, $(p^{(2)})^2=0$, 
Before compactification the matrix $T^{(1)}$ is a generator in the adjoint of the $E_8$ gauge group associated with the gauge particle labelled $r$.   After compactification  on $S^1$ with the insertion of an appropriate Wilson line, the gauge group is broken to $SO(16)$.  In that case the external Yang--Mills states are either massless $SO(16)$ gauge particles or massive $SO(16)$ spinor states, as will be discussed in more detail in section~\ref{sec:YMtrees}.

\subsubsection*{Yang--Mills Chern-Simons three-point function}

The linearised Yang--Mills Chern-Simons interaction in $S_{CS}$ \eqref{boundint} is given  by  
\begin{eqnarray}
\frac{\lepl^2}{2\raele}A_\mu^A\partial_\nu A_\rho^A\Big(\partial^\mu  C^{\nu \rho 11}+ \partial^{\nu} C^{\rho \mu 11}+ \partial^\rho C^{\mu \nu 11} \Big)\,.
\end{eqnarray}
 Recall that the ten-dimensional antisymmetric tensor potential is identified with the zero Kaluza--Klein charge of the three-form, $B_{\mu\nu} = C_{\mu\nu 11}\left|_{p_{11} = 0} \right.$.  This gives rise to a three-point function of the same  form as \eqref{ymgravvert} but with an antisymmetric polarisation tensor wave function for the $C_{\mu\nu11}$ field, so that  
 \begin{eqnarray} 
A^{\textrm{YM CS}}= \frac{\lepl^2}{\raele} \Big[(\epsilon^{(1)} k^{(2)})(k^{(3)}\zeta^{(2)}\epsilon^{(3)})
	+(\epsilon^{(3)} k^{(2)})(k^{(1)}\zeta^{(2)}\epsilon^{(1)})\Big]\textrm{tr}\Big(T^{(1)}T^{(2)}\Big)\,.
\label{ymcsvert}
\end{eqnarray}  

\sm

Although knowledge of the bosonic components  of the propagators and vertices is sufficient to construct the gauge and graviton  tree diagrams,  when we discuss loop diagrams we will make implicit use of a supersymmetric formalism.  This will be based on an extension of the first-quantised light-cone formalism used to describe the $S^1$ compactification of eleven-dimensional supergravity \cite{Green:1999by} to the Ho\u rava--Witten background.  The  linearised forms of the interactions  in $S$ \eqref{totact} arise in this formalism as expectation values of vertex operators that  describe the emission of single particle states from a world-line in a manner that is modelled on the vertex operator construction of light-cone gauge closed superstring theory.  For present purposes we need to extend the formalism to include vertex operators localised in either boundary, acting on states that may be in either the boundary Yang--Mills supermultiplet or the components of the bulk graviton that couple to the boundary.  For example,   
the three-point interactions between a pair of Yang--Mills particles and the bulk graviton or $C$-field  are given by a matrix element of the form
\begin{eqnarray}
A^{\textrm{YM bulk}}= 
\langle\epsilon^{(1)},k^{(1)}|V^{YM}_{bulk}(k^{(2)},\zeta^{(2)})
|\epsilon^{(3)},k^{(3)}\rangle\, \tr\Big(T^{(1)}T^{(2)}\Big) \,,
\label{firstcs}
\end{eqnarray}    
where $V_{bulk}(k^{(2)},\zeta^{(2)})$ is a vertex operator describing the emission of an on-shell bulk state from the world-line of a Yang--Mills superparticle  embedded in light-cone superspace. The emitted state is a  graviton when $\zeta^{(2)\mu\nu}$ is symmetric and an on-shell $C_{\mu\nu 11}$ state when it is antisymmetric. This reproduces the three-particle interactions in  \eqref{ymgravvert} and \eqref{ymcsvert}.  This vertex operator is a function of bosonic and fermionic light-cone superspace coordinates $x^i$ and $S^A$, which are similar to the zero modes of the coordinates that enter the light-cone description of the heterotic string.  Since we will only use very general features of this formalism we will not present the details, which can be reconstructed from \cite{Green:1999by}.

\sm

\subsubsection*{Lorentz Chern-Simons, $\curv^2$ and $(\partial H)^2$ interactions}

The on-shell amplitude for three bulk tensor bosons  (the graviton or $C$-field) interacting on a boundary is given by 
\begin{eqnarray} 
A^{bulk}\left|_{boundary} \right.=
\left(
t_{\mu_1 \mu_2 \mu_3}+
\frac{\lepl^2}{\raele}
k^{(1)}_{ \mu_3}k^{(2)}_{\mu_1}k^{(3)}_{\mu_2}\right)
t_{\nu_1 \nu_2 \nu_3}
\zeta^{(1)\mu_1\nu_1}\zeta^{(2)\mu_2\nu_2}\zeta^{(3)\mu_3\nu_3}\,,
\label{gravcsvertvert}
\end{eqnarray} 
where 
$t_{\nu_1 \nu_2 \nu_3}$  is given by 
\begin{eqnarray}
	t_{\mu_1 \mu_2 \mu_3}=\eta_{\mu_1\mu_2}k^{(1)}_{\mu_3}+\eta_{\mu_2\mu_3}k^{(2)}_{\mu_1}+\eta_{\mu_3\mu_1}k^{(3)}_{ \mu_2}	\,,
\end{eqnarray}
 and we have specialised to the situation in which the only non-zero components of both the metric tensor and the $C_{\mu\nu 11}$ field are those parallel to the boundaries. 
 
 \sm
 
\subsubsection*{\it The Lorentz Chern--Simons interaction}
 
When one of the  three polarisation tensors is antisymmetric in $(\mu,\nu)$ the $t_{\dots} t_{\dots}$ term in \eqref{gravcsvertvert} does not contribute and this expression describes the  linearised gravitational Chern--Simons interaction in $S_{CS}$ in \eqref{boundint}.  This is given by 
\begin{eqnarray}
A^{CS\, grav.}=\frac{\lepl^2}{\raele }\ t_{\nu_1 \nu_2 \nu_3}\,
k^{(1)}_{ \mu_3}k^{(2)}_{\mu_1}k^{(3)}_{\mu_2}\,
\zeta^{\{\mu_1\nu_1\}}_1\zeta^{[\mu_2\nu_2]}_2\zeta^{\{\mu_3\nu_3\}}_3\,,
\label{gravcsvertverty}
\end{eqnarray}
where state number 2 is the $C$-field and the other two states are gravitons (and $\zeta^{\{\mu \nu\}}$ denotes a symmetrised tensor).

\sm

This interaction is reproduced by replacing the Yang--Mills states in \eqref{ymcsvert} by graviton states (which entails replacing the gauge states and their colour factors by the tensor graviton states) and once again using an antisymmetric polarisation tensor for particle 2.  The resulting matrix element is
\begin{eqnarray}
A^{CS\, grav.} = \langle\zeta^{(1)},k^{(1)}|V_{bulk}(k^{(2)}, \zeta^{(2)})
|\zeta^{(3)},k^{(3)}\rangle  \,.
\label{firstgrav}
\end{eqnarray}
  
\sm
 
\subsubsection*{\it The $\curv^2$ and $(\partial\, H)^2$ interactions}

The two-point functions for gravitons may be obtained by linearising the curvature tensors in \eqref{grav3pt1}.  The two gravitons have momenta $(k_\mu, p_{11}^{(1)})$ and $(-k_\mu, p_{11}^{(2)})$, where we have used conservation of the ten-dimensional momentum ($k^{(1)} = - k^{(2)}= k$).  The mass-shell condition requires that  $k^2 + (p_{11}^{(1)})^2 =k^2 + (p_{11}^{(2)})^2=0 $ so that $p_{11}^{(1)} = \pm p_{11}^{(2)}$.  This leads to a two-particle on-shell vertex given by
\begin{eqnarray} 
A^{2h} = \frac{1}{2(2\pi)^9\lepl^6 } \zeta^{(1)}_{ \mu\nu} \zeta^{(2)\mu\nu} \, (k^{(2)})^2 \,,
\label{lineartwo}
\end{eqnarray} 
where we have used the physical state condition $k_\mu^{(r)} \zeta^{(r)\mu\nu}=0$ and $\zeta^{(r)\mu\nu}$ is symmetric for external gravitons.  The  $(\partial \, H)^2$ interaction gives rise to the same on-shell two-point function with the graviton polarisations replaced by antisymmetric $\zeta^{(1)\mu\nu}$ and $\zeta^{(2)\mu\nu}$, which are the $C^{\mu\nu 11}$ polarisation tensors. 

\sm

The three-point functions for three gravitons or two $C$-states and one graviton can be extracted from  \eqref{gravcsvertvert}  by a suitable choice of polarisation tensors.  The $t_{\dots} t_{\dots}$  term simply reproduces the gravitational interaction of a graviton with two gravitons or with a pair of $C$'s.   The remaining part of the interaction in  \eqref{gravcsvertvert}  is quartic in momenta  and corresponds to  the three-field terms in the expansions of the effective interactions of the form $\curv^2$ and $(\partial\, H)^2$. 
The on-shell three-point function resembles that discussed in \cite{Garousi:1996ad,Bachas:1999um, Garousi:2006zh}  in the context of the interactions of string ground states in the background of the D8-brane.

\sm

These three-point functions are reproduced in the world-line first quantised formalism by replacing the antisymmetric polarisation tensor $\zeta^{(2)}$ in \eqref{firstgrav} by the symmetric  graviton polarisation.  Choosing the external states to either be gravitons or $C$-states leads to the two kind of three-point functions described in the previous paragraph.

\sm  
 
While the perturbative rules for constructing amplitudes follow from the local action as described above, the string theory interpretation requires an extrapolation of these rules into a regime in which perturbation theory may be questionable.  This is well-illustrated by the interpretation of the amplitudes that involve Kaluza--Klein modes in the $x^{11}$ direction (the orbifold direction) that have masses given by  $p_{11}=m/(\lepl  \raele )$ ($m \in \mathbb{Z})$.   Such modes decay since $p_{11}$ is non conserved because of the orbifold boundary conditions.   The spectrum of particle states in the Ho\u rava--Witten background compactified on $\mathbb{M}^9 \times S^1/\mathbb{Z}_2 \times S^1$ is briefly reviewed in appendix~\ref{sec:particlestates}.

\sm

In later sections, we will encounter amplitudes in which these unstable Kaluza--Klein modes contribute to propagators in a manner that accounts for certain low order terms in the effective action that are presumably protected by supersymmetry.   In particular, we will encounter interesting  instanton terms in the  HO and type I amplitudes, where the instantons correspond to winding configurations of euclidean world-lines of  these non-perturbative states of the HE and type IA theories.

\section{Some features of four-particle Yang--Mills amplitudes in $\mathcal{N}=1$ string theories}
\label{sec:YMamps}  

 Before describing the Yang--Mills amplitude calculations in the Ho\u rava --Witten background we will summarise some features of the amplitudes that arise in different string theory limits.  The following are some of the features that we expect to reproduce.  Although most of these have been noted before (see, in particular, \cite{Tseytlin:1995fy,Tseytlin:2000sf}) there are some subtleties that will be explained in more detail in \cite{Rudra:2016xx}.

\begin{itemize}
\item Upon compacitifcation to nine dimensions on a circle of radius $\rahe=1/\raho$ (in the presence of appropriate Wilson lines) the $SO(16) \times SO(16)$ tree-level amplitudes in the HO and HE  theories are equivalent under T-duality, which equates $\rahe/\ghe^2$ with $\raho/\gho^2$.  The HO/HE tree level heterotic expression has a Yang--Mills pole  $\ghet^{-2}\,t_8\,\tr F^4/st$, where the symbol $\tr$  again indicates the trace in the  fundamental representation of either $SO(16)$ subgroup (and $g_{het}$ is either of the heterotic coupling constants).  This is produced by a disk diagram in the type I/IA theories.   The tree level terms with a graviton pole in the HO or HE theories have the form  $t_8(\tr F^2)^2/s$,  which is produced by an annulus (one-loop) diagram in the type I/IA theories.

\item  The first non-pole term in the low-energy expansion of the  tree amplitudes in both the heterotic theories has the form $g_{het}^{-2}\,t_8 (\tr F^2)^2$.  This arises as a {\it three}-boundary (i.e. {\it  two} open-string loop) term proportional to $\gib$  in the ten-dimensional type I theory (and vanishes in the ten-dimensonal type IA theory), which is a striking illustration of the way the perturbation expansion of the heterotic theory is reorganised by its type I parameterisation\footnote{  
The leading term in the low-energy expansion of the  three-boundary open string diagram naively has the form $s\, t_8(\tr_i F_i^2)^2$
but there is closed-string (graviton) propagator that cancels the factor of $s$, which accounts for the agreement with the heterotic expression.}. Note further that the open-string one-loop (annulus) contribution to  $t_8(\tr F^2)^2$ in the type I theory vanishes at large  $\raib$, which is consistent since otherwise S-duality would require a term of order $1/\gho$ in the HO theory, which does not exist.  However the annulus contribution to $t_8\,(\tr_i F_i^2)^2$ in the type IA theory (i.e., in the large $\raia$ limit)  is non-vanishing  and is dual to a one-loop term in the HE theory.

\item The next order in the low-energy expansion of the tree amplitudes in the HO theory contributes a term of the form  $\gho^{-2}\,\zeta(3) s\, t_8\tr F^4$   (where we have emphasised the occurrence of a $\zeta(3)$ coefficient).  Likewise, in the type I theory the tree-level disk diagram contributes $\gib^{-1}\zeta(3) s\, t_8\tr F^4$.  These two expressions ought to be related by HO/type I duality, but this cannot act term by term since it would transform the type I coefficient $\gib^{-1}$ into $\gho $, which is not a possible power of $\gho $.  These tree-level terms must be part of a function of the coupling that is transforms appropriately under the $S$ transformation ($S: \gib \to \gho= \gib^{-1}$), which can be viewed as a remnant of the $SL(2,\mathbb{Z})$ duality of the type IIB theory.  Furthermore, the next perturbative contribution to $s\, t_8\tr F^4$  in the HO theory is at one loop, whereas the one-loop term vanishes in type I and the next contribution is of order $\gib$ and is associated with world-sheets with the geometry of a torus with a single boundary.  This is the same order in $\gib$ as the two-loop disk diagram (which has three boundaries), which contributes to  $\gib\, t_8(\tr F^2)^2$  as stated earlier (but not to  $\gib\, t_8 \tr F^4$ or  $\gib  s\, t_8\tr F^4$ \cite{Rudra:2016xx}).   We will see in section~\ref{sec:higherloop} how these general features are reproduced by loop contributions to the four gauge particle amplitude, although since $s\, t_8\tr F^4$  is not protected from loop corrections, we do not expect our analysis to give the complete expression for its coupling constant dependent coefficient. 

\item The one-loop amplitudes for the two ten-dimensional heterotic theories are independent of the couplings and are  interchanged by the  identification $\raho = 1/\rahe$.  However, the low-energy limits of the theories look rather different.  In the large-$\rahe$ limit the leading behaviour in the HE theory is $t_8\left(\sum_{i=1,2} (\tr_i F_i^2)^2 -  \tr_1 F_1^2  \tr_2 F_2^2\right) $,  whereas in the large-$\raho$ limit of the HO theory the  leading behaviour is $\tr_{SO(32)} F^4= \tr_1 F_1^4 + \tr_2 F_2^4$ \cite{Ellis:1987dc}.  In both cases the traces are evaluated in the fundamental representation of either  $SO(16)$ sub-group.

\sm

The  type I theory has a $t_8\tr F^4$ tree (disk) interaction but does not generate a one-loop (annulus) contribution to $t_8\tr F^4$ even in  $D=9$ (i.e., for finite $\raib$) \cite{Rudra:2016xx}.  There is a contribution from the annulus diagram to an interaction that is suppressed by a power of $\raib$ of the form $t_8\,  (\sum_{i=1,2} ( \tr_i F_i^2)^2 -  \tr_1 F_1^2\,  \tr_2 F_2^2) /\raib$, which leads to a contribution to the type IA  theory at large $\raia$ \cite{Rudra:2016xx} that agrees with the expression in the HE theory.

\sm

Whereas in the HE and HO theories the $t_8(\tr F^2)^2$ interaction arises at tree level, in the ten-dimensional type I theory it arises at two loops  (i.e., at order $\gib$) and is unrenormalised.  

\item The preceding description of parity-conserving terms in the low-energy action has its counterpart in the parity-violating sector.  These parity-violating terms are important for ensuring the absence of chiral anomalies.  The absence of chiral anomalies in the ten-dimensional heterotic theories is attributed to the presence of anomaly cancelling terms   \cite{Green:1984sg}  of the form $B\wedge \xgs(F,\curv)\equiv \epsilon_{10}\, B \ygs (F,\curv)$, which arises as one-loop effects associated with the interaction of the Neveu--Schwarz/Neveu--Schwarz antisymmetric tensor with a total of four gauge bosons  and gravitons on a toroidal world-sheet.   The notation, which is reviewed in appendix~\ref{sec:A}, emphasises that these terms are related by supersymmetry to the parity conserving terms contained in $t_8\,\ygs $.

\sm

As explained in \cite{Horava:1996ma}, in the Ho\u rava--Witten description of the HE theory the eight-form  $\ygs$ is naturally expressed as the sum of three pieces\footnote{Note that since  anomaly cancellation is only of relevance in ten dimensions where $E_8$ is unbroken, the symbol $\Tr$ here refers to trace in the adjoint representation of the unbroken $E_8$ group.}
\begin{eqnarray}
\ygs(F_1,F_2,\curv) = 2\yvw(\curv)  +2 \left(\frac{1}{30}\Tr_1 F_1^2  - \half \tr \curv^2\right)^2 + 2 \left(\frac{1}{30}\Tr_2 F_2^2  - \half \tr \curv^2\right)^2 ,
\label{hegsterm}
\end{eqnarray}
where $\yvw(\curv)$ (implicitly defined by \eqref{vwdef} and \eqref{vwterm})  comes from the bulk interaction, and $(\frac{1}{30}\Tr_i F_i^2 - \half \tr \curv^2)^2$ arises as an effect of either boundary (labelled $i$).\footnote{\label{fn:6} The definition of $\xgs$ in \cite{Horava:1996ma} is a  factor of $8$ greater than our expression   and  the definition of $\xvw$ is a factor of $4$ greater than our expression.}  Upon compactification this has coefficient $\rahe= 1/\raho$ and therefore vanishes in the large $\raho$ limit. The HO anomaly-cancelling interaction  in the large-$\raho$ limit, where the gauge group $SO(32)$ is unbroken, is a one-loop interaction of the form $\epsilon_{10}\, B\, \ygs (F,\curv)$, where
\begin{eqnarray}
\ygs(F,\curv) =8\, \tr_{SO(32)} F^4-\tr F^2 \tr \curv^2+\left(\tr \curv^4 + \frac{1}{4}(\tr \curv^2)^2\right) 	\,.
\label{hogsterm}
\end{eqnarray}
This is related by S-duality to a similar expression in the type I theory which is associated with an amplitude coupling a Ramond--Ramond $B$-field and a total of four gravitons and gauge bosons to a disk world-sheet.

\end{itemize}

\sm

In the course of indicating how these features arise from the analysis of supergravity coupled to Yang--Mills  in the Ho\u rava --Witten background we will be led to several insights into possible non-perturbative effects that seem to be required required for their consistency.

\section {Yang--Mills four-particle tree amplitudes}
\label{sec:YMtrees}

We will here consider some contributions of tree-level four-particle gauge  amplitudes, which are relatively straightforward to evaluate. 
The external scattering states have null Minkowski momenta, $k^{(r)}_\mu$ ($r=1,2,3,4$), with $k^{(r)}\cdot k^{(r)}=0$.   The Yang--Mills polarization vectors, $\epsilon_{\mu}$ satisfy $k_\mu \, \epsilon^\mu=0$.   The resulting amplitude is proportional to a function of the Mandelstam invariants multiplying four powers of the linearised field strength, $(\hat F_{\mu,\nu})_{A B}$ ($\mu,\nu =1, \dots,10$ and $A\,,B=1, \dots,496$), which belongs to the $496$-dimensional adjoint representation of $E_8\times E_8$.  The  Mandelstam invariants are defined by 
\begin{eqnarray}
s =- (k^{(1)}+ k^{(2)})^2\,,\qquad  t= - (k^{(1)} +k^{(4)})^2\,,\qquad  u   = -  (k^{(1)}+ k^{(3)})^2\,.
\label{mandelstamdef}
\end{eqnarray}
The linearised field strength for the particle labelled $r$ has the form\footnote{Here and in the following we indicate a linearised approximation by a hat.}
\begin{eqnarray}
(\hat F^{(r)}_{\mu \nu})_{B C} &\equiv& \hat F_{\mu\nu}\, \epsilon^{(r)A_r}\,T^{A_r}_{BC}  \nonumber\\
&=&  (k^{(r)}_\mu\, \epsilon^{(r)}_\nu - k^{(r)}_\nu\, \epsilon^{(r)}_\mu)\, \epsilon^{(r)\,A_r}\, T^{A_r}_{BC}  \,.
\label{gravipol}
\end{eqnarray}

\sm

In the uncompactified ten-dimensional gauge theory $T^{A}$ ($A=1,\dots,496$)   are $ 496\times 496 $-dimensional matrices in the adjoint representation of the Lie algebra of $E_8 \times E_8$ and $\epsilon^{r\, A_r}$ is the polarization vector in the internal gauge group space, which specifies the quantum numbers of the particle labelled $r$. This may be written as the sum of the field strength in each $E_8$ factor of $E_8 \times E_8$ in the form
\begin{eqnarray}
F =F_1 \oplus F_2\, ,
  \label{gaugef1}
\end{eqnarray}
where $F_1$ and $F_2$ are $248 \times 248$ matrices in the adjoint representation of each of the $E_8$'s (and two-forms in $\mu,\nu$).  Denoting the generators of the adjoint representation for the $E_8$ labelled $i =1,2$ by the matrices $(T_i^A)_{BC}$ (where $A,B,C= 1,\dots,248$), with $[T_i^A, T_j^B] =\delta_{ij}\, f^{AB}_{\ \ \ C} \, T_i^C$, we have
\begin{eqnarray}
 F_1= F_1^A\, T_1^A\, , \qquad F_2 =  F_2^A\, T_2^A\, ,
  \label{gaugef2}
 \end{eqnarray}
and so
\begin{eqnarray}
\Tr_{E_8\times E_8} F^4 =\frac{1}{100}\, \Tr_1(F_1^2)\, \Tr_1 (F_1^2)\,+  \frac{1}{100}\,\Tr_2(F_2^2)\, \Tr_2 (F_2^2) \,,\label{eeighttrace}
\end{eqnarray} 
where we have used
\begin{eqnarray}
\Tr_i (F_i^4) =\frac{1}{100}\, \Tr_i(F_i^2)\, \Tr_i (F_i^2)\,,
\label{fouthrelate}
\end{eqnarray}
for $i=1,2$.

\sm

In the following we will discuss dualities that relate amplitudes in the heterotic and type I/IA  theories when one direction is compactified on a circle of radius $\lepl \raten$.    
 These connections are most straightforward for amplitudes in which the scattering gauge particles in nine dimensions are in a $SO(16)\times SO(16)$ subgroup of  $E_8\times E_8$, which transforms under T-duality along the $x^{10}$ direction  into the same subgroup of $Spin(32)/\mathbb{Z}_2$. 

\sm
 
 The  breaking of $E_8\times E_8$ is achieved by considering Wilson lines in both boundaries.  Recall that the $E_8$ adjoint weights comprise the union of the $SO(16)$ adjoint weights and the $SO(16)$ spinor weights.  The Wilson line that breaks a boundary $E_8$ gauge symmetry to $SO(16)$ is an element of the Cartan subalgebra of the form  \cite{Ginsparg:1986bx} 
\begin{eqnarray}
A^I_{E_8} =\frac{1}{\lepl \raten} {\rm diag}(1, 0^7)\,,
\label{cartane8}
\end{eqnarray}
where $I$ is the index labelling the Cartan sub-algebra.
The compactified theory then contains a tower of Kaluza--Klein $SO(16)$ adjoint states with square of the masses 
\begin{eqnarray}
\frac{n^2}{\lepl^2\raten^2}= \frac{n^2}{\lesth^2 \rahe^2}\,=\frac{n^2}{\lesti^2 \gia^2}\,,
\label{adjointhe}
\end{eqnarray}
with integer $n$, which includes the massless gauge  potentials. These correspond to D-particles in type IA theory \cite{Kachru:1996nd}. There is also a tower of massive $SO(16)$ spinor states with square of the masses given by 
\begin{eqnarray}
\frac{(n-1/2)^2}{\lepl^2\raten^2}=\frac{(n-1/2)^2}{\lesth^2 \rahe^2 }=\frac{(n-1/2)^2}{\lesti^2 \gia^2}\,,
\label{spinormass}
\end{eqnarray} 
which  correspond to {\it stuck} D-particles \footnote{In type IA theory, a single D-particle is necessarily stuck to the O8 planes and it is sub-threshold BPS bound state.} in type IA theory \cite{Kachru:1996nd}. In considering the duality between the HE and HO theories we need to break both $E_8$ subgroups to $SO(16)$, which involves the  Wilson line in $E_8 \times E_8$ of the form $A^I_{E_8\times E_8} = {\rm diag}(1, 0^7,1, 0^7)/(\lepl\raten)$, which gives masses to the $SO(16)$ spinor states in both subgroups.   
Although the massive $SO(16)$ spinors will not be relevant to the gauge theory trees with massless external gauge states that we will consider below, they are an essential ingredient in the discussion of loop amplitudes in section~\ref{sec:loopamp}.

\sm

The loop amplitude discussion in appendix~\ref{sec:holoopamp} will also involve a discussion of the  breaking of the HO theory with  $SO(32)$ broken to $SO(16)\times SO(16)$.  We are here interested in the limit of M-theory of relevance to the HO string, which is the limit in which $ \raho \lesth= \lepl /(\raten \raele)\to \infty$.  In this case the Wilson line is the element of the Cartan subalgebra of $SO(32)$ of the form \cite{Ginsparg:1986bx} 
\begin{eqnarray}
A^I_{ho} =\frac{1} {\raho \lesth} {\rm diag}\left(\half^8, 0^8 \right) = \frac{\raten \raele}{\lepl} {\rm diag}\left(\half^8, 0^8 \right) \,,
\label{cartanho}
\end{eqnarray}
and gives rise to Kaluza--Klein tower of massive bi-fundamental states of $SO(16) \times SO(16)$   with masses given by 
\begin{eqnarray}
\frac{\raten^2 \raele^2}{\lepl^2}   (n-1/2)^2 =
\frac{(n-1/2)^2}{\lesth^2\raho^2}\,,
\label{bifunmass}
\end{eqnarray}
in addition to the Kaluza--Klein tower of adjoint states with the masses given in \eqref{adjointhe} (with $\rahe$ replaced by $\raho$).

\sm

An important point to note in considering the following expressions for scattering amplitudes is that the momentum conservation delta functions will not be explicitly included in the amplitudes.  However, 
 in the compactified theory momentum conservation in the compact $x^{10}$ dimension   involves the replacement of the continuous momentum conservation delta function 
$\delta(\sum_{r=1}^4 k^{(r)})$ by  $2\pi \raten \lepl \,\delta_{\sum_{r=1}^4 l^{(r)}}$, where the Kronecker delta imposes conservation of the discrete (Kaluza--Klein) momenta of the external particles.  Although in this paper we will be setting $l^{(r)}=0$ it is obviously important to keep the volume factor, $2\pi\lepl  \raten $.   In the following this factor will always be included in the expression for a compactified amplitude.  

\subsection{The Yang--Mills tree amplitude in a single boundary}

\begin{figure}[h]
\begin{center}
\begin{tikzpicture}[scale=1.0]
\begin{scope}[shift={(0,0)}]
\draw [hwbou](-2,-3) -- (-2,2);
\draw [hwbou] (-2,2)-- (0,3);
\draw [hwbou] (-2,-3)-- (0,-2);
\draw [hwbou] (0,3) -- (0,-2); 
\begin{scope}[shift={(-1,-.2)}]
\draw [suym, ultra thick](0,0) -- (0,.85);
\draw [suym, ultra thick](0,0) -- (-.3,-.65);
\draw [suym, ultra thick](0,0) -- (.4,-.45);
\begin{scope}[shift={(0,.85)}]
\draw [suym, ultra thick](0,0) -- (-.3,.65);
\draw [suym, ultra thick](0,0) -- (.4,.45);
\end{scope}
\end{scope}
\end{scope}
\begin{scope}[shift={(5,0)}]
\draw [hwbou](-2,-3) -- (-2,2);
\draw [hwbou] (-2,2)-- (0,3);
\draw [hwbou] (-2,-3)-- (0,-2);
\draw [hwbou] (0,3) -- (0,-2);
\end{scope}
\end{tikzpicture}
\end{center}
\caption{ The Yang--Mills tree amplitude  localised on one boundary (i.e. in a single $E_8$).   }
\label{fig:ymtree1}
\end{figure}
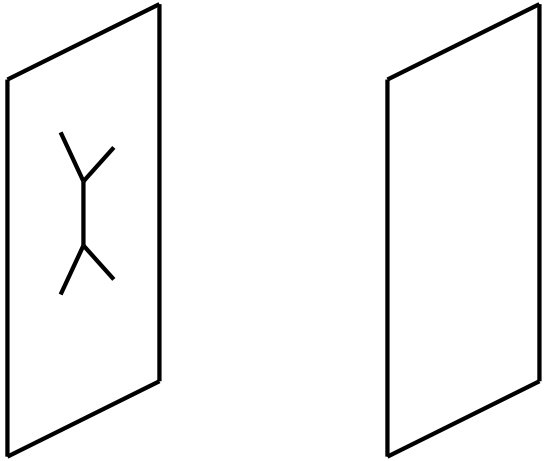

In the following we will discuss Yang--Mills four-point amplitudes in which pairs of particles may be in either of the Ho\u rava --Witten boundaries, and therefore in either of the $SO(16)$ subgroups of the two $E_8$'s.  The amplitude will therefore be written as a matrix, $A_{ij}=A_{ji}$, where $i,j=1,2$ label the two subgroups.

\sm

The simplest example is the sum of tree amplitudes for  the scattering of four gauge bosons that are all in a particular $SO(16) \subset E_8$ sub-group of $E_8\times E_8$ associated with one of the ten-dimensional boundaries, compactified on a circle in the $x^{10}$ direction.   The lowest order contribution  is given by  the sum of   poles in $s$, $t$ and $u$ channels, each corresponding to the  propagation of an intermediate gauge boson.   If the gauge particles are in the $E_8^{(1)}$ factor (where the superscript indicates which boundary we are considering),  the amplitude is proportional to
\begin{eqnarray}
A_{11}^{\textrm{YM-pole}}&=&\frac{\raten}{(2\pi)^5 \lepl^5}\,\left[ \frac{1}{tu}\  \, \tr (T_1^{(1)}\, T_1^{(2)}\, T_1^{(3)}\, T_1^{(4)}) +  \frac{1}{us}\  \, \tr( T_1^{(1)}\, T_1^{(4)}\,  T_1^{(2)}\, T_1^{(3)})  \right. \nonumber \\ 
&&\qquad \qquad \qquad \left. +\frac{1}{st}\  \, \tr( T_1^{(1)}\, T_1^{(3)}\, T_1^{(2)}\, T_1^{(4)}) \right] \, t_8 \hat F^4 \,,
\label{treeinone}
\end{eqnarray}
where the superscript ${}^{(r)}$  ($r=1,2,3,4$)  labels the scattering particle, $T_1^{(r)} = \epsilon^{r\,A_r}\, T_1^{A_r}$ encodes the colour dependence and $T_1^{A_r}$ is a matrix in the $16\times 16$ representation of $SO(16)$.  The coefficient is proportional to $\kappa_{11}^2/\lambda^2$ together with the factor of $2 \pi\, \raten\, \lepl$ to account for the compactification of the tenth dimension.  Clearly, the expression for the amplitude localised in the other boundary is  $A_{22}^{\textrm{YM-pole}}$ is obtained by replacing  $T_1^{(r)}$ by $T_2^{(r)}$.
The expression \eqref{treeinone} simply reproduces the tree-level Yang--Mills amplitude in the low-energy limit of any of the $\mathcal{N}=1$ string theories by the following  straightforward interpretation of the M-theory parameters  in terms of those of the heterotic and the type I/IA superstring theories, as follows.  

\sm

The identities in \eqref{Idua1}-\eqref{Idua3} imply that the coefficient in the HE description is $\lesth^{-5}\, \rahe \, \ghe^{-2}$ and so \eqref{treeinone}  is proportional to the leading term in the low-energy expansion of the Yang--Mills tree amplitude in the HE theory.  The fact that this is proportional to $\rahe$ implies that it has has a sensible ten-dimensional limit as $\rahe\to \infty$.

\sm

Since  T-duality implies that  $\rahe/\ghe^2= \raho/\gho^2$ the expression \eqref{treeinone} transforms consistently to the corresponding expression for the scattering of gauge particles in the $SO(16)\times SO(16)$ subgroup of $SO(32)$ in the  HO theory.

\sm

Similarly, using \eqref{Idua3} the prefactor in \eqref{treeinone}  is interpreted in terms of the parameters of the type IA theory by noting that $ \lepl^{-5} \,\raten= (\lesti)^{-5} \gia^{-1}$, describing the four gauge boson  amplitude for scattering in the eight  $D8$ branes and their mirrors that are coincident with one of the orientifold $O8$ planes.   Note, in particular, that this is independent of $\raia$ since the scattering is entirely within one of the orientifold planes and is insensitive to the radius of the eleventh dimension.   

\sm

Finally, T-duality converts the type IA amplitude into the type I amplitude by the replacement $\gia^{-1}= \raib\, \gib^{-1}$ using \eqref{Idua5}.  This description involves a factor proportional to the radius of the eleventh dimension, $\raib$, since T-duality is non-local along that direction and the type I amplitude depends on $\raib$.  

\sm

We see, therefore, the first (and rather simple)  example of an amplitude that has a consistent description in all four versions of the $D=10,\ \mathcal{N}=1$ string theory.

\subsection{The Yang--Mills/gravity tree amplitude in a single boundary}

\begin{figure}[h]
\begin{center}
\begin{tikzpicture}[scale=0.8]
\begin{scope}[shift={(0,0)}]
\draw [hwbou](-2,-3) -- (-2,2);
\draw [hwbou] (-2,2)-- (0,3);
\draw [hwbou] (-2,-3)-- (0,-2);
\draw [hwbou] (0,3) -- (0,-2);
\begin{scope}[shift={(-1,-1)}]
	\draw [sugra, ultra thick,domain=-90:90] plot ({cos(\x)},{1+ sin(\x)});
\filldraw (0,2) circle (.05) ; 
\filldraw (0,0) circle (.05) ; 
		
\draw [suym, ultra thick](0,0) -- (-.5,-1);
\draw [suym, ultra thick](0,0) -- (.6,-.7);
\begin{scope}[shift={(0,2)}]
\draw [suym, ultra thick](0,0) -- (-.5,1);
\draw [suym, ultra thick](0,0) -- (.6,.7);;
\end{scope}
\end{scope}
\end{scope}
\begin{scope}[shift={(5,0)}]
\draw [hwbou](-2,-3) -- (-2,2);
\draw [hwbou] (-2,2)-- (0,3);
\draw [hwbou] (-2,-3)-- (0,-2);
\draw [hwbou] (0,3) -- (0,-2); 
\end{scope}
\end{tikzpicture}
\end{center}
\caption{Tree level four gauge boson  amplitude generated via boundary Yang--Mills Chern-Simons  and gravitational interactions}
\label{fig:ymtree2}
\end{figure}
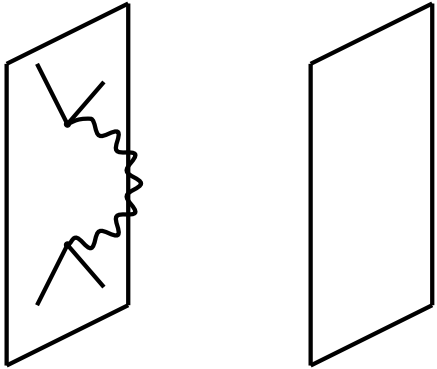

\sm

The first higher derivative contribution to the four-particle Yang--Mills amplitude  arises from tree diagrams with an intermediate graviton or antisymmetric potential propagating between pairs of Yang--Mills particles on the same boundary.  
In this case the intermediate particle propagates in the eleven-dimensional bulk and so the propagator involves the sum over the quantised momentum $p_{11}$ as in (\ref{Iprop2}).  This is  analogous to the case considered in  \cite{Dasgupta:2000xu}, where the amplitude described a pair of  scattering gauge particles in each boundary  joined by a propagator for a graviton or $C$ field, which will be reviewed later (see \eqref{totall}).  In the present case the gauge particles are scattering  in a  single boundary. The $C$ field couples to the boundary gauge fields via the (two-derivative) Chern--Simons interaction described earlier while the graviton couples via its minimal coupling (which also has two derivatives).      

\sm

An important difference from the case considered in \cite{Dasgupta:2000xu} is  that the boundary conditions require the use of the propagator $\mathcal{G}(p^M; 0,0)$ defined in \eqref{Iprop8} and \eqref{Iprop2}.  In addition, when all four gauge particles are in a single $E_8$ it is necessary to include the contact interactions coming from the square of the gauge Chern--Simons three-form  that is contained in the $|d \hat C|^2$ term in the boundary action \cite{Horava:1996ma}.  Applying the Feynman rules to this process, taking into account the factor of $\kappa_{11}^2$ in the propagator and $\lambda^{-6}$ for each vertex,  and using \eqref{gravigauge}  to express these parameters in terms of $\lepl $,
gives the Yang--Mills four-particle amplitude with an intermediate graviton propagator and all gauge particles in $E_8^{(1)}$, 
\begin{eqnarray} 
 \label{fullamp}
 A_{11}^{YM-gravity} &=&   \frac{ 2 \lepl\raten }{(2\pi)^6\, \lepl^4\, \raele }\,   \Biggl( \ \tr (T_1^{(1)}\, T_1^{(2)})  \, \tr( T_1^{(3)}\, T_1^{(4)})  \sum_{m=-\infty}^{\infty}\frac{1}{ -s+\frac{\pi^2m^2}{\ciele^2}} +\nonumber\\
&& \qquad\qquad\qquad\quad \tr( T_1^{(1)}\, T_1^{(4)})  \, \tr( T_1^{(2)}\, T_1^{(3)})  \sum_{m=-\infty}^{\infty}\frac{1}{-t +\frac{\pi^2m^2}{\ciele^2}}+\nonumber\\
&&\quad\qquad\qquad\qquad\ \tr( T_1^{(1)}\, T_1^{(3)})  \, \tr (T_1^{(2)}\, T_1^{(4)}) \sum_{m=-\infty}^{\infty}\frac{1}{-u +\frac{\pi^2m^2}{\ciele^2}}\Biggr)
 t_8 \hat F^4  \,.
 \end{eqnarray}  
 We have again inserted a factor of $2 \pi\,  \raten\, \lepl$ to account for the volume of the compactification of the tenth dimension, as mentioned earlier. The standard kinematic factor $t_8 \hat F^4$ is defined in \eqref{t8def}.    In writing this expression we have used the form of the propagator expressed as a sum over Kaluza--Klein modes   in \eqref{Iprop8} since this will be useful for comparison with the analogous expression in type I string theory.   

\sm

We may now expand the terms in \eqref{fullamp}  in the low-energy limit, $s \raele^2\lepl^2= s \ghe^2 \ell_H^2<<1$, using 
\begin{eqnarray} 
\frac{1}{\lepl^4\, \raele}\,  \sum_{m =-\infty}^{\infty}\frac{1}{ -s+\frac{m^2}{\lepl^2 \raele^2}}  &=& - \frac{1} {\lepl^4 \raele\, s} +
 2 \frac{\raele }{\lepl^2} \sum_{m=1}^{\infty} \frac{1}{m^2}
\frac{1}{1 - \frac{\raele^2 \lepl^2 }{m^2}s}	\nonumber\\
&=& - \frac{1} {\lepl^4 \raele\, s} +
2 \frac{\raele}{\lepl^2}\sum_{m=1}^\infty  \frac{1}{m^2} \, \left(1+ \frac{\raele^2 \lepl^2}{m^2}\, s +O( \raele^4\lepl^4\, s^2)\right)\nonumber\\
&=&  	 \frac{1} {\lesth^2}\left( - \frac{1}{\lesth^2\ghe^2 s} +
\frac{\pi^2}{3} + \frac{\pi^4}{45} \, \ghe^2 \lesth^2\,s +O( \ghe^4 \lesth^4\,s^2) \right)\,,\nonumber\\
\label{lowen}
\end{eqnarray} 
which is  an expansion in  powers of $\ghe^2\lesth^2 \,s$.   Although we started with the expression in \eqref{Iprop8} as a sum over Kaluza--Klein modes, the same result obviously arises starting from the winding number expression \eqref{Iprop2}, where we have
\begin{eqnarray}
\frac{1}{\sqrt{-s}}\,\frac{1}{\tanh(\sqrt{-s}\ciele)}=\ciele\left[ - \frac{1}{\ciele^2 s\, } + \frac{1}{3}  + \frac{\ciele^2 s}{45}+ O(\ciele^4 s^2)\,\right]\,,
\label{expantree}
\end{eqnarray}   
with $L= \pi \raele \lepl$.
We will see that  the sum over Kaluza--Klein charges  makes a direct connection with the form of the one-loop amplitude in the type I description of the amplitude.

\subsection*{\it The lowest order term in the low-energy expansion}

The leading term in the low-energy expansion of \eqref{fullamp} using \eqref{lowen} contributes the pole term,  
\begin{eqnarray}
\frac{ 2 \lepl\raten }{(2\pi)^6\, \lepl^4\, \raele }\, \Big(\frac{1}{s} \, \tr (T_1^{(1)}\, T_1^{(2)})  \, \tr( T_1^{(3)}\, T_1^{(4)}) +{\rm perms.} \Big)\, t_8 \,\tr \hat F^4\,.
\end{eqnarray}
Its coefficient is interpreted in the HE string theory by using  the M-theory/string theory dictionary, giving the identification  
\begin{eqnarray}
 \frac{ \raten} {\raele\,\lepl^3 }  =  \frac{\rahe}{ {\lesth^3\, \ghe^2}}\,.
\label{dicttrans}
\end{eqnarray}
Therefore, the leading term in the low-energy  expansion reproduces the low-energy limit of the tree-level heterotic $E_8 \times E_8$ theory.  This again transforms into the corresponding HO tree amplitude under T-duality, using $\rahe/\ghe^2=\raho/\gho^2$.

\sm

The prefactor in the type I theory is obtained from the relation $\raho\,\lesth^{-3}\,\gho^{-2} = \raib\, (\lesti)^{-3} $,
which reproduces the fact that the gravity/$C$-field pole arises in the one-loop amplitude (the annulus diagrams) in the type I theory.   T-duality implies that $\raib= (\raia)^{-1}$,
so the type IA amplitude vanishes in the $\raia\to \infty$ limit.  This is the limit in which the non-zero winding numbers of the bulk propagator are suppressed so the implication is that the zero winding number contribution to the amplitude also vanishes.
  
  \subsection*{\it Higher order terms}
  
The next term in the low-energy expansion of the amplitude in \eqref{fullamp} is the term of order $s^0$ in (\ref{lowen}), so the coefficient \eqref{dicttrans} is multiplied by a factor of $-s\, \pi^2\, \raele^2/3$, giving a contribution to the amplitude  \eqref{fullamp} of the form
\begin{eqnarray}
A_{11}^{\textrm{YM-gravity}} \Big|_{t_8(\tr_1 F^2_1)^2} &=&\frac{\raten\, \raele}{12\,(2\pi)^{4} \, \lepl}\,  t_8 \hat F^4 \,  \tr (T_1^{(1)}\, T_1^{(2)})  \, \tr( T_1^{(3)}\, T_1^{(4)}) \,,
 \label{ymtreeone}
 \end{eqnarray}
where we are again only displaying terms that are $SO(16)$ singlet in the $s$-channel. 
This  is interpreted as a local contribution to the effective action of order $t_8\,(\tr_1 F_1^2)^2$ (with a similar term involving $t_8\,(\tr_2 F_2^2)^2$). 
In the parameterisation of the various $\mathcal{N}=1$ string theories, the  coefficient of the amplitude is  proportional to
\begin{eqnarray}
\frac{\raten \raele}{\lepl}=\frac{\rahe }{\lesth} =  \frac{1}{\raho \lesth}= \frac{\raia }{\lesti}  = \frac{1}{\raib \lesti}\,.
 \label{ymtreeone1}
\end{eqnarray}
We therefore see that this interaction is associated with a one-loop effect in the HE  and IA theories in $D=10$ (the large $\rahe$ or $\raia$ limit) but vanishes in HO and type I in $D=10$ (as $\raho$ or $\raib$ $\to \infty$).  

\sm 
 
The next term in the low-energy expansion of (\ref{fullamp}) is
\begin{eqnarray}
A_{11}^{\textrm{YM-gravity}} \Big|_{s\,t_8(\tr_1 F_1^2)^2} &=&\frac{\raten\, \raele^3 \lepl}{4720\,(2\pi)^{2}}\, s\,  t_8 \hat F^4 \,  \tr (T_1^{(1)}\, T_1^{(2)})  \, \tr( T_1^{(3)}\, T_1^{(4)}) \,,
\label{nextamp}
\end{eqnarray}
which is expected to be a protected interaction that has no contributions beyond two loops in the heterotic theories.
Using the relations
\begin{eqnarray}
\raten \raele^3 \lepl=\lesth \rahe\, \ghe^2= \lesth \frac{\gho^2}{\raho^3}=\lesti \raia^3 = \lesti \frac{1}{\raib^3}\,.
 \label{ymtreeone2}
\end{eqnarray} 
This interaction is interpreted as  a two-loop term in the ten-dimensional HE theory,  which could be (but has not been) checked by analysing the low-energy limit of the explicit genus-two  amplitude in  HE perturbation theory \cite{D'Hoker:2005ht, D'Hoker:2001nj, D'Hoker:2002gw, D'Hoker:2005jc}. This interaction is not present in the ten-dimensional limit  of the HO theory since it is suppressed by a factor of $1/\raho^3$.  It is also interpreted as a one-loop contribution in the the  open string theories.  
We will shortly see (in  \eqref{typeiloop}) that the curious-looking dependence of this term on $\raia = 1/\raib$  arises explicitly from the expansion of the type I annulus amplitude.  

\sm
 
More generally, the low-energy expansion of the propagator given in (\ref{lowen}) produces a  sequence of terms of the form $(\ghe^2 \lesth^2\, s)^n\, t_8(\tr_i F_i^2) (\tr_i F_i^2) $ that are interpreted as $(n+1)$-loop terms of  order $s^n$ in the low-energy expansion of the HE theory. 
We do not expect terms with $n>1$ to be protected against higher loop contributions and there are sure to be other contributions to these higher order terms in the low-energy expansion.
 In  the type I interpretation this is an expansion in powers of $\lesti^2s\, \raia^2=\lesti^2s\,  /\raib^2$, which is  independent of the type I coupling constant, $\gib$, and therefore all such terms should  originate from the one-loop (annulus) diagram in the type I or IA theory, which we will now describe.

\subsection*{\it The type I annulus diagram in $D=9$ dimensions \cite{Rudra:2016xx}}  

We can see how the structure of the low-energy expansion changes when $n>1$ from the explicit form of the 
 type I annulus contribution to  the double-trace terms in  the four gauge particle amplitude.
 \begin{figure}[h]
\begin{center}
\begin{tikzpicture}[line width=1 pt, scale=1]

\begin{scope}[shift={(3,0)}]
\filldraw [light-gray] (0,15pt)--(3,15pt)--(3,-15pt)--(0,-15pt)--(0,15pt);
\filldraw [white] (0,0) ellipse (10pt and 15pt);
\filldraw [white] (3,0) ellipse (10pt and 15pt);
\draw [ultra thick] (0,-15pt)--(3,-15pt);
\draw [ultra thick] (0,0) ellipse (10pt and 15pt);
\draw[ultra thick] (3,0) ellipse (10pt and 15pt);
\draw [ultra thick] (0,15pt)--(3,15pt);
\draw [ultra thick] (0,-15pt)--(3,-15pt);
\node[ultra thick] at (8.5pt,8pt) {\Large $\times$};	
\node at (-8.5pt,-8pt) {\Large $\times$};
\node at (-15pt,-8pt) {\Large $1$};
\node at (5pt,7pt) {\Large $2$};
\begin{scope}[shift={(3,0)}]
\node at (8.5pt,-8pt) {\Large $\times$};	
\node at (-8.5pt,8pt) {\Large $\times$};
\node at (15pt,-8pt) {\Large $3$};
\node at (-5pt,7pt) {\Large $4$};
\end{scope}	
\end{scope}
\end{tikzpicture}
\end{center}
\caption{One loop four gauge boson  amplitude in type I string theory}
\label{fig:nonplanar}
\end{figure}
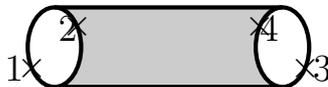
Figure~\ref{fig:nonplanar} represents  the amplitude, which has a low-energy expansion proportional to 
\begin{eqnarray}
&&\frac{1}{2(2\pi)^7\lesti}\sum_{\hat n=-\infty}^{\infty} \frac{\raib}{s\, \lesti^2  + \hat n^2\raib^2}
	\Bigg[\tr (T_1^{(1)}\, T_1^{(2)}) \tr (T_1^{(3)}\, T_1^{(4)})    +\tr (T_2^{(1)}\, T_2^{(2)}) \tr (T_2^{(3)}\, T_2^{(4)})\nonumber\\
&& \qquad\qquad\qquad\quad +
	(-1)^{\hat n} \tr (T_1^{(1)}\, T_1^{(2)}) \tr (T_2^{(3)}\, T_2^{(4)}) 
\Bigg]\Big(1 + O(\lesti^6\, s^3)\Big)  \, t_8 \hat F^4\,,
\label{typeiloop}
\end{eqnarray}
where we are specialising to the case where the quantum numbers of the external states have been chosen so that $T_i^{(1)} T_i^{(2)}$ and $T_i^{(3)} T_i^{(4)}$ contain an $SO(16)$ singlet and therefore couple to the gravitational sector.\footnote{The expression includes the terms associated with the second $SO(16)$ subgroup as well as the first. In particular, the term proportional to $\tr_1 (T_1^{(1)}\, T_1^{(2)}) \tr_2 (T_2^{(3)}\, T_2^{(4)})$ contains a factor from each $SO(16)$ subgroup and is obtained in Ho\u rava --Witten supergravity from a propagator stretching between the two boundaries, as  will be discussed in the next subsection.}  The  sum over $\hat n$  is a sum over the winding numbers around the $x^{11}$ circle of the closed type I string propagating in the cylinder channel.  This is interpreted as the sum over Kaluza--Klein momentum of the type I open string when the world-sheet is evaluated as an open-string loop (there is no winding number for the type I open string). In the type IA description the open strings satisfy Dirichlet boundary conditions in the $x^{11}$ direction so such strings carry no $p_{11}$ momentum and the sum translates into a sum over open-string winding modes.  This, in turn, transforms into a sum over $p_{11}$ Kaluza--Klein  momentum modes in the type IA closed-string description (while the Dirichlet boundary conditions on the cylinder boundaries imply that the type IA closed string has no winding around the $x^{11}$ circle).

\sm 
 
The graviton pole arises as the  $\hat n=0$ term in \eqref{typeiloop}.  The factor  $(s\, \lesti^2 + \hat n^2\raib^2)^{-1}$  also contains the same infinite sequence  of massive poles  as the field theory propagator in \eqref{lowen}, and expanding it in powers of $s\, \lesti^2/\raib^2$ gives the same infinite sequence of higher-derivative  terms as in the expansion of the supergravity tree diagram. The higher order terms in the last parentheses arises from the Koba--Nielsen-like factor associated with excited string states. This is a sign that the supergravity expression is not valid for interactions of order $s^2 (\tr F^2)^2$ and  higher (taking into account the fact the $\hat n=0$ term in the sum cancels one power of $s$).  We see that these interactions,  which are not protected by supersymmetry,  receive contributions from higher string modes, which are not captured by the supergravity approximation.

\subsection{Tree stretching between distinct  boundaries}

When particles $1$ and $2$ are in one $E_8$ subgroup and particles $3$ and $4$ are in the other,  the ends of the tree are on distinct Ho\u rava--Witten boundaries, which was the example considered in \cite{Dasgupta:2000xu}.   The amplitude again consists of the sum of Feynman diagrams with a graviton or the third-rank potential, $C$ propagating between the boundaries, (but with no contact term).  The resulting amplitude has the form (\cite{Dasgupta:2000xu})
\begin{equation}
\label{totall}
A_{12}^{\textrm{YM-gravity}} = \frac{\raten}{(2\pi)^5\lepl^2}\, \tr_1 (T_1^{(1)}\, T_1^{(2)})  \, \tr_2( T_2^{(3)}\, T_2^{(4)})\,
\frac{1}{\sqrt{-s}}
{1 \over \sinh(\sqrt{-s}\ciele)} t_8 \hat F^4 \,,
\end{equation}
where we have again included a factor of $2 \pi\,  \raten\, \lepl$ to account for the volume of the compactification of the tenth dimension.  The low-energy expansion can be obtained by using
\begin{eqnarray}
\frac{1}{\sqrt{-s}}\, \frac{1}{\sinh(\sqrt{-s}\ciele)} =\ciele \left[- \frac{1}{s\, \ciele^2} - \frac{1}{6}  - \frac{7 \ciele^2 s}{360}+ O(\ciele^4s^2)\,\right] \, ,
\label{loesinh}
\end{eqnarray}
or, equivalently, by expanding the expression for the propagator as a sum over Kaluza--Klein modes (the first equation in \eqref{Iprop7})

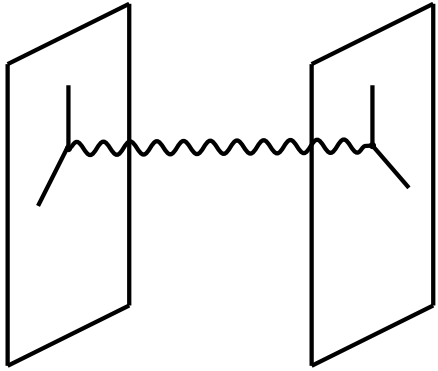
\begin{figure}[h]
\begin{center}
\begin{tikzpicture}[scale=.8]
\begin{scope}[shift={(0,0)}]
\draw [hwbou](-2,-3) -- (-2,2);
\draw [hwbou] (-2,2)-- (0,3);
\draw [hwbou] (-2,-3)-- (0,-2);
\draw [hwbou] (0,3) -- (0,-2);
\begin{scope}[shift={(-1,-.2)}]
\begin{scope}[shift={(0,.85)}]
\draw [suym, ultra thick](0,0) -- (-.5,-1);
\draw [suym, ultra thick](0,0) -- (0,1);
\end{scope}
\end{scope}
\end{scope}
\begin{scope}[shift={(5,0)}]
\draw [hwbou](-2,-3) -- (-2,2);
\draw [hwbou] (-2,2)-- (0,3);
\draw [hwbou] (-2,-3)-- (0,-2);
\draw [hwbou] (0,3) -- (0,-2);
\draw [sugra, ultra thick](-6,.6) -- (-1,.65);
\filldraw (-6,.6) circle (.05) ; 
\filldraw (-1,.65) circle (.05) ; 
\begin{scope}[shift={(-1,-.2)}]
\begin{scope}[shift={(0,.85)}]
\draw [suym, ultra thick](0,0) -- (-0,1);
\draw [suym, ultra thick](0,0) -- (.6,-.7); 
\end{scope}
\end{scope}
\end{scope}
\end{tikzpicture}
\end{center}
\caption{Tree-level four gauge boson amplitude generated via boundary Yang--Mills Chern-Simons and gravitational  interactions}
\label{fig:ymtree3}
\end{figure}

\subsection*{\it The lowest order term in the low-energy expansion}

The leading behaviour of the amplitude in the low-energy limit $\lesth^2s,\lesth^2t,\lesth^2 u\ll 1$ reduces, after using the M-theory/heterotic string theory dictionary to
\begin{eqnarray}
A_{12}^{\textrm{YM-gravity}} = - \frac{2\rahe}{\, (2\pi)^6 \lesth^{3} \, \ghe^{2}}\tr_1(T^{(1)}_1 T^{(2)}_1)\, \tr_2(T^{(3)}_2 T^{(4)}_2)
\frac{1}{s} \, t_8 \hat F^4 \,,
\label{leadtree}
\end{eqnarray}
which again agrees with the corresponding term in the HE tree-level amplitude as noted in \cite{Dasgupta:2000xu}.
The correspondence with the HO, type I and type IA theories follows as in the discussion following \eqref{dicttrans}.

\subsection*{\it Higher order terms}

The next term in the expansion of  \eqref{totall} following from \eqref{loesinh} 
\begin{eqnarray}
A_{12}^{YM-gravity} = -\frac{\raten\raele}{24\, (2\pi)^4 \lepl}\, \tr (T_1^{(1)}\, T_1^{(2)})  \, \tr( T_2^{(3)}\, T_2^{(4)})\,
t_8 \hat F^4 \,,
\label{aym21}
\end{eqnarray}
which differs from that of \eqref{ymtreeone} by a factor of $-1/2$.   This relative factor of $-1/2$ between the coefficient of $  \tr (T_1^{(1)}\, T_1^{(2)})  \, \tr( T_1^{(3)}\, T_1^{(4)})$  in $A_{(1,1)}^{YM-gravity}$ and $A_{(1,2)}^{YM-gravity} $  arises from the factor of $(-1)^m$ in \eqref{Iprop7} as can be seen from the identity
\begin{eqnarray}
\sum_{m\ne 0} \frac{(-1)^m}{m^2}=	-\frac{1}{2}\sum_{m\ne 0}\frac{1}{m^2}\,.
\label{relfact}
\end{eqnarray} 
The  relative factor of $-1/2$  is in accord with the  computation of the annulus loop diagram in type I string perturbation theory, where the coefficient \eqref{relfact} is obtained from the factor   $\sum_{\hat n \ne 0}(-1)^{\hat n}\,(s\, \lesti^2  + \hat n^2\raib^2)^{-1}$  in \eqref{typeiloop}  in the $s\to 0$ limit.
  This relative factor is  also in accord with the analysis  given in equation (3.16) of  \cite{Abe:1988cq}, where, in the low-energy limit,  the one-loop effective action in HE string  perturbation theory was found to have the form
\begin{eqnarray}
\frac{\rahe }{ 96(2\pi)^4\lesth }
\left[
\sum_{i=1}^2 t_8\left( \tr_i F_i^2\right)^2 - t_8\tr_1 F_1^2 \tr_2 F_2^2\right]  \,.
\label{oneloophe}
\end{eqnarray}
 This expression is the parity conserving partner of the parity-violating interaction that serves to cancel the chiral gauge anomalies when decompactified  to the ten-dimensional heterotic  $E_8 \times E_8$ limit.
 By contrast, the tree-level HE effective action for the double-trace terms has the form \cite{Gross:1986mw, Cai:1986sa}
\begin{eqnarray}
\frac{\rahe }{ 2^{10}(2\pi)^4\lesth \ghe^2 }
\left[\sum_{i=1}^2 t_8\left( \tr_i F_i^2\right)^2 +2\ t_8 \tr_1 F_1^2 \tr_2 F_2^2\,\right]\,,
\label{treehe}
\end{eqnarray}
which we will obtain from supergravity in the Ho\u  rava--Witten background in section \ref{subsec:itetree}.

\sm

The terms that arise at  the next order in the expansion of the propagator in powers of $(\raele^2\ell_{11}^2 s)$ following \eqref{oneloophe}  include the $\ell_{H}\, \ghe^2 s\,t_8(\tr_1 F^2)^2$  interaction in \eqref{nextamp} together with terms related by permutations of the external particles and involving both $SO(16)$ subgroups. These are described by a two-loop effective action of the form 
\begin{eqnarray}
\frac{\lesth \rahe\, \ghe^2 }{ 2880\, (2\pi)^2 }\, 
\left[
\sum_{i=1}^2 t_8\left( \tr_i F_i^2\right)d^2 \left( \tr_i F_i^2\right)-\frac{7}{4} t_8 (\tr_1 F_1^2) d^2 (\tr_2 F_2^2)\right]  \, ,
\label{morehigh}
\end{eqnarray}
which should agree with the low-energy limit of the genus-two contribution to HE superstring theory. 

\sm

To summarise, the  tree amplitudes illustrated in figures~\ref{fig:ymtree2} and \ref{fig:ymtree3} that have a single gravitational propagator capture the low order terms of the form $s^n\, t_8(\tr F^2)^2$ in the  low-energy expansion of the ${\cal N}=1$ superstring four-point amplitude compactified on a circle.     Indeed the field theoretic  amplitudes  in \eqref{fullamp} and  \eqref{totall}  
precisely  reproduce the corresponding factors associated with closed-string ground states with arbitrary winding numbers in the contribution of the annulus (one open string loop) diagram to the four-point function of the  type I  string  theory \eqref{typeiloop}.  For $n<2$ the agreement is exact, whereas stringy corrections enter into the low-energy expansion of \eqref{typeiloop}  at order $s^2\, t_8(\tr F^2)^2$.  Other arguments (for example, see \cite{Berkovits:2009aw}) suggest that this is a non-BPS interaction that is not protected against higher loop corrections.  

\subsection{``Iterated'' Yang--Mills tree diagrams}
\label{subsec:itetree}

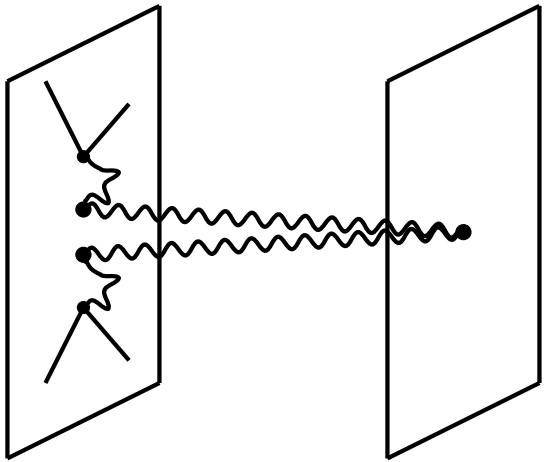
\begin{figure}[h]
\begin{center}
\begin{tikzpicture}[scale=1.0]
\begin{scope}[shift={(0,0)}]
\draw [hwbou](-2,-3) -- (-2,2);
\draw [hwbou] (-2,2)-- (0,3);
\draw [hwbou] (-2,-3)-- (0,-2);
\draw [hwbou] (0,3) -- (0,-2);
\begin{scope}[shift={(-1,-1)}]
	\draw [sugra, ultra thick,domain=-90:90] plot ({.3*cos(\x)},{(.3+.3*sin(\x))});
	\draw [sugra, ultra thick,domain=-90:90] plot ({.3*cos(\x)},{(1.7+.3*sin(\x))});
	\end{scope}
	\begin{scope}[shift={(-0.5,-1)}]
\draw [sugra, ultra thick](-0.5,.7) -- (4.5,1);
\draw [sugra, ultra thick](-0.5,1.3) -- (4.5,1);
\filldraw (4.5,1) circle (.10) ; 
\filldraw (-0.5,.7) circle (.10) ; 
\filldraw (-0.5,1.3) circle (.10) ; 
\filldraw (-0.5,2) circle (.08) ; 
\filldraw (-0.5,0) circle (.08) ; 
		
\draw [suym, ultra thick](-0.5,0) -- (-1,-1);
\draw [suym, ultra thick](-0.5,0) -- (.1,-.7);
\begin{scope}[shift={(0,2)}]
\draw [suym, ultra thick](-0.5,0) -- (-1,1);
\draw [suym, ultra thick](-0.5,0) -- (.1,.7);;
\end{scope}
\end{scope}
\end{scope}
\begin{scope}[shift={(5,0)}]
\draw [hwbou](-2,-3) -- (-2,2);
\draw [hwbou] (-2,2)-- (0,3);
\draw [hwbou] (-2,-3)-- (0,-2);
\draw [hwbou] (0,3) -- (0,-2);
\end{scope}
\end{tikzpicture}
\end{center}
\caption{A contribution to  the low-energy expansion of the double-trace amplitude with pairs of gauge particles in the same $SO(16)$ subgroup.}
\label{fig:iterone}
\end{figure}

\begin{figure}[h]
\begin{center}
\begin{tikzpicture}[scale=1.0]
\begin{scope}[shift={(0,0)}]
\draw [hwbou](-2,-3) -- (-2,2);
\draw [hwbou] (-2,2)-- (0,3);
\draw [hwbou] (-2,-3)-- (0,-2);
\draw [hwbou] (0,3) -- (0,-2);
\begin{scope}[shift={(-1,-.2)}]
\begin{scope}[shift={(0,-1)}]
\draw [suym, ultra thick](0,0) -- (-.5,-1);
\draw [suym, ultra thick](0,0) -- (0,1);
\end{scope}
\end{scope}
\end{scope}

\begin{scope}[shift={(5,0)}]
\draw [hwbou](-2,-3) -- (-2,2);
\draw [hwbou] (-2,2)-- (0,3);
\draw [hwbou] (-2,-3)-- (0,-2);
\draw [hwbou] (0,3) -- (0,-2);
\draw [sugra, ultra thick](-6,.6) -- (-1,.65);
\draw [sugra, ultra thick](-6,-1.2) -- (-1,-1.15);
\draw [sugra, ultra thick](-6,.6) -- (-1,-.55);
\begin{scope}[shift={(-1,-1.15)}]
	\draw [sugra, ultra thick,domain=-90:90] plot ({.3*cos(\x)},{(.3+.3*sin(\x))});
\end{scope}
\filldraw (-1,-.55) circle (.10) ; 
\filldraw (-1,-1.15) circle (.10) ;  
\filldraw (-6,.6) circle (.10) ; 
\filldraw (-6,-1.2) circle (.08) ;
\filldraw (-1,.65) circle (.08) ; 
\begin{scope}[shift={(-1,-.2)}]
\begin{scope}[shift={(0,.85)}]
\draw [suym, ultra thick](0,0) -- (-0,1);
\draw [suym, ultra thick](0,0) -- (.6,-.7); 
\end{scope}
\end{scope}
\end{scope}
\end{tikzpicture}
\end{center}
\caption{A contribution to  the low-energy expansion of the double-trace amplitude with pairs of gauge particles in distinct $SO(16)$ subgroups.}
\label{fig:itertwo}
\end{figure}
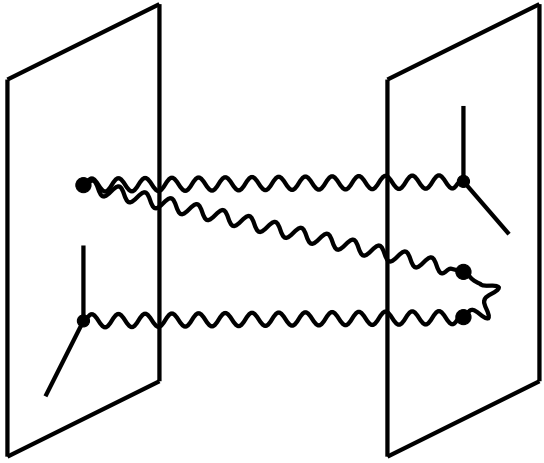

We will now consider an infinite class of  generalised tree diagrams that are illustrated in figures~\ref{fig:iterone} and \ref{fig:itertwo}.  

\sm

The black dots in these diagrams represent two-point functions  induced by the presence of the $\tr \curv^2$ and $(\partial H)^2$ terms that enter as boundary interactions in  \eqref{grav3pt1}.  These vertices have the form  $\coc\, s^2\, \lepl^{-6}$, where the dimensionless constant $\coc$ can be determined from \eqref{lineartwo} to have the value  $\coc=1/2(2\pi)^7$.     At any given order, the low-energy expansion of the amplitude involves the sum over all ways of arranging a given number of propagators to form a chain linking the pairs of external gauge particles.  Each propagator ($D$) introduces a factor of $\kappa_{11}^2 D= (2\pi)^8\,\lepl^9\, D/2$.
The pairs of gauge particle at the ends of the chains may lie in the same $SO(16)$ subgroup or in distinct subgroups, as is represented by the two figures. and each chain of given length may have both endpoints  on the same boundary (and is of the form  of the form $D_{11}$ or $D_{22}$ defined in \eqref{dmatrixdefa}  or it may stretch between boundaries (and is of the form $D_{12}$ or $D_{21}$).  All such possibilities are to be summed over.  In figure \ref{fig:iterone} there is an even number of propagators linking the boundaries while in figure \ref{fig:itertwo} there is an odd number.

\sm

The amplitude obtained by summing over all possible ways of joining the initial and final pairs of gauge particles is given by the matrix
\begin{eqnarray}
 A_{ij} = 
 \frac{\raten }{4(2\pi)^5\lepl^2}\,  \tr (T_i^{(1)}\, T_i^{(2)})  \, \tr( T_j^{(3)}\, T_j^{(4)})  \,t_8  \, \hat F^4 \, \left( \frac{D}{1- \pi\, \lepl^3 s^2 D/4} \right)_{ij}\label{twovert}\,.
\label{matrixsum}
\end{eqnarray}
The indices $i$ and $j$ are not to be summed on the right-hand side of the above equation.  This expression can be written as an expansion in powers of  $\pi \lepl^3\,s^2 D/4$.  Each factor of $D$ can itself be expanded as a power series in $\raele^2\lepl^2 s$ (with the leading term $\sim 1/(\pi \raele \lepl s)$).  In terms of string theory parameters, the former is an expansion in powers of $ \gia  \lesti^2 s/\raia $ and the latter is an expansion in $\ghe^2 \lesth^2 s$.
The term of zeroth order in the two-graviton vertex manifestly reproduces the result obtained earlier due to the exchange of a single  $D_{ij}$  between the pairs of external gauge particles, which we previously related to the contribution of closed-string winding number states  to the annulus diagram in type IA string theory.  If we keep the leading term in the expansion of $D_{ij}$, which is $ - (\pi \raele \lepl s)^{-1} $  (independent of $i,j$), we have 
\begin{eqnarray}
 A_{ij} &=&\frac{\raten }{2(2\pi)^6\, \raele\, \lepl^3 } \,  \tr (T_i^{(1)}\, T_i^{(2)})  \, \tr( T_j^{(3)}\, T_j^{(4)})  \,t_8  \, \hat F^4 \,\frac{1}{s}\, \left( \frac{1}{1  +\frac{\lepl^2}{4\raele }s } \right)(1+ O(\lepl^2\raele^2\, s )) \nonumber
\\
&=& -\frac{2\rahe }{(2\pi)^6\, \ghe^2\, \lesth^3}\,  \tr (T_i^{(1)}\, T_i^{(2)})  \, \tr( T_j^{(3)}\, T_j^{(4)})  \,t_8  \, \hat F^4 \,\frac{1}{s}\, \left( \frac{1}{4  + \lesth^2 s} \right) (1+O(\lesth^2\ghe^2\, s))\,.
\label{leading}
\end{eqnarray}
Thus, keeping only the massless Kaluza--Klein modes in $D_{ij}$ gives an expression that is interpreted as a tree-level expression in the HE theory.  

\sm 

We may compare \eqref{leading} with the expression for the tree-level  four gauge particle amplitude in the heterotic string given in  equation (4.4) of \cite{Gross:1986mw}.  This is proportional to\footnote{In \cite{Gross:1986mw} the heterotic string scale was chosen to be  $\lesth^2=1/2$. }
\begin{eqnarray}
&&\frac{\lesth^3 \rahe\, }{(2\pi)^6\ghe^2}\Bigg(\frac{tu}{2^4 (4+\lesth^2\, s ) } \tr (T_1^{(1)}\, T_1^{(2)})   \tr( T_1^{(3)}\, T_1^{(4)}) - \frac{s\, \lesth^2}{4} \tr(T_1^{(1)}\, T_1^{(2)}T_1^{(3)}\, T_1^{(4)})\Big) \nonumber\\
&& + {\rm non-cyclic\ perms.} \Bigg) \left(\frac{2^5}{ \lesth^6\, stu} + \zeta(3) + O( \lesth^2 s)\right)\,  \,t_8  \, \hat F^4 \,.
\label{grosssloan}
\end{eqnarray}
We see that the supergravity amplitude in \eqref{leading} is the component of the double-trace part of this  string theory tree amplitude that involves the  $1/(stu)$ term in the last parenthesis\footnote{We have not attempted to compare the  overall normalisations   of these expressions, but the agreement of the residue at $s=0$ in \eqref{leading} with the HE string pole term, as noted earlier, guarantees the agreement of the rest of the expression.}.  The dependence on string theory factors begins at order   $s^2\, (\tr F^2)^2$, which is not protected by supersymmetry. and is the order at which  the $\zeta(3)$ factor in the last parenthesis enters in the expansion.  Similarly, 
the single-trace term of order  $ s\, \tr F^4$ is   not expected to be protected by supersymmetry and also has a prefactor proportional to $\zeta(3)$.   We will  see in section \ref{sec:higherloop} how these $\zeta(3)$ terms can be motivated from the effect of loop amplitudes in supergravity in the Ho\u rava--Witten background - although in these cases we do not expect to reproduce the exact coefficients.

\sm 

Note that the  apparent pole at $\lesth^2s=-4$ in \eqref{grosssloan}  is cancelled by stringy corrections that are subsumed in the terms of $O(s \ell_{H}^2)$ in the last parenthesis.  Such a cancellation is not captured by the Feynman diagram expression \eqref{matrixsum}.  This is consistent with the fact that we do not expect to reproduce the exact expressions for interactions of order $s^2 t_8(\tr F^2)^2$ and beyond.  

\sm 

Furthermore,  the expansion of \eqref{leading} in powers of $ \lesth^2\,s$ can be interpreted in the type I theory as an expansion in powers of $\gib\, \lesti^2 \,s$ (using  \eqref{Idua5}).  Each power of $\gib$ is interpreted as the insertion of a boundary  or cross-cap in the open string world-sheet.  In this way we see that a contribution to the tree-level HE  amplitude is associated with an infinite series of higher order terms in type I perturbation theory.

\sm 

Thus, we have described the HE and HO tree level contributions to  the interaction $t_8(\tr_1 F_1^2+\tr_2 F_2^2)^2$ and the one-loop contribution in the HE theory of the form  $s\  t_8 \left[ \sum_{i=1}^2 \left( \tr_i F_i^2\right)^2 - \tr_1 F_1^2 \tr_2 F_2^2 \right]$ (which has yet to be verified by a direct string theory calculation).

\section{Yang--Mills one-loop amplitudes}
\label{sec:loopamp}

We will now turn to consider one-loop Feynman integrals for supergravity in the Ho\u rava--Witten background compactified on a circle.  We need to include the complete supermultiplet of states circulating in the loop, which can be expressed in a simple manner by using the world-line light-cone superspace  procedure described earlier.    In the following we will be interested in determining local terms induced by the loop amplitudes and will not discuss the non-local effects associated with non-analytic parts of the amplitude, which can be separated from the analytic terms in unambiguous fashion\footnote{It is far from obvious that such a separation of analytic and non-analytic parts of the amplitude is possible at higher orders in the low-energy expansion.}.

\subsection{A loop of gauge particles on one boundary }
\label{sec:heloopamp}

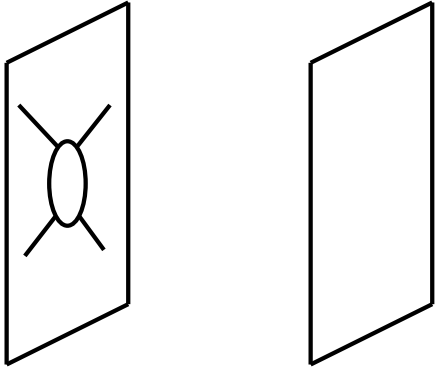
\begin{figure}[h]
\begin{center}
\begin{tikzpicture}[scale=.8]
\begin{scope}[shift={(0,0)}]
\draw [hwbou](-2,-3) -- (-2,2);
\draw [hwbou] (-2,2)-- (0,3);
\draw [hwbou] (-2,-3)-- (0,-2);
\draw [hwbou] (0,3) -- (0,-2);
\draw [suym, ultra thick](-0.3,1.3) -- (-.85,.60);
\draw [suym, ultra thick](-0.4,-1.1) -- (-.80,-.55);
\draw [suym, ultra thick](-1.80,1.3) -- (-1.15,.6);
\draw [suym, ultra thick](-1.70,-1.2) -- (-1.20,-.55);
\draw  [suym, ultra thick](-1,0) ellipse (.3cm and .7cm);
\end{scope}

\begin{scope}[shift={(5,0)}]
\draw [hwbou](-2,-3) -- (-2,2);
\draw [hwbou] (-2,2)-- (0,3);
\draw [hwbou] (-2,-3)-- (0,-2);
\draw [hwbou] (0,3) -- (0,-2);
\end{scope}
\end{tikzpicture}
\end{center}
\caption{Four gauge boson  amplitude in the $E_8\to SO(16)$ boundary gauge theory compactified to nine dimensions on $S^1$. The particles circulating in the loop are Kaluza--Klein modes of $SO(16)$ adjoint and spinor states. }
\label{fig:oneloopgauge}
\end{figure}

We will first  consider the four gauge boson  one-loop amplitude in $E_8$ gauge theory compactfied to nine dimensions on a circle of radius $\lepl \raten $ in the presence of a Wilson line that breaks the gauge symmetry to $SO(16)$.\footnote{This amplitude can be determined in an efficient manner  by means of a first-quantised world-line formalism (modelled on string theory calculations), in which vertex operators describe the emission of massless gauge particles from a circulating ${\cal N}=1$ gauge supermultiplet.  We omit the details here, but see section~\ref{subsec:susyloops} for a discussion of the vertex operator construction of the gravitational loop amplitude.}
We will consider the loop amplitude with external nine-dimensional massless states  that transform in the $SO(16)$ adjoint representation. These states couple to the Kaluza--Klein towers of particles circulating in the compactified loop, which are both the $SO(16)$ adjoint states and the $SO(16)$ spinor states. 

\sm 

The value of the gauge potential associated with such a Wilson line is given in \eqref{cartane8} and the corresponding  loop momentum is quantised in integer units when the circulating states are in the adjoint representation of $SO(16)$, so $p_{10}=n/(\lepl\raten)$, where $n \in \mathbb{Z}$ and the $n=0$ states are the massless $SO(16)$ gauge bosons. The circulating $SO(16)$ spinor states are those for which $p_{10}=(n-1/2)/(\lepl\raten)$, which  have masses given by \eqref{spinormass} and there are no massless states in this sector.

\sm 

The complete loop amplitude consists of the sum of the contributions from the circulating adjoint Kaluza--Klein tower and the spinor Kaluza--Klein tower 
\begin{eqnarray}
A^{1-loop} = A_{adj} + A_{spin}\,.
\label{fulloop}
\end{eqnarray}

\sm
The contribution of  the SO(16) adjoint states in the loop is given by
\begin{eqnarray}
A_{adj} &=&\frac{2}{3(2\pi)^{10}}t_8 \hat F^4 \, \mathcal{C}_{adj}\, I_{adj}(s,t,u; \raten ) 	\,,
 \label{gluoncal1} 
\end{eqnarray}
where $ \mathcal{C}_{adj}$ is the colour factor for the loop amplitude of $SO(16)$ adjoint states (and the overall factor of $t_8\, \hat F^4$ is determined by maximal Yang--Mills supersymmetry).  This is given by setting $N=16$ in the $SO(N)$ colour factor (where $N$ is even) that has the following form, for a particular colour ordering:
\begin{eqnarray}
 \mathcal{C}_{adj}
&=&
(N-8) \tr(T^{a_1}T^{a_2}T^{a_3}T^{a_4}) +\{ \tr(T^{a_1}T^{a_2})\tr(T^{a_3}T^{a_4})+\textrm{perms.}\}\,.
 \label{gluoncal2}
\end{eqnarray}
We have left $N$ as a free parameter in order to emphasise later the special features of the value $N=16$.
The dynamical part of the loop amplitude is contained in $I_{adj}(s,t,u;\raten)$ that is given by a scalar box Feynman diagram compactified on the circle of radius $\lepl\raten $ with the loop momentum $p_{10}$ replaced by the sum over integer Kaluza--Klein charges, as will be discussed below.
 
 \sm

The other piece of the four gauge boson  amplitude, where only the SO(16) spinor states circulate in the loop, is given by
\begin{eqnarray}
A_{spin}&=&\ \frac{2}{3(2\pi)^{10}} t_8\hat F^4\, \mathcal{C}_{spin}\,  I_{spin}(s,t,u; \raten ) \,,
 \label{gluoncal4}
\end{eqnarray}
where $\mathcal{C}_{spin}$ is the colour factor for the loop of spinor states and is given, for any $SO(N)$ group (with even $N$), and as before,  for a particular colour ordering,  by
\begin{eqnarray}
\mathcal{C}_{spin}&=& - 2^{\frac{N}{2}-7} \,
\left[4\, \tr(T^{a_1}T^{a_2}T^{a_3}T^{a_4}) -\{ \tr(T^{a_1}T^{a_2})\tr(T^{a_3}T^{a_4})+\textrm{perms.}\}\right]\,.
 \label{gluoncal5}
\end{eqnarray} 
The quantity $I_{spin}(s,t,u; \raten )$ is the dynamical part of the amplitude, which is again given by a Feynman box diagram with the integral over the $p_{10}$ component of the loop momentum replaced by a sum over half-integer Kaluza--Klein charges.  This will also be discussed below.

\sm 

It is an important fact that when $N=16$ the sum of the  adjoint colour factor and the spinor colour factor satisfies
\begin{eqnarray}
\mathcal{C}_{adj}+\mathcal{C}_{spin} =  3\, \tr(T^{a_1}T^{a_2})\,\tr(T^{a_3}T^{a_4})+\textrm{perms.} \,,
\label{cassum}
\end{eqnarray}
and therefore does not contain any fourth order Casimir invariant in the $SO(16)$ fundamental representation.  This, of course, is connected with the fact that the amplitude is inherited from the ten-dimensional amplitude in which the circulating states are the massless states in the adjoint of $E_8$, which has no independent fourth order Casimir.  
Another combination of the colour factors that will prove important below is 
 \begin{eqnarray}
 \mathcal{C}_{adj}-\frac{1}{2} \mathcal{C}_{spin} =  12\, \tr(T^{a_1}T^{a_2}T^{a_3}T^{a_4})\,,
 \label{diffcas}
 \end{eqnarray}
which  is a purely fourth order invariant.

\subsection{Evaluation of the lowest order terms in the loop amplitude}

In the following we will make use of the standard Poisson summation formulae
\begin{eqnarray}
\sum_m e^{-\pi a^2 m^2} = \frac{1}{|a|} \sum_{\hat m} e^{-\pi a^{-2} \hat m^2}\,,\qquad\ \ \sum_m e^{-\pi a^2 \left(m-\frac{1}{2}\right)^2} = \frac{1}{|a|} \sum_{\hat m}(-1)^{\hat m}  e^{-\pi a^{-2} \hat m^2}\,.
\label{poisson}
\end{eqnarray}

As mentioned above, the Feynman integral reduces to a kinematic prefactor $t_8 \hat F^4 \, \mathcal{C}_{adj}$ multiplying the compactified scalar box diagram with integer Kaluza--Klein charges.  The box diagram evaluated in nine dimensions contains non-analytic threshold terms of order $\sqrt s$, which do not concern us here and are, in any case, subleading in the low-energy expansion. The lowest term in the low-energy expansion is obtained by setting $s=t=u=0$, in which case it is simple to show that
\begin{eqnarray}
I_{adj}(0,0,0; \raten )=
2\pi^{11/2}
\int_0^\infty 
\frac{d\tau}{\tau^{3/2}}
\sum_{m\in \mathbb{Z}} 
e^{-\tau\left(\frac{m}{\lepl\raten}\right)^2} \,,
\label{gluoncal3}
\end{eqnarray}
where the sum in the lattice factor is over the Kaluza--Klein charge, $m$.   Each term in the $m$ sum  obviously possesses the ultraviolet divergence of the ten-dimensional theory.    This will be dealt with by defining the loop in the winding number basis by performing a Poisson summation (using \eqref{poisson})  that converts \eqref{gluoncal3} into
\begin{eqnarray}
I_{adj}(0,0,0; \raten) &=&
2\pi^6\lepl\raten 
\int_0^\infty d \hat \tau
\sum_{\hat m\in \mathbb{Z}}
 e^{-\hat \tau  \hat m^2\, \left(\pi \lepl\raten\right)^2}\nonumber\\
&=& 2\pi \left[ C_1\frac{\raten}{\lepl} + 
\frac{\pi^3}{ \raten \lepl}
\zeta(2)\right] \,,
 \label{gluoncal3a}
\end{eqnarray}
where $\hat \tau=1/\tau$.   Here we have separated the divergent zero winding  ($\hat m=0$) term, that depends on the cut-off and is represented by 
$C_1 = \int_0^\Lambda d\hat \tau$, where $\Lambda$ is an arbitrary dimensionless constant that we need not specify. 
The non-zero winding terms have precisely determined coefficients. 

\sm 

The leading term in the low-energy expansion of $I_{spin}$ is again obtained by setting $s=t=u=0$ and evaluating the box diagram with the appropriate lattice factor to describe the circulating  spinor states, giving,
\begin{eqnarray}
I_{spin}(0,0,0; \raten )&=&
2\pi^{11/2}
\int_0^\infty 
\frac{d\tau}{\tau^{3/2}}
\sum_{m\in \mathbb{Z}
} e^{-\tau\left(
\frac{m-1/2}{\lepl\raten}\right)^2}
\nonumber\\
&=& 
2\pi^6\lepl\raten
\int_0^\infty d \hat \tau
\sum_{\hat m\in \mathbb{Z}}(-1)^{\hat m}
 e^{-\hat \tau  \hat m^2\, \left(\pi \lepl\raten\right)^2}
\nonumber\\
&=&
  2\pi \left[ C_2\frac{\raten }{\lepl} -
\frac{1}{2}\frac{\pi^3}{\raten \lepl}
\zeta(2)\right] \,,
 \label{gluoncal6}
\end{eqnarray}
where we have used  \eqref{poisson} and the fact that $\sum_{\hat m>0} 1/\hat m^2 = - 2\sum_{\hat m>0} (-1)^{\hat m}/\hat m^2 = \zeta(2)$.  The dimensionless constant, $C_2$, again arises from the zero winding mode ($\hat m=0$) and is cut-off dependent.  

\sm 

The cut-off dependent terms proportional to $C_1\raten/\lepl$ and $C_2\raten/\lepl$ should be renormalised by the addition of counterterms.  However, since $\raten/\lepl = \rahe /\ghe^{2/3}\lesth$ there is no consistent perturbative string theory interpretation of such terms so we will choose the counterterms so that $C_1=C_2=0$.
In that case the total contribution to the amplitude in the low-energy limit is given by adding 
\eqref{gluoncal3a} and \eqref{gluoncal6}), which gives
\begin{eqnarray}
A^{1-loop} &=&\frac{2}{3(2\pi)^{10}}\,  \frac{2\pi^4}{\lepl\raten } \zeta(2)\left(\mathcal{C}_{adj} -\frac{1}{2} \, \mathcal{C}_{spin}  \right)  \,t_8 \hat F^4 \nonumber\\
&=& \frac{1}{12(2\pi)^6} \frac{1}{\lepl\raten }\zeta(2)\, \tr(T^{a_1}T^{a_2}T^{a_3}T^{a_4})\, t_8 \hat F^4 \,.
\label{totloop}
\end{eqnarray}
We see therefore that the amplitude vanishes in the $\raten\to \infty$ limit, which is consistent with fact that there is no $t_8\tr F^4$ term in the $E_8$ gauge theory.

\sm 

The preceding results translate into the following effective action in the various string theories. (assuming that the gauge potentials are in the $SO(16)$ subgroup labelled $1$ and ignoring the overall normalisation)
\begin{eqnarray}
	\, t_8 \tr_1F_1^4\frac{1}{(\lepl\raten )} \zeta(2) &=&t_8 \tr_1F_1^4 	
	\frac{1}{\lesth \rahe }\zeta(2) = 	t_8 \tr_1F_1^4 	
	\frac{\raho}{\lesth }\zeta(2)	 \nonumber\\
&=&t_8 \tr_1F_1^4 \frac{1}{\lesti\gia}\zeta(2)  = t_8 \tr_1F_1^4 \frac{\raib}{\lesti \gib}\zeta(2) \, .
\label{heterf4}
\end{eqnarray}
From the expressions  on the first line we see that the ten-dimensional limit of the HE theory ($\rahe\to \infty$) has no one-loop contribution to $t_8 \tr_1F_1^4$ whereas the $\raho\to \infty$ limit of the HO theory is non-zero.  Recall that we earlier found an explanation for a ten-dimensional contribution to the one-loop $t_8 (\tr_1 F_1^2)^2$ interaction in the HE theory (in terms of a tree-level supergravity amplitude), which vanished in the ten-dimensional HO limit.  From the second line of \eqref{heterf4} we see that both the type IA and type I theories have tree-level contributions to   $t_8 \tr_1F_1^4$ in the ten-dimensional limit.  In the type I case the presence of the requisite volume factor of $\raib$ indicates that the limit involves the distance between the Ho\u rava --Witten walls.  In  the type IA theory there is no volume factor because the interaction is localised in one wall independent of the value of $\raia$.  These type I contributions come from  disk diagrams, which only generate the single-trace $t_8\tr F^4$. 

\sm

We have thus accounted for the string theory result that the one-loop amplitude in the HO theory is proportional to $\tr_1 F_1^4+ \tr_2 F_2^4$ with the trace in the fundamental representation of either $SO(16)$ subgroup, whereas the standard (UV divergent) $SO(16) \times SO(16)$ gauge theory loop amplitude would be proportional to $\Tr_1 F_1^4+ \Tr_2 F_2^4$, with the trace in the adjoint representation of either $SO(16)$.  This is also in agreement with the form of the type I open-string tree amplitude (defined on a world-sheet disk) that has a group theory Chan--Paton factor.   
 
 \subsection*{One-loop amplitude in the compactified $SO(32)$ gauge theory}

It is of interest to see how the above results  are complemented by starting from a one-loop four-particle amplitude in the compactified $SO(32)$ theory.  However, since this is not a Feynman diagram that arises in  supergravity in the Ho\u rava --Witten background we have relegated the detailed argument to appendix \ref{sec:holoopamp}.  There we consider the  one-loop amplitude compactified on a circle of radius $\lepl/(\raten \raele)  = \lesth  \raho$ with $SO(32)$ gauge symmetry broken to $SO(16) \times SO(16)$.  
The discussion parallels that of the $E_8$ gauge theory loop in section \ref{sec:heloopamp}.  In this case we find that when the non-zero winding number configurations of the loop around the dual $x^{10}$ direction are interpreted as Kaluza--Klein modes of the HE theory, and with the identification $\rahe =1/\raho$, the  loop amplitude reproduces the correct effective action involving a combination of $t_8\, (\tr F^2)^2$ interactions obtained earlier (see \eqref{oneloophe})  by considering tree amplitudes in supergravity in the Ho\u rava--Witten background.  
 

\section{Some features of four-graviton amplitudes in ${\cal N}=1$ string theories}
\label{sec:graviamps}

As in the case of the Yang--Mills amplitude it is useful to describe some features that arise in considering four-graviton amplitudes in the various kinds of ${\cal N}=1$ supersymmetric  string theory amplitudes before describing the corresponding features in supergravity in the Ho\u rava--Witten background.

\sm 

Recall that in the maximally supersymmetric ${\cal N}=2$ case the only kinematic structures that contribute to the low-energy expansion of the four-graviton scattering amplitude have the form of derivatives acting on $t_8 t_8 \curv^4$ and $\epsilon_{10}\epsilon_{10} \curv^4$.  In the type IIA theory there is a single parity-violating term (the Vafa--Witten term \cite{Vafa:1995fj}) of the form $ B\wedge \curv \wedge \curv \wedge \curv \wedge \curv$ (or $\epsilon_{10} B \yvw $ in the notation of  \eqref{vwterm} in appendix \ref{sec:gravityreview}).  This structure is a consequence of the very strong constraints of maximal supersymmetry.   However, in ${\cal N}=1$  theories two other structures arise.  These are parity conserving terms of the form  $ t_8 \tr \curv^4$ and $t_8 (\tr \curv^2)^2$, where the trace is over the ten-dimensional tangent-space group, $SO(9,1)$.  These parity conserving interactions are not independent since they satisfy the identity \cite{Tseytlin:1995bi} 
\begin{eqnarray}
t_8t_8\, \curv^4 - 24\, t_8\, \tr \curv^4 + 6\, t_8\, (\tr \curv^2)^2 =0\,.
\label{rfourrelations}
\end{eqnarray}
This linear relationship means there is an ambiguity in the choice of basis for these terms.  We will see below that a natural basis is defined in the Ho\u rava--Witten description since $t_8t_8\, \curv^4$ is produced entirely by a bulk effect while a particular combination of the other two terms is localised on the boundaries.  As reviewed in appendix~\ref{sec:gravityreview} these two localised  $\curv^4$ interactions  are related by supersymmetry to corresponding parity-violating interactions terms, $\epsilon_{10} B \tr \curv^4$ and $\epsilon_{10} B (\tr \curv^2)^2$.

\sm 

In the following it will be useful to  recall the origin of $\curv^4$ terms in $\mathcal N=1$ perturbative  superstring theory, which we will now summarise.

\begin{itemize}
\item  The HE and HO theories have leading tree-level pole contributions of the form $  t_8 t_8 \curv^4/stu$,  which is simply a compact way of expressing the sum of all the tree level four-graviton amplitudes in Einstein gravity.  There are also poles in amplitudes of higher order in the low-energy expansion  that correspond to terms of the form $t_8 \tr \curv^4/st$ and  $t_8 (\tr \curv^2)^2/s$.  
 
\item  At the next order of the tree-level expansion of the heterotic theories there are terms of the form  $\zeta(3)\, t_8 t_8 \curv^4$  and  $t_8 (\tr \curv^2)^2$, where we have again explicitly indicated the occurrence of a notable  factor of $\zeta(3)$.   Although \eqref{rfourrelations} implies an ambiguity in how the combination of $\curv^4$ terms is expressed,  the coefficient of $\zeta(3)$ in the $\zeta(3)\, t_8 t_8 \curv^4$ term suggests that this particular term is singled out from the other one in an unambiguous fashion.

\sm 

  In the type I theory, the term $\zeta(3)\, t_8 t_8 \curv^4$ again arises from the spherical world-sheet diagram and is of order $1/\gib^2$, but $t_8 (\tr \curv^2)^2$  comes from diagrams with  two open-string loops (three boundaries/cross-caps) and are of order $\gib$ (which is analogous to the origin of the $(\tr F^2)^2$ terms considered earlier).
 
\item The one-loop  HE or HO contributions again have form $ t_8 t_8\curv^4$ and  $t_8 (\tr \curv^2)^2$.  But it is now natural to express this (using \eqref{rfourrelations}) as the sum of 
  $ t_8 t_8 \curv^4$ and  $t_8 \ygs $  (where $t_8\, \ygs $ contains the combination of $t_8 \tr \curv^4$ and  $t_8 (\tr \curv^2)^2$ defined in \eqref{gsfull}).  As described in the context of the Yang--Mills action, this is motivated by supersymmetry with the one-loop anomaly cancelling terms, which have the form $\epsilon_{10} B\, \ygs $.  
We will describe later how these observations fit with HO - type I duality.
 
 \end{itemize}
 
 {\it $\mathcal{N}=1$ effective $\curv^4$ actions at tree-level and one loop}
 
We will here summarise the duality relationships between the $\curv^4$ interactions in the various  $\mathcal{N}=1$ string theories in terms of their  effective actions compactified to nine dimensions.  These can be expressed with the help of the $\mathcal{N} = 1$ superinvariants $J_0$ (defined in \eqref{j0def}), $\mathcal{I}_2$ (defined in \eqref{supdef}) and $X_1$ and $X_2$ (defined in  \eqref {x1x2def}), where the notation is based on \cite{Tseytlin:2000sf}.   The string frame effective action for the terms of order $R^4$ in the nine-dimensional HO theory at tree-level and one loop, obtained by combining expressions in \cite{Gross:1986mw,Ellis:1987dc,Abe:1988cq}, is\footnote{We are grateful to Michael Haack for pointing out errors in this equation in the first version of this paper.} 
\begin{eqnarray}
S_{\textrm{HO}}\left |_{\curv^4}	\right.&=&
 \frac{\raho }{2^9 (2\pi)^6\, 4! \lesth } \int_{\mathcal{M}_{9}} d^{9}x
\sqrt{-G}\left[
\frac{2\zeta(3)}{\gho^2}\left(t_8t_8\curv^4-\frac{1}{8}\epsilon_{10}\epsilon_{10}\curv^4
\right)-\frac{1}{2\gho^2}t_8 (\textrm{tr}\curv^2)^2 
\right.
\nonumber\\
&& 
\left.
+\frac{2\pi^2}{3} \left( 48t_8\ygs(\curv,0) -12\epsilon_{10}B \ygs (\curv,0)
\right)\left(1+ \frac{1}{\raho^2}\right)\right] \nn\\
&=&
\frac{\raho}{2^9 (2\pi)^6\, 4!  \lesth }
 \int_{\mathcal{M}_{9}} d^9 x
\sqrt{-G}\left(
	\frac{2\zeta(3)}{\gho^2}J_0 - \frac{1}{2\gho^2} (\tr \curv^2)^2 \right. \nn\\
&&\left. \qquad+\frac{2\pi^2}{3} \left(J_0 -  \mathcal{I}_2 +24X_1 + 18 X_2\right) \left(1+ \frac{1}{\raho^2}\right)\right)
\, .
\label{hoeffstring}
\end{eqnarray}
Here the contributions of the massless Kaluza--Klein scalar associated with the compact $x^{10}$ direction should be included in the definition of the superinvariants, although they  have been ignored in the first two lines of this equation. 

\sm

For later comparison with the type I theory  it is useful to write the HO one-loop contribution to the parity-conserving terms in the last line of \eqref{hoeffstring} using the identities in appendix \ref{sec:gravityreview}
\begin{eqnarray}
48t_8\ygs  &=& J_0 -(J_0 - 48\,t_8\, \tr \curv^4 - 12 \, t_8\,(\tr \curv^2)^2)\nn\\
  &=&  J_0 + (24 \,t_8\, \tr \curv^4 + 18 \, t_8\,(\tr \curv^2)^2+ \frac{1}{8}\,\epsilon_{10}\epsilon_{10}\, R^4)\,.
\label{looprelate}
\end{eqnarray}
The combination of terms in parentheses will be  identified with disk-level contributions to the amplitude in type I theory whereas the first term arises from the torus diagram.  Further understanding of these points will emerge from the analysis in section \ref{sec:gravloop}.

\sm 

The following comments concerning the effective action are of note:

\begin{itemize}

\item  The parity-violating anomaly cancelling term $-12\epsilon_{10}B \ygs$ in the second line is contained in the combination of the invariants $X_1$, $X_2$ and $ \mathcal{I}_2 $ in the fourth line.   These terms are one-loop exact in the HE and HO theories.  The tree-level  interaction in the first line  $t_8(\tr \curv^2)^2$  is part of the expression $t_8(\tr F^2 - \tr \curv^2)^2$ when the gauge fields are included. This receives no loop corrections since it is related by supersymmetry to the three-point interactions in $t_8 (\tr F^2 - \tr \curv^2)$ and these are unrenormalised beyond tree level.  
  The relationship \eqref{rfourrelations} means that there is an ambiguity in the coefficients of $t_8t_8 R^4$, $t_8\tr R^4$ and $t_8 (\tr R^2)^2$ in the one-loop  terms in the second line of \eqref{hoeffstring}.   However, there is no ambiguity in the expression written in terms of superinvariants,  which is exhibited in the last two lines of the equation.

\item 
The effective action for the HE theory is obtained simply by performing the heterotic T-duality relation \eqref{hettdual1}.  The expression \eqref{hoeffstring}  is invariant under this transformation since the four-graviton amplitudes in  HO and HE theories on a spherical or a toroidal world-sheet are identical - they are insensitive to the details of the gauge group \footnote{The group theory lattice factor is same for the two gauge groups. }.

\item

The nine-dimensional type I superstring effective action is proportional to 
\begin{eqnarray}
S_{\curv^4}&=&
\frac{\raib}{2^9(2\pi)^6 4!\lesti}\int_{\mathcal{M}_{9}} d^9x
\sqrt{-G}  \left(\frac{2\zeta(3)}{\gib^2}J_0
+\frac{2\pi^2}{3\gib} \left(-  \mathcal{I}_2 +24X_1 + 18 X_2\right)\right. \nn\\
&&\left. \qquad\qquad\qquad+ \frac{2\pi^2}{3} J_0
-\frac{1}{2}\,\gib (\tr \curv^2)^2
\right) \,.
\label{typeIr4}
 \end{eqnarray} 
 As expected the interactions that are tree-level or one-loop exact, translate straightforwardly from the HO theory using the S-duality relations.    The $(2\zeta(3) /\gib^2+2 \pi^2/3) J_0$ terms arise from the diagrams with spherical and toroidal world-sheets, just as in the HO theory.  S-duality cannot simply act on these terms in isolation, since they would transform into  terms of order $\gho$ and $1/\gho$, respectively, which are powers that do not make sense in the  HO theory.  This is a signal that the  coupling constant-dependent coefficient of $t_8 t_8 \curv^4$ in the HO/type I theories is a non-trivial function of the coupling constant that transforms under S-duality in a manner that preserves these first two perturbative terms.   We will later find a candidate for such a function motivated by supergravity in the Hora\u va--Witten background, in a manner analogous to the modular function that enters as the coefficient of $t_8t_8 \curv^4$ in the  type IIB theory.  
 
 \item
Whereas in the HO expression \eqref{hoeffstring} the $\epsilon_{10}\epsilon_{10}\curv^4$ interaction only arises at tree level, in the type I expression \eqref{typeIr4} there are  three distinct terms containing this interaction.  Two of these  arise from the spherical and toroidal world-sheets in the same manner as  in the type IIB theory (and  are proportional to $1/\gib^2$ and $\gib^0$).  According to  \eqref{typeIr4} there should also be a disk contribution proportional to $1/\gib$, which has not been determined directly from the string theory.  The  $\epsilon_{10}\epsilon_{10}\curv^4$ interaction does not contribute to the graviton four-point function so it would be necessary to evaluate an amplitude with $N$ gravitons coupling to a disk, with $N\ge 5$, in order to verify its presence. 
 
  \item
A certain amount is known concerning higher derivative terms in $\mathcal{N}=1$ supersymmetric string perturbation  theory that will also  be discussed in the context of supergravity in the  Ho\u rava--Witten background.

\end{itemize}

\section{Graviton tree amplitudes in the Ho\u rava--Witten background }
\label{sec:gravitrees}
 
 We will now turn to the explicit calculations of graviton amplitudes in the Ho\u rava--Witten background.   
 This will extend the earlier analysis of the gauge theory amplitudes. 
 The graviton polarization tensor  satisfies $k^\mu \, \zeta_{\mu \nu} =k^\nu \, \zeta_{\mu \nu} =0$.    The symmetric part of $\zeta_{\mu\nu}$ describes the graviton polarisation while the antisymmetric part describes the polarisation of an antisymmetric tensor potential.  The  curvature tensor linearised around Minkowski space has the form
\begin{eqnarray}
\hat \curv_{\mu\nu\rho\sigma} = k_\mu \, k_\rho\, \zeta_{\nu \sigma} \,.
\label{lincurve}
\end{eqnarray} 
We will again  only consider amplitudes in which the  external momenta   and polarization tensors are oriented in the nine non-compact dimensions, so we will not consider the scattering of the scalar states arising from Kaluza--Klein compactification. Since much of the analysis is a simple extension of the analysis of the scattering of  gauge particles the following exposition will be brief. 
  
\subsection*{Graviton tree amplitudes}
Four graviton tree amplitudes in the Ho\u rava--Witten background  arise from the four-graviton vertex operator on sphere in the heterotic theories. But they arise at different orders of type I perturbation theory.

\sm 

There are tree amplitudes with conventional gravitational vertices in the eleven-dimensional bulk, as well as trees with either one or both vertices localised in the boundary induced by the boundary Chern--Simons and $\curv^2$ interactions.      
In the following we will adapt the notation used to label the Yang--Mills tree amplitudes by denoting the gravitational amplitudes by 
\begin{eqnarray}
A^G_{IJ} = (A^G_{00}\,,A^G_{i0}\,, A^G_{ij}) \,,
\label{gravcon}
\end{eqnarray}
where $i,j =1,2$ labels the boundary in which a vertex is localised and $0$ denotes a bulk vertex operator.

\subsection{The bulk tree}

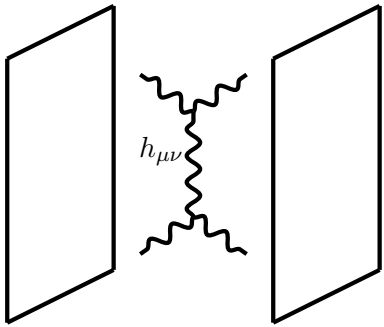
\begin{figure}[h]
\begin{center}
\begin{tikzpicture}[scale=.7]
\begin{scope}[shift={(0,0)}]
\draw [hwbou](-2,-3) -- (-2,2);
\draw [hwbou] (-2,2)-- (0,3);
\draw [hwbou] (-2,-3)-- (0,-2);
\draw [hwbou] (0,3) -- (0,-2);
\end{scope}
\begin{scope}[shift={(5,0)}]
\draw [hwbou](-2,-3) -- (-2,2);
\draw [hwbou] (-2,2)-- (0,3);
\draw [hwbou] (-2,-3)-- (0,-2);
\draw [hwbou] (0,3) -- (0,-2);
\draw (-4.1,.25) node{  ${h_{\mu \nu}} $}; 
\begin{scope}[shift={(-3.5,1)}]
\draw [sugra, ultra thick](0,0) -- (-1,.7);
\draw [sugra, ultra thick](0,0) -- (1,.7);;
\end{scope}
\draw [sugra, ultra thick](-3.5,-1) -- (-3.5,1);
\begin{scope}[shift={(-3.5,-1)}]
\draw [sugra, ultra thick](0,0) -- (-1,-.7);
\draw [sugra, ultra thick](0,0) -- (1,-.7);;
\end{scope}
\end{scope}
\end{tikzpicture}
\end{center}
\caption{Tree level four-graviton amplitude with graviton exchange  in the bulk}
\label{superbulk}
\end{figure}

The supergravity tree amplitude  (together with the standard four-graviton contact interaction) has the form 
\begin{eqnarray}
A^G_{00}= \frac{2\raele\, \raten} {(2\pi)^6\lepl^7}\, t_8t_8\, \hat \curv^4\frac{1}{stu} &=& \frac{2\rahe}{(2\pi)^6\, \ghe^2\lesth^7} \, t_8t_8\, \hat \curv^4\frac{1}{stu}=\frac{2\raho}{(2\pi)^6\, \gho^2\lesth^7} \, t_8t_8\, \hat \curv^4\frac{1}{stu}
\nonumber\\
&=& \frac{2\raib}{(2\pi)^6\, \gib^2\lesth^7} \, t_8t_8\, \hat \curv^4\frac{1}{stu}=\frac{2\raia }{(2\pi)^6\, \gia^2\lesth^7} \, t_8t_8\, \hat \curv^4\frac{1}{stu} \,,
\label{ehtree}
\end{eqnarray}
where we have displayed the interpretation of the supergravity tree amplitude in all this $\mathcal{N}=1$ string theory. This amplitude is the tree level amplitude of ordinary Einstein gravity where the intermediate particle can only be graviton. Since the external states are assumed to have $p_{11} = 0$ this is identical to the expected tree-level interaction of ten-dimensional gravity coupled to a dilaton. This amplitude arises from a spherical world-sheet in any of the $\mathcal{N}=1$ string perturbation expansions. 

\subsection{Tree amplitude with one vertex on a boundary }

\begin{figure}[h]
\begin{center}
\begin{tikzpicture}[scale=.7]
\begin{scope}[shift={(0,0)}]
\draw [hwbou](-2,-3) -- (-2,2);
\draw [hwbou] (-2,2)-- (0,3);
\draw [hwbou] (-2,-3)-- (0,-2);
\draw [hwbou] (0,3) -- (0,-2);
\begin{scope}[shift={(-1,-1)}]
\filldraw (0,0) circle (.05) ; 		
\draw [sugra, ultra thick](0,0) -- (-.5,-1);
\draw [sugra, ultra thick](0,0) -- (.6,-.7);
\end{scope}
\end{scope}
\begin{scope}[shift={(5,0)}]
\draw [hwbou](-2,-3) -- (-2,2);
\draw [hwbou] (-2,2)-- (0,3);
\draw [hwbou] (-2,-3)-- (0,-2);
\draw [hwbou] (0,3) -- (0,-2);
\draw [sugra, ultra thick](-6,-1) -- (-3,0);
\draw (-4.3, -1.20) node{  ${h_{\mu \nu}} $}; 
\begin{scope}[shift={(-3,0)}]
\draw [sugra, ultra thick](0,0) -- (-.5,1);
\draw [sugra, ultra thick](0,0) -- (.6,.7);;
\filldraw (0,0) circle (.05) ; 		
\end{scope}
\end{scope}
\end{tikzpicture}
\end{center}
\caption{Graviton tree amplitude with graviton exchange coupling to one vertex localised on a boundary}
\label{grtreefix}
\end{figure}
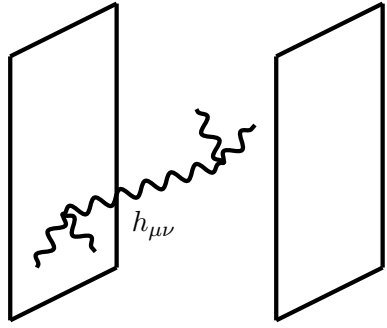

This amplitude is obtained by joining a bulk graviton vertex to the cubic contribution to the $\curv^2$ vertex localised on either boundary as in figure~\ref{grtreefix},  together with the four-point contact term that arises from $\curv^2$. This is a higher derivative pole contribution of order $s \lesth^2$ relative to the Einstein gravity tree. This gives a contribution to the amplitude that is independent of $\raele$  since it is localised on only one of the boundary, and has the following form 
\begin{eqnarray}
A^G_{10} =  \frac{\raten} {12(2\pi)^6\lepl^5 }  \, t_8\tr  \hat \curv^4\, \frac{1}{tu} + {\rm perms.}& =& \frac{\rahe}{12(2\pi)^6 \ghe^2\lesth^5} \, t_8\,\tr  \hat \curv^4\frac{1}{tu} + {\rm perms.} \nonumber\\
&=&   \frac{\raib}{12(2\pi)^6 \gib\lesti^5} \, t_8\,\tr  \hat \curv^4\frac{1}{tu} + {\rm perms.}\nonumber\\ 
&=&   \frac{1}{12(2\pi)^6 \gia\lesti^5} \, t_8\,\tr  \hat \curv^4\frac{1}{tu} + {\rm perms.}\,,
 \label{onebound}
\end{eqnarray} 
where we have translated the amplitude into the heterotic, type I and type IA  parmeterisations.    
The fact that the amplitude is localised on a single Ho\u rava--Witten boundary means that its type IA description is independent of $\raia$, although it is proportional to $\raib$ in the  T-dual type I description. 

\sm 

While this amplitude arises as a higher order term in the expansion of the tree amplitude in the HE and HO theories,  in type I perturbation theory,  it arises from the sum of the amplitudes on the disk and $\mathbb{RP}^2$ world-sheets with four closed string vertex operators. 

\sm 

The amplitude  $A^G_{20}$ has an identical form and the total amplitude is $A^G_{10}+A^G_{20}$.

\subsection{Tree amplitude with both vertices on boundaries}

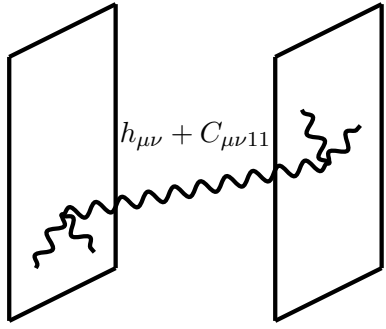
\begin{figure}[h]
\begin{center}
\begin{tikzpicture}[scale=.7]
\begin{scope}[shift={(0,0)}]
\draw [hwbou](-2,-3) -- (-2,2);
\draw [hwbou] (-2,2)-- (0,3);
\draw [hwbou] (-2,-3)-- (0,-2);
\draw [hwbou] (0,3) -- (0,-2);
\begin{scope}[shift={(-1,-1)}]
\filldraw (0,0) circle (.05) ; 		
\draw [sugra, ultra thick](0,0) -- (-.5,-1);
\draw [sugra, ultra thick](0,0) -- (.6,-.7);
\end{scope}
\end{scope}
\begin{scope}[shift={(5,0)}]
\draw [hwbou](-2,-3) -- (-2,2);
\draw [hwbou] (-2,2)-- (0,3);
\draw [hwbou] (-2,-3)-- (0,-2);
\draw [hwbou] (0,3) -- (0,-2);
\draw [sugra, ultra thick](-6,-1) -- (-1,0);
\draw (-3.5,.5) node{  ${h_{\mu \nu}+C_{\mu \nu 11}}$};
\begin{scope}[shift={(-1,0)}]
\draw [sugra, ultra thick](0,0) -- (-.5,1);
\draw [sugra, ultra thick](0,0) -- (.6,.7);;
\filldraw (0,0) circle (.05) ; 		
\end{scope}

\end{scope}
\end{tikzpicture}
\end{center}
\caption{Graviton tree amplitude with graviton and $C$ exchange with vertices localised on one or both boundaries}\end{figure}

In this case the vertices are either the $\curv^2$ vertex or the gravitational Chern--Simons vertex, both of which are localised on the boundaries.  The exchanged particle can now be either a graviton or the potential $C_{\mu \nu 11}$. 
This case is analogous to the Yang--Mills amplitudes depicted in figure \ref{fig:ymtree2} and figure \ref{fig:ymtree3}.  The form of the amplitude is 
\begin{eqnarray} 
 \label{gravamp}
 A_{ij}^G &=&   \frac{ \raten  }{2^4 (2\pi)^6\,  \lepl^3 \,\raele }\, \sum_{m=-\infty}^{\infty}\frac{ (-1)^{m(i-j)}}{ -s+\frac{\pi^2m^2}{\ciele^2}} \,  \, t_8 (\tr \hat \curv^2)^2    \,, 
 \end{eqnarray}  
 where the numerator factor is $1$ when the vertices are on the same boundary or $(-1)^m$ when they are on different boundaries. The term with $m=0$ gives the pole contribution
\begin{eqnarray}
\label{twobound}
 \frac{ \raten  }{2^4 (2\pi)^6\,  \lepl^3 \,\raele }\,t_8(\tr \hat\curv^2)^2\, \frac{1}{s}
  &=&\frac{\rahe\,}{2^4(2\pi)^6\,\lesth^3 \ghe^2 } t_8(\tr \hat\curv^2)^2\, .\frac{1}{s}
\nonumber\\ 
 &=&\frac{\raho\,}{2^4(2\pi)^6\,\lesth^3 \gho^2 } t_8(\tr \hat\curv^2)^2\, .\frac{1}{s}
\end{eqnarray}

\sm

The next term in the low energy expansion is given by setting $s=0$ in \eqref{gravamp}.  Combining the contributions from $A^G_{11}$, $A^G_{22}$, $A^G_{12}$ and $A^G_{21}$ gives 
\begin{eqnarray}
\frac{ 3\, \raten  }{2^4 (2\pi)^6\,  \lepl^3 \,\raele}  \zeta(2) \, t_8(\tr \hat \curv^2)^2
=  \frac{\rahe\,}{2^7(2\pi)^4\, \lesth } t_8(\tr \hat \curv^2)^2
	\label{moreeqn5}
\end{eqnarray}
This corresponds to a $t_8 \, (\tr \curv^2)^2$ contribution to the one-loop HE effective action that is the parity-conserving partner of the parity-violating term $\epsilon_{10}\, B\, (\tr \curv^2)^2$, which is part of the GS anomaly cancelling term.  As we will see in section \ref{sec:gravloop} the other part of the parity-conserving partner of the anomaly-cancelling term in the HE theory, which is proportional to $t_8t_8 \, \curv^4$, emerges from the contribution of a graviton loop propagating in the bulk in the Ho\u rava--Witten background and has the coefficient $\hat C$ in  \eqref{Idual7} and \eqref{hor4}.
 
 \sm
 
The discussion of the expansion of the non-zero KK terms in powers of $\raele^2 \lepl^2 s$ is similar to the discussion of the higher derivative $t_8(\tr F^2)^2$ interactions in section~\ref{sec:YMtrees}.  As in that case all the terms in the expansion arise from contributions of order $\gib^0$ in the type I description, which are associated with a world-sheet  cylinder, M\"obius strip and Klein bottle.  The integer $m$ is the type I closed string winding number around the compact dimension.

\subsection{``Iterated'' graviton tree diagrams}
\label{subsec:itergrav}
 Similarly  there are gravitational tree amplitudes with the same structure as the Yang--Mills amplitudes discussed in subsection \ref{subsec:itetree}, in which there are chains of propagators joining the vertices with the external particles.  These are gravitational analogues of the Yang--Mills processes shown in  figures \ref{fig:iterone} and \ref{fig:itertwo}.    These possibilities generate higher derivative contributions to the tree processes described in section~\ref{sec:gravitrees}, in much the same way as that discussed in the context of the Yang--Mills interactions earlier.
 
\section{Graviton one-loop amplitudes}
\label{sec:gravloop}

We turn now to consider the one-loop  four-graviton amplitude. There are two kinds of loop amplitudes that contribute to leading terms in the low-energy expansion.  
In  section~\ref{subsec:gravloop2} we will consider the loop with circulating super-gauge particles localised in either 
ten-dimensional boundary and compactified on a circle of radius $R_{10}$.  This can be constructed by use of a  light-cone gauge world-line vertex operator formalism based on the vertex in \eqref{firstcs}, which describes the emission of a graviton from a super Yang--Mills world-line.  The resulting loop amplitude  is identical to the contribution that arises from the gauge loop in  ten-dimensional ${\cal N}=1$ supergravity compactified to nine dimensions.   Much as in the case of the Yang--Mills four-particle loop amplitude discussed in section~\ref{sec:loopamp},  after transforming to the winding number basis by performing a  Poisson summation over the Kaluza--Klein modes in the $x^{10}$ direction, we are able to make contact with various $\curv^4$ terms in the string theory effective action. 

\sm

The other loop contribution is the ``bulk'' gravity loop  with circulating supergravity particles propagating in the eleven-dimensional space compactified on the interval of length $L=  \pi \raele \lepl$ and a circle of radius $\raten\lepl$, which is the subject of sections~\ref{subsec:susyloops} and  \ref{subsec:gravloop1}.
In section~\ref{subsec:susyloops}   we will make use of an extension of the  light-cone gauge vertex operator construction  used in the description of the four-graviton loop in eleven-dimensional supergravity compactified on $S^1$ \cite{Green:1997as, Green:1999by}.  Implementing the $\mathbb{Z}_2$ orbifold condition that defines the Ho\u rava--Witten background raises some subtleties connected with the breaking of supersymmetry.  The amplitude that results from this construction is discussed in section~\ref{subsec:gravloop1}.  Its low energy limit  contains a $t_8t_8\, \curv^4$ contribution that manifests the  strong coupling duality relating the HO and type I theories in an interesting manner.

\smallskip

\subsection{Four gravitons coupled to gauge particle loop on the boundary}
\label{subsec:gravloop2}

\begin{figure}[h]
\begin{center}
\begin{tikzpicture}[scale=1.2]
\begin{scope}[shift={(0,0)}]
\draw [hwbou](-2,-3) -- (-2,2);
\draw [hwbou] (-2,2)-- (0,3);
\draw [hwbou] (-2,-3)-- (0,-2);
\draw [hwbou] (0,3) -- (0,-2);
\draw [sugra, ultra thick](-0.2,2.0) -- (-.65,.85);
\draw [sugra, ultra thick](-0.2,-1.6) -- (-.55,-.6);
\draw [sugra, ultra thick](-1.8,1.8) -- (-1.41,.7);
\draw [sugra, ultra thick](-1.8,-2) -- (-1.41,-.7);
\draw  [suym, ultra thick](-1,0) ellipse (.5cm and 1.2cm);
\end{scope}

\begin{scope}[shift={(5,0)}]
\draw [hwbou](-2,-3) -- (-2,2);
\draw [hwbou] (-2,2)-- (0,3);
\draw [hwbou] (-2,-3)-- (0,-2);
\draw [hwbou] (0,3) -- (0,-2);
\end{scope}
\end{tikzpicture}
\end{center}
\caption{ One-loop four-graviton amplitude localised in a Ho\u{r}ava-Witten boundary.}
\label{oneloopgrav}
\end{figure}
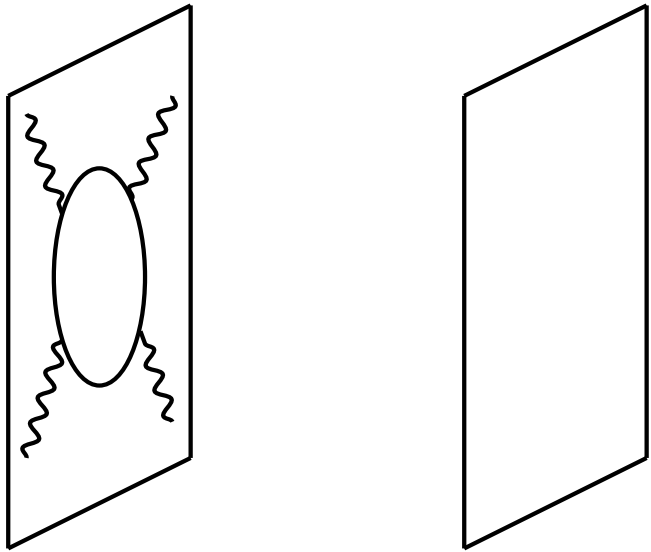

The one-loop  amplitude with gravitons coupled to a circulating supermultiplet of  $\mathcal{N}=1$  gauge  fields in one boundary is depicted in figure \ref{oneloopgrav}.    Since we are restricting the graviton polarisations and momenta to lie in the boundary directions the bulk propagation plays no r\^ole in this calculation and the amplitude could, in principle, be obtained from the Feynman rules of ten-dimensional $\cN=1$ supergravity compactified to nine dimensions in the $x^{10}$ direction. 
The expression for this amplitude could also be determined by making use of a world-line formalism analogous of that described in the case of the bulk loop in sections~\ref{subsec:susyloops} and \ref{subsec:gravloop1}.  This involves the product of four vertex operators of the form $V_{bulk}$ \eqref{firstcs} that couple the external gravitons to a $\cN=1$ gauge supermultiplet.
 We will denote the low energy limit of this amplitude by $A^{gauge\ loop}_i$, where the subscript $i=1,2$ labels the boundary.

The complete ten-dimensional $\cN=1$ one-loop supergravity amplitude also involves a contribution from the supergravity multiplet circulating in the loop.   As we will see in section~\ref{subsec:gravloop1} this contribution is naturally thought of as a $p_{11}=0$ contribution to the ``bulk loop''.  We will here denote it by $A^{gravity\ loop}_{p_{11}=0}$.

\sm

Although we have not performed the explicit  gauge loop calculations in detail we know that when added to $A^{gravity\ loop}_{p_{11}=0}$ the result will contribute to the parity conserving part of a linear sum of $\cN=1$ invariants, which also contains parity violating pieces of these $R^4$ terms.  Therefore, the precise combination of parity conserving $R^4$ interactions is determined from knowledge of the anomaly cancelling terms.
So we conclude that  the  amplitude must be proportional to the one-loop  kinematic factor   $t_8\left(\tr \hat \curv^4 + (\tr \hat \curv^2)^2/4 \right)= t_8\, \ygs(\hat \curv,0)$   multiplying a scalar box integral.  In other words, after adjusting the normalisation to agree with \eqref{hoeffstring}, the low energy limit of the loop amplitude is given by
\begin{eqnarray}
A^{total}_{{\cal N}=1}&=&   A^{gauge\ loop}_1+A^{gauge\ loop}_2 + A^{gravity\ loop}_{p_{11}=0}\nn\\
  &=&\frac{3}{(2\pi)^{10} }t_8\left(\tr \hat \curv^4 + \quart (\tr \hat \curv^2)^2 \right) I(0,0,0; \raten ) 	\,.
 \label{Ical7.2}   
\end{eqnarray} 

\sm 

Using  \eqref{gluoncal3a} and  \eqref{gluoncal6} we have
\begin{eqnarray}
A^{total}_{{\cal N}=1}=&&\frac{3}{(2\pi)^9 }
t_8\left(\tr \hat \curv^4 +  \quart (\tr \hat \curv^2)^2 \right)
\left(
\hat C\,\frac{\raten}{\lepl}
+
\frac{\pi ^3}{\lepl\raten}
\zeta(2)
\right) \,.
\label{Ical7.5}  
\end{eqnarray}
As in the case of the gauge theory loop amplitude, we will set  the coefficient of the renormalised divergence to zero, $\hat C=0$, since the quantity $\raten/\lepl$ translates (using the relations \eqref{Idua2ho}) into $(\raho^{4/3}\, \gho^{-2/3})/ \lesth$ in the HO description, which involves a nonsensical power of the string coupling.

\sm

The expression \eqref{Ical7.5} corresponds to an effective action in terms of the HO string theory parameters of the form
\begin{eqnarray}
\frac{3\, \raho}{4!\, 2^3(2\pi)^6\lesth}\zeta(2)
\int_{\mathcal{M}_{9}} d^9x \sqrt{-G} \, 
t_8\, \ygs(\hat \curv,0)\,.
\label{efftrr4}
\end{eqnarray}
The analogous expression in the HE theory is proportional to $1/\rahe$. 
In type I string theory this interaction is of order $1/\gib$  and arises from  four graviton vertex operators coupled to a world-sheet disk and to the projective plane $\mathbb{RP}^2$ (a sphere with a cross-cap).   However, as pointed out following \eqref{looprelate}, the type I theory also has a term proportional to $t_8t_8  \curv^4$ (contained in the $\cN=2$ invariant $J_0$), which is crucial for understanding how HO/type I duality is realised, as we will see following the discussion of the bulk loop in section~\ref{subsec:gravloop1}.

\sm

{\it Comments on supersymmetry connection with chiral anomaly cancelling terms.}

The above argument gave the parity-conserving  part of the combination of superinvariants $\left(- \mathcal{I}_2 +24X_1 + 18 X_2\right)$ in the HO theory (where the invariants are defined \eqref{x1x2def} and \eqref{i2comb}).  This combination also contains the parity-violating  anomaly-cancelling term $B\wedge \xgs (\curv)$ and is protected from higher loop corrections.  There is no ten-dimensional HE contribution from \eqref{efftrr4}  in the $\rahe\to \infty$ limit.  However,  we earlier found the HE one-loop  contributions to $t_8\,(\tr \curv^2)^2$ arising from the tree-level supergravity graphs obtained in \eqref{moreeqn5} in the previous section.   This  is part of the same superinvariant as the parity-violating ten-form, $\epsilon_{10} B (\tr \curv^2)^2$.  
 It was argued in \cite{Horava:1996ma} that the remaining part of the anomaly cancelling term in the HE theory is provided by the bulk Vafa--Witten ten-form of the type IIA theory $\epsilon_{10} B\, \yvw(\curv)$  (reviewed in appendix~\ref{sec:gravityreview}).  We will see  that the superpartner of the Vafa--Witten term ($t_8\, \yvw(\curv) = t_8 t_8\, \curv^4$)  is generated by the bulk loop calculation in section~\ref{subsec:gravloop1} (the term with coefficient $\hat C$ in  \eqref{Idual7} and \eqref{hor4}).
 
 \sm 
 
   In other words,  the tree amplitudes of supergravity in the Ho\u rava--Witten background combine with the Vafa--Witten interaction to give the anomaly cancelling terms in the HE theory, whereas in the HO theory these terms arise from the loop of gauge particles localised in either boundary.  
\sm

\subsection{Supersymmetry and the bulk one-loop amplitude.} 
\label{subsec:susyloops}

We will here describe the supersymmetric world-line formalism that will be used in the next subsection to determine  the properties of the four-graviton loop amplitude in which the circulating particles are bulk supergravitons.

  M-theory compactified on a $x^{11}$ circle is invariant under eleven-dimensional supersymmetry, associated with a 32-component $SO(10,1)$ Majorana spinor,  ${\cal Q}$, which decomposes into two 16-component $SO(9,1)$ spinors of opposite chirality,
${\cal Q} =( Q_1\,, Q_2 )$
which satisfy the chirality conditions
\begin{eqnarray}
	\Gamma_{11}\, Q_r = (-1)^r\, Q_r
\label{chiralspin}
\end{eqnarray}
where $r =1,2$ and $\Gamma_{11} = \Gamma_1\dots \Gamma_{10}$ is the product of the gamma matrices of the ten-dimensional theory.   The supercharges $Q_{1,2}$ are those of the type IIA theory and their anti-commutation relation takes the form 
\bea
\label{superalgebra}
\{Q_r,\, \bar Q_s\}= \delta_{rs} \, \Gamma^\mu  p_\mu + \epsilon_{rs} \, p_{11}\,
\qquad\qquad \mu =1,...,10
\eea
 where the Kaluza--Klein momentum, $p_{11}$, enters as the  central extension and  is the signal of $1/2$-BPS D0-brane states in type IIA string theory.    As discussed in \cite{Horava:1995qa, Horava:1996ma}, in the Ho\u rava--Witten  background the boundary conditions at $x^{11}=0$ and $x^{11}=L$ break the supersymmetry so that only  $Q_2$  survives and the theory possesses ${\cal N}=1$ ten-dimensional  supersymmetry.  

\smallskip

In constructing the bulk loop amplitude we will adapt the eleven-dimensional light-cone vertex operator formalism \cite{Green:1999by}, which was used to discuss one-loop amplitudes in  eleven-dimensional supergravity compactified on a $d$-torus, $T^d$.
In 11-dimensional Minkowski space we choose the light-cone gauge with $x^{\pm} = (x^1 \pm x^2)/2$.  The world-line fields comprise the transverse bosonic coordinates $x^I$ ($I=3,\dots,11$), which form a $SO(9)$ vector, and the fermionic coordinates, $S^A$, which form a 16-component $SO(9)$ spinor.  After compactification on a $x^{11}$ circle the light-cone coordinates naturally decompose into $x_i, x^{11}$, where $x^i$ is a $SO(8)$ vector  ($i=3,\dots,10$) and $S^A=(S_1^{\dot a},\, S_2^{ a})$, where $S_r$ ($r=1,2$) are eight-comonent  $SO(8)$ spinors of opposite chiralities (indicated by undotted and dotted indices).

\smallskip

Single-particle states are labelled by the vector and spinor indices appropriate for the ``left-moving'' and ``right-moving'' sectors of the type IIA  string theory, together with the value of the Kaluza--Klein charge $m$,
\bea
\label{stesdef}
|i,\tilde j; m\rangle  \,, \qquad  |i, \dot b; m\rangle  \,, \qquad |a, \tilde j; m\rangle \,, \qquad |a,\dot b;m\rangle \,.
\eea
The states in the massless supermultiplet of the IIA theory are the $m=0$ states and the Kaluza--Klein recurrences (the D0-brane states in the type IIA theory) have $m\ne 0$.
The type IIA supersymmetry generators can be expressed in terms of the $SO(8)$ spinors, $S_r$, which relates fermionic and bosonic states in  the following manner
\bea
\label{saction}
S_1^{\dot a} \, | i \rangle =  \gamma_i^{\dot a b}\, |  b \rangle\,, \qquad S_1^{\dot a} \, |  b \rangle =  \gamma_i^{\dot a  b}\, | i \rangle\,,\qquad
S_2^{a} \, | \tilde i \rangle =  \gamma_{\tilde i}^{a  \dot b}\, | \dot b \rangle\,, \qquad S_2^{a} \, |\dot b \rangle =  \gamma_{\tilde i}^{a \dot b}\, |\tilde i \rangle
\eea
(where we have suppressed the vector/spinor labels that are not affected by the action of $S_r$ on a state).

\smallskip

In considering the Ho\u rava-Witten background we need to identify states under the action of the orbifold  $\IZ_2$, which identifies $x^{11}$ with $-x^{11}$, which is represented by the action of an operator $\Omega$.   This reverses the sign of $m$ and changes the dotted spinor by a minus sign, giving
\bea
\label{omegct}
&&\Omega\, | i,\tilde j\, : m\rangle  =   | i,\tilde j\, ; -m\rangle   \,, \qquad   \Omega\, | i, \dot b\, ; m \rangle=  -| i, \dot b\, ; -m \rangle\,, \nn\\
&& \Omega\,   | a, \tilde j \, ; m\rangle= | a, \tilde j \, ; - m\rangle \,, \qquad \Omega\,  | a, \dot b\, ;m\rangle = - | a, \dot b \, ; -m\rangle\,.
\eea

\smallskip

The states that are invariant under $\Omega$ are those obtained by the action of the projection operator $(1+\Omega)/2$, which gives
\bea 
\label{[rojectop}
&&\frac{1}{2}(1+\Omega) \,  | i,\tilde j;m\rangle = \frac{1}{2}(| i,\tilde j; m\rangle +| i,\tilde j; - m\rangle )\,, \nn\\
&&
\frac{1}{2}(1+\Omega) \,   | a, \tilde j; m\rangle =\frac{1}{2}(| a, \tilde j ; m\rangle + | a, \tilde j ; -m\rangle) \,,\nn\\
&&
\frac{1}{2}(1+\Omega) \, | i, \dot b;m \rangle=  \frac{1}{2}(| i, \dot b;m \rangle -(| i, \dot b; -m \rangle)\,,\nn\\
&&
 \frac{1}{2}(1+\Omega) \,  | a, \dot b;m\rangle =  \frac{1}{2}(| a, \dot b;m\rangle -(| a, \dot b; -m\rangle)\,.
 \eea
 
 \smallskip

 So when $m=0$ we have
\bea
\label{zerosect}
&&\frac{1}{2}(1+\Omega) \,  | i,\tilde j; 0\rangle = (| i,\tilde j; 0\rangle \,, \qquad  \frac{1}{2}(1+\Omega) \,   | a, \tilde j ; 0\rangle =| a, \tilde j \, ; 0\rangle\,, \nn\\
&&
\frac{1}{2}(1+\Omega) \, | i, \dot b\, ; 0 \rangle=  0\,,\qquad\quad \qquad \frac{1}{2}(1+\Omega) \,  | a, \dot b\, ;0\rangle = 0\,.
\eea
and therefore the dotted space is killed by this projection.  The physical states are therefore spanned by 
\bea
\label{nonestates}
 | i,\tilde j; 0\rangle,\,\qquad  | a, \tilde j ; 0\rangle\,,
 \eea
 which are the states of ${\cal N}=1$ supergravity.  The matrix elements of the undotted spinor, $S_2^{\dot a}$, vanish between these projected states. 
 On the other hand when $m\ne 0$ both chiralities contribute with equal weight (the relative minus signs cancel out).
 \sm
 
\begin{figure}[h]
\begin{center}
\begin{tikzpicture}[scale=1]
\begin{scope}[shift={(0,0)}]
\draw [hwbou](-2,-3) -- (-2,2);
\draw [hwbou] (-2,2)-- (0,3);
\draw [hwbou] (-2,-3)-- (0,-2);
\draw [hwbou] (0,3) -- (0,-2);
\end{scope}
\begin{scope}[shift={(2.5,0)}]
\draw [sugra, ultra thick] (0,0) circle (40pt);
\draw [sugra, ultra thick](2,2) -- (1,1);
\draw [sugra, ultra thick](2,-2) -- (.9,-.9);
\draw [sugra, ultra thick](-2,-2) -- (-.9,-.9);
\draw [sugra, ultra thick](-2,2) -- (-1,1);
\end{scope}

\begin{scope}[shift={(7.5,0)}]
\draw [hwbou](-2,-3) -- (-2,2);
\draw [hwbou] (-2,2)-- (0,3);
\draw [hwbou] (-2,-3)-- (0,-2);
\draw [hwbou] (0,3) -- (0,-2);
\end{scope}
\end{tikzpicture}
\end{center}
\caption{The bulk contribution to the four-graviton amplitude  loop.}
\label{fig:oneloop1}
\end{figure}
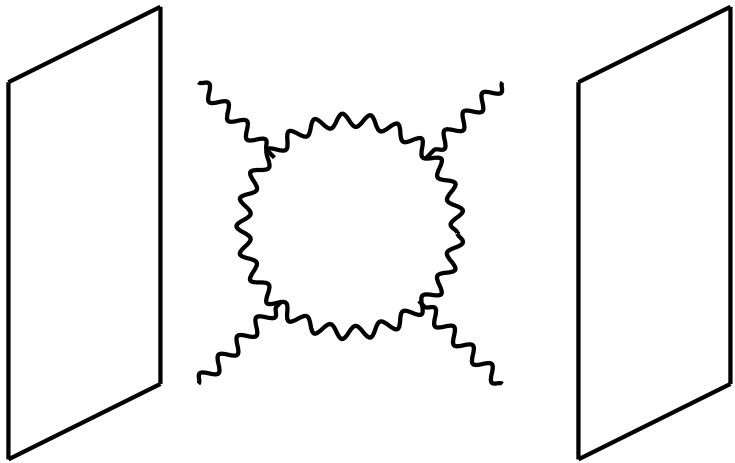

 With these preliminaries we can now proceed to evaluate the one-loop four-graviton amplitude in the bulk  shown in figure~\ref{fig:oneloop1}.
 
\subsection{The bulk one-loop four-graviton amplitude}  
\label{subsec:gravloop1}

 The four-graviton loop amplitude for eleven-dimensional supergravity in the compactified background, $\mathbb{M}^{9} \times S^{1}\times S^1$,  can be constructed in terms of a trace over the product of four vertex operators attached to the loop (as in equations (5.1) and (5.2) in \cite{Green:1999by})
\bea
\label{loopamp} 
\int_0^\infty  \frac{dt}{t}\int
d^9 {\rm\bf p} \, \sum_m
   e^{-t\left({\rm\bf p}^2+  \frac{m_1^2}{\lepl^2R_{10}^2} +\frac{m_2^2}{\lepl^2R_{11}^2}\right)} \,
\Tr\left\langle \prod_{r=1}^n \big(\int dt^{(r)}V_h^{(r)}(t^{(r)})\big)
\right\rangle\, 
\eea
(where  ${\bf p}$ is the continuous nine-dimensional loop momentum and $m_1$, $m_2$ are Kaluza--Klein charges) in the theory compactified in the $x^{10}$ and $x^{11}$ directions, and the proper times of the vertex operators, $t^{(r)}$ are integrated over the range $0\le t^{(r)} \le t$.
Each vertex operator has the form (at $t^{(r)}=0$) 
\bea
\label{vertrxdef}
V_h(0) = \zeta_{ij} \, \left(\dot x^i - \half\, S_1 \gamma^{il} S_1 \, k_l\right)\,  \left(\dot x^j - \half\, S_2 \gamma^{jm} S_2 \, k_m\right)\,
e^{i k \cdot x}\,,
\eea
which describes the emission of a graviton with polarisation $\zeta_{ij}$ and momentum $k_i$, where we choose $i,j = 3,\dots, 9$ which are directions transverse to the light-cone directions and to the compact directions, $x^{10}$  and $x^{11}$.  
In the type IIA theory the trace over the eight $S_1$ and eight $S_2$ fermionic modes arising from the product of four vertex operators in \eqref{loopamp} gives a kinematic factor of $t_8t_8\, \curv^4$ multiplying a scalar field theory box diagram in $\mathbb{M}^{9} \times S^1\times S^{1}$ \cite{Green:1997as, Green:1999by}.\footnote{The $\epsilon_{10}\epsilon_{10} \curv^4$ pieces  of the type IIA effective action is not captured by the four-graviton amplitude since it vanishes on shell.}

\smallskip

The orbifold condition of relevance to the Ho\u rava--Witten background compactified on a circle, $\mathbb{M}^{9} \times S^{1} \times S^{1}/\mathbb{Z}_{2}$,  is implemented by inserting a factor of $(1+\Omega)/2$ between the vertex operators in \eqref{loopamp}. The trace in the $m_2\ne0$ and $m_2=0$ sectors in the sum over $m_2$ must be treated separately since the space of states in the $m_2=0$ sector  is reduced by the conditions \eqref{zerosect} to the ${\cal N}=1$ states \eqref{nonestates}.   

\begin{itemize}

 \item[(i)]  $m_2\ne 0$ 

In this case all of the ${\cal N}=2$ states circulating in the loop survive the projection in \eqref{[rojectop}.  The trace over the components of $S_1$ and $S_2$ again leads to the kinematic prefactor $t_8t_8\, \curv^4$ of the type IIA theory.   The sum over $m <0$ is equivalent  to the sum over $m>0$, which leads to a factor of $1/2$ in the overall normalisation of these terms relative to the type IIA case.  
 
 \item[(ii)]  $m_2=0$ 

In this case the  circulating states are those in \eqref{nonestates} and we can set $S_2=0$ in the vertex operator acting on this projected space, reducing it to
\bea
\label{vertrxreddef}
V_{h}^{m=0}(0) = \zeta_{ij} \, \dot x^i \,  \left(\dot x^j - \half\, S_1 \gamma^{jm} S_1 \, k_m\right)\,
e^{i k \cdot x}\,,
\eea
which is the zero mode piece of the graviton vertex in the heterotic string acting on the supergravity multiplet.  The fermionic trace only involves the dotted spinors $S_1^{\dot a}$, which produces a factor of $t_8$, leading to a prefactor that is a linear combination of $t_8\, \tr \curv^4$ and $t_8\, (\tr(\curv^2))^2$.  The complete  one-loop amplitude is obtained by adding this contribution to that of the boundary gauge loop considered in section~\ref{subsec:gravloop2}.  
 \end{itemize}

\sm

 The dynamical factors in the amplitude are given in terms  
of the integral of the product of four Green functions in the orbifold background.
From \eqref{productfour} we see that, apart from a factor of $1/2$ in its normalisation,  the $p_{11}\ne 0$  ($m_2\ne 0$) contribution has precisely the same form as the expression that enters the loop amplitude of eleven-dimensional supergravity compactified on a torus.  In order to make heterotic/type I duality manifest it is very useful to write the  resulting expression for the sum of the $p_{11} \ne 0$ terms in the low energy limit of the loop amplitude in the form
 \begin{eqnarray}
 A^{bulk\ loop}= \frac{\lepl^2\raten \raele}{3\cdot 2^5(2\pi)^{6}} t_8t_8\hat \curv^4\, \left(I^{bulk}(0,0,0 \raten, \raele)-I_{p_{11}=0}^{bulk}(0,0,0; \raten, \raele)\right)\,,
 \label{Idef}
 \end{eqnarray} 
 where we have included a $p_{11}=0$ contribution to the loop momentum in the box diagram amplitude, $I^{bulk}(s,t,u; \raten, \raele)$,  and subtracted it again in the term $-I_{p_{11}=0}^{bulk}(s,t,u; \raten, \raele)$.
 
 \sm
 
 \subsection*{Properties of the bulk loop amplitude}
 
 The coefficient of $t_8t_8\, \hat \curv^4$ in the $p_{11}=0$ sector, $I_{p_{11}=0}^{bulk}(s,t,u; \raten, \raele)$, has the same form as for the gauge theory loop in \eqref{gluoncal3} and is interpreted as a one-loop contribution in the HO theory by the same Poisson summation over $m_1$ argument that leads from \eqref{gluoncal3} to \eqref{gluoncal3a}. 
 
  The contribution $I^{bulk}(s,t,u; \raten, \raele)$, which involves the sum of all values of $p_{11}$  is very similar to the expression for the loop amplitude  of  eleven-dimensional supergravity  compactified on a torus \cite{Green:1997as} with purely imaginary complex structure.   
The leading order term in the low-energy expansion is obtained by setting the momenta to zero in the factors of $e^{i k\cdot x}$ in the vertex operators in \eqref{loopamp}.  After performing the integral over the nine-dimensional momentum we obtain
\begin{eqnarray}
I_0(\raten, \raele) \equiv I^{bulk}(0,0,0; \raten, \raele)=
\frac{3\pi ^{9/2}}{\lepl^2\raten\, \raele}
\int_0^\infty 
\frac{dt}{t^{3/2}}
\sum_{m_1\in \mathbb{Z} \,,m_2\in \mathbb{Z}
}  e^{- \frac{t}{\lepl^2} \left( \frac{m_1^2}{\raten^2} +  \frac{m_2^2}{\raele^2}\right)}\,
\label{Idetail}
\end{eqnarray} 
 The integral for each term in the sum over Kaluza--Klein charges is divergent in the ultraviolet region (the $t\to 0$ limit).  However, the total integrand may be expressed in terms of the winding numbers $\hat m_1$ and $\hat m_2$ by  Poisson summation over $m_1$ and $m_2$.  The result is 
\begin{eqnarray}
I_0(\raten, \raele)&=&
3
 \pi ^{11/2}
\int_0^\infty  d \hat t \,\hat t^{1/2}
\sum_{\hat m_1\in \mathbb{Z},\, \hat m_2\in \mathbb{Z}
} e^{-\pi^2 \hat t\  \lepl^2 \,(\raten^2\,\hat m_1^2 + \raele^2\, \hat m_2^2)}\,,
\label{poissonres}
\end{eqnarray}
where $\widehat t =1/t $.   The divergence is now entirely in the $(\hat m_1, \hat m_2)=(0,0)$ term while every term with $(\hat m_1, \hat m_2)\ne (0,0)$  is convergent.   Performing the integral  for each winding number and separating the divergent $(0,0)$ term  gives
\begin{eqnarray}
I_0(\raten, \raele)= \frac{3}{2^4\lepl^3\raten^{3/2}\raele^{3/2} } \, \, \sum_{(\hat m_1,\hat m_2)\ne (0,0)}  \frac{\left(\raten/\raele\right)^{\frac{3}{2}}}
 {\left( \hat m_1^2 \,(\raten/\raele)^2 + \hat m_2^2 \right)^{\frac{3}{2}}} + \widehat C \,. 
\label{eisenform}
\end{eqnarray} 
The quantity $\widehat C$ represents the regulated zero winding piece.    The divergence can be subtracted by introducing a one-loop counterterm  that leaves  a finite but undetermined contribution, $\widehat C$.  

\sm

The contribution of $I_0(\raten, \raele)$ to the nine-dimensional  low-energy  supergravity amplitude can be translated into a term in the effective M-theory action of the form 
\begin{eqnarray}
S_{t_8t_8 \curv^4}=\frac{1}{2^{9}\, (2\pi)^6\, 4!\, \lepl}\int_{\mathcal{M}_{9}} d^9x \sqrt{-G}\, \, t_8t_8\curv^4 \left( 
\frac{1}{\raten^{1/2}\raele^{1/2}}
\sum_{(\hat m_1,\hat m_2)\ne (0,0)} \frac{\Omega_2^{\frac{3}{2}}}{(\hat m_1^2\,  \Omega_2^2 + \hat  m_2^2)^{\frac{3}{2}}}
+
\raten\raele\, 
\widehat C
\right)\,.\nonumber\\
\label{Idual7}
\end{eqnarray} 
 In the limit ${\mathcal{V}}= 2\pi^2\,\raten\raele \to \infty $, the expression describes an action in eleven non-compact dimensions and only the $\hat C$ term survives, while the first term in parentheses is suppressed by the factor of ${\mathcal{V}}^{-\frac{3}{2}}$.    
 
 \subsubsection*{Relation to the $\mathcal{N}=1$ string  theories}

The expression for the $t_8t_8\, \curv^4$ M-theory action in (\ref{Idual7}) can be translated into the language of string theory by using the dictionary  in appendix~\ref{sec:dictionary}, so that the parameters $\raten$ and $\raele$ are related to the parameters of the HO theory in $D=9$ dimensions via the relations
\begin{eqnarray}
\Omega_2\equiv \frac{\raten}{\raele}=\gho^{-1}\,,\qquad \quad\frac{1}{\lepl \raten^{1/2}\raele^{1/2}}=	\frac{\raho }{\lesth }\, \gho^{-\frac{1}{2}} \,.
\label{Idual8}
\end{eqnarray}
Thus, the contribution to the  M-theory effective action in \eqref{Idual7} translates into the HO effective action
\begin{eqnarray}
S^{HO}_{t_8t_8 \curv^4}=\frac{\raho }{2^9 (2\pi)^64!\,\lesth }\, \int_{\mathcal{M}_{9}} d^9x \sqrt{-G} \, \gho^{-\half}\, t_8t_8\,\curv^4 \left( 
\sum_{(\hat m_1,\hat m_2)\ne (0,0)}\frac{\gho^{-\threeh}}{(\hat m_1^2\,  \gho^{-2} +\hat m_2^2)^{\frac{3}{2}}}
+\frac{\gho^{\frac{1}{2}}} {\raho^2}\,\widehat C\
\right)\,.\nonumber
\\
\label{hor4}
\end{eqnarray}

\sm

The first term in parentheses s proportional to $\raho$ and has a finite ten-dimensional limit as $\raho \to \infty$.  This term is closely related to the Eisenstein series $E_\frac{3}{2}(\Omega)$, which is the $SL(2,\mathbb Z)$-invariant function that arises as the coefficient of the $ t_8t_8\,\curv^4 $ in the ten-dimensional type IIB superstring.  Whereas $\Omega= \Omega_1+ i  \Omega_2$ in the  type IIB theory, the Ramond--Ramond axial scalar does not arise in the heterotic theories, so $\Omega_1=0$.  
In other words, the first term in parentheses in (\ref{hor4}) is identified with $E_\frac{3}{2}(\gho^{-1})$.  
The second term in parentheses is proportional to the regulated quantity $\hat C$ and vanishes as $\raho \to \infty$.  We will later argue that $\hat C=4\pi^2/3$ in order to reproduce the HE one-loop effective action (the term proportional to $1/\raho =\rahe$ in \eqref{hoeffstring}).
\sm 

Taking the limit $\raho\to \infty$, it follows that the coefficient of the $t_8t_8\, \curv^4$ interaction in the ten-dimensional HO effective action has the form
\begin{eqnarray}
S^{HO}_{t_8t_8 \curv^4}=\frac{ \gho^{-\half}}{2^9\, (2\pi)^7\, 4!\,\lesth^2 }\,\int_{\mathcal{M}_{10}} d^{10}x \sqrt{-G} \, \, t_8t_8\,\curv^4\, E_{\frac{3}{2}}(\gho^{-1})\,.
\label{hotend}
\end{eqnarray}
We may now make use of the standard expression for the Fourier modes of the $SL(2,\mathbb Z)$ Eisenstein series,
\begin{equation}
E_s(x+i y) = \sum_{(m_1,m_2)\,\neq\,(0,0)}\frac{y^s}{|m_1(x+i y)+m_2|^{2s}}    =  \sum_{n\in\mathbb Z} {\cal F}_{n,s}(y) \, e^{2\pi i  n x}\,, 
\label{eisenfourier}
\end{equation}
where  (see, for example, \cite{Terras:1985})  the zero mode consists of two power behaved terms,
\begin{equation}
{\cal F}_{0,s}(y)   = 2 \zeta(2s)\, y^s \  +  \ \frac{2 \sqrt \pi \,\Gamma(s-{\frac{1}{2}}) \zeta(2s-1)}{\Gamma(s)}\, y^{1-s} \,,
\label{eisenzero}
   \end{equation}
and the non-zero modes  are proportional to $K$-Bessel functions,
\begin{equation}
{\cal F}_{n,s}(y) \ \ = \ \  \frac{4\,\pi^s}{\Gamma(s)}\,  |n|^{s-{\frac{1}{2}}} \, \sigma_{1-2s}(|n|)
\sqrt{y}\,K_{s-{\frac{1}{2}}}(2\pi |n|y) \,, \  \ \ n\neq 0 \,,
\label{nonzeroeisen}
\end{equation}
and the divisor sum is defined by $\sigma_{s}(n)= \sum_{d|n} d^s$.

\sm 

 Using the large-$z$ expansion $K_1 (z) = \sqrt \pi e^{-2z}/\sqrt{2z}\, (1 + O(z^{-1}))$, the small $\gho$ (or large $\Omega_2$) expansion of \eqref{eisenfourier} with $s=3/2$  takes the form
\begin{eqnarray}
E_{\frac{3}{2}}( \gho^{-1}) &=& \sum_{(\widehat m_1,\widehat m_2)\ne (0,0)}\frac{\gho^{-\threeh}}{( \widehat  m_1^2\, \gho^{-2} +  \widehat  m_2^2)^{\threeh}}\nonumber\\&=&  2\zeta(3)\, \gho^{-\threeh}+2 \zeta(2)\, \gho^{\half} + \sum_{n\in \mathbb{Z}^+ } 8\pi\,  \sigma_{-1}(|n|)\, 
e^{-\frac{2\pi |n|}{\gho}} \left(1+O(\gho)\right) \,.
\label{Idual6} 
\end{eqnarray}
Substituting this expression in \eqref{hotend} gives the perturbative expansion of the $t_8 t_8 \curv^4$ interaction in the ten-dimensional HO theory.  
The following features of the resulting expression are worth noting.

\begin{itemize}
\item
  The expression \eqref{hotend} contains two perturbative terms, a tree-level term of order $\gho^{-2}$  and a one-loop contribution of order $\gho^0$.  These have the same coefficients as in the type IIB  theory.   
Invariance of the Eisenstein series under the transformation $\Omega\to - 1/\Omega$ implies 
\begin{eqnarray}
E_\threeh (i\gho^{-1}) = E_\threeh (i\gib^{-1})\,,
\label{hettypeI}
\end{eqnarray}
which is a manifestation of HO/type I S-duality.  In particular, the perturbative contributions  to $t_8t_8\, \curv^4$ in the HO theory from spherical and toroidal world-sheets are identical to contributions from spherical and toroidal world-sheets in the type I theory.

\item
  In addition,  \eqref{hotend} contains an infinite set of non-perturbative terms that appear as D-instanton contributions proportional to $\exp(-2\pi |k|/\gho)$ in the HO parameterisation.   The instanton action is identified with the action of the euclidean world-line of the $m$'th Kaluza--Klein mode in the $x^{11}$ interval, winding $\hat n$ times around the $x^{10}$ circle, where $k=m \hat n$. In the HE description the contribution of such an object  is $\exp(-2\pi |k| \rahe/\ghe)$, which vanishes in the ten-dimensional heterotic limit, $\rahe \to \infty$.  
The possible r\^ole of such D-instantons is intriguing since they do not arise in HO string theory in any obvious manner.   However, it is worth recalling that  the original argument for the existence of  D objects in closed string theories by Shenker \cite{Shenker:1990uf}  was based on a counting argument that applies to any closed-string theory and does not distinguish between heterotic and the type II theories, which allows for the possibility that  D-instantons might indeed contribute to heterotic amplitudes.
Moreover, the fact that  the instantonic contributions might be present in the HO theory, but not the HE theory, is reminiscent of Polchinski's observation \cite{Polchinski:2005bg} concerning the possible r\^ole of open heterotic strings in the HO theory.
We also note the peculiarity that   the perturbative expansion around each instanton in \eqref{Idual6} is an expansion in powers of $\gho$ and not $\gho^2$.

\item
The total contribution to $A^{bulk\ loop}$ in \eqref{Idef} includes the term $-I^{bulk}_{p_{11}=0}\, t_8t_8\, \hat\curv^4$, which subtracts the one-loop contribution from $I^{bulk}\, t_8t_8\, \hat\curv^4$.  As a result, there is no one-loop $t_8t_8\, \hat\curv^4$ contribution in the  HO theory and the complete one-loop contribution gives the effective action \eqref{efftrr4} described in section~\ref{subsec:gravloop2}.  This agrees with the expression \eqref{looprelate}.   However the $-I^{bulk}_{p_{11}=0}$ contribution is interpreted in the  type I theory via HO/type I duality as a disk diagram contribution (of order $1/\gib$).

\item  Since (in the Einstein frame) the coefficient of the $t_8t_8 \curv^4$ contribution contained in \eqref{hotend}  in the HO  theory  is identical to the coefficient   in the type I theory after the replacement $\gho \to \gib^{-1}$, the  type I coefficient   also contains effects due to D-instantons.  In contrast to the HO description these type I D-instantons are required by symmetry considerations.  As argued in \cite{Witten:1998cd} the type I theory would have gauge group $O(32)$ were it not for the presence of type I $\mathbb{Z}_2$ D-instantons that break the invariance under the transformation transformations in $O(32)/\mathbb{Z}_2$ to transformations in $SO(32)$ since $\pi_9(SO(32)) =\mathbb{Z}_2$.    In our discussion of  supergravity  in the compactified Ho\u rava--Witten background, these non-BPS type I D-instantons  have an interpretation, via T-duality,  in terms of pairs of euclidean world-lines of type IIA D-particles winding  (with opposite orientations) around the $x^{11}$ orbifold.\footnote{A more complete discussion of these type I $\mathbb{Z}_2$ D-instantons is given in section 4 of  \cite{Dasgupta:2000xu}.}  Note also that the fact that D-instantons do not contribute to the HE theory in the $\rahe\to \infty$ limit is consistent with the fact that $\pi_9(E_8) =0$. 

\item
We know that the parity conserving one-loop effective actions of  both the heterotic theories are equal, which is consistent with T-duality on the $x^{10}$ circle.  In this section we have seen how this is obtained in the HO theory by summing boundary and bulk loop contributions to supergravity in the Ho\u rava--Witten background.  In HE coordinates
the one-loop effective action is given by 
\begin{eqnarray}
	\frac{2\pi^2}{3}\, \frac{\rahe  }{2^8 (2\pi)^6\, \lesth } t_8\ygs\,,
	\label{moreeqn2}
\end{eqnarray}
where $t_8\, \ygs = t_8 \, \tr \curv^4 + \quart (\tr \curv^2)^2$.  We found in \eqref{moreeqn5} that the $t_8\, (\tr \curv^2)^2$ part of this expression arises from the expansion of a gravity tree diagram and contributed a term 
\begin{eqnarray}
	\frac{2\pi^2}{3}\, \frac{\rahe  }{2^8 (2\pi)^6\, \lesth } t_8\,(\tr R^2)^2\,, 
		\label{moreeqn2}
\end{eqnarray}
to the HE effective action.  We also see from \eqref{hor4}, after replacing $\raho$ by $\rahe^{-1}$, that the 
bulk loop calculation produces a term of the form
\begin{eqnarray}
	\frac{\rahe  }{2^8 (2\pi)^6\, 48\, \lesth }\, \hat C\, t_8t_8R^4\,.
	\label{moreeqn3.1}
\end{eqnarray}
 It follows that in order to ensure that the complete one-loop term in the HE theory has the effective action \eqref{moreeqn2}
we need to set the renormalised value of the $\hat C$ to the value
\begin{eqnarray}
\hat C =\frac{4\pi^2}{3}	\,.
	\label{moreeqn4}
\end{eqnarray}

\item   Finally, it is not at all obvious why the expression we have deduced from the bulk supergravity loop should give the exact form of the $\curv^4$ interactions.  Unlike the type II theories, in which this interaction is 1/2-BPS, in the half-maximally supersymmetric theories the $\curv^4$ interactions do not preserve any supersymmetry in any obvious manner.  
However, the situation is a little murky since this statement also suggests that there should be a three-loop $\curv^4$ ultraviolet divergence in four-dimensional $\mathcal{N}=4$ supergravity.   The obvious local counterterm for such a divergence is the volume of superspace, but this was shown to vanish in \cite{Bossard:2011tq}, where an alternative and less obvious  counterterm was determined.     However, explicit supergravity  calculations in  \cite{Bern:2012cd} demonstrate that this ultraviolet divergence  is absent.  In addition, as pointed out in  \cite{Tourkine:2012ip,Tourkine:2012vx}, the absence of a $\curv^4$ interaction at two loops in the heterotic string suggests that there is no renormalisation of $\curv^4$ beyond one loop.

 Even though we have not analysed higher-loop amplitudes in detail we know that these have low energy limits that start with at least two derivatives on $\curv^4$.  This adds weight to the suggestion that $\curv^4$ is not renormalised beyond one loop.

\end{itemize}

Since our non-perturbative expression contains the correct perturbative terms for both the heterotic and type I theories it is of interest to further understand the significance of the non-perturbative  contributions.

\section{Higher order contributions from other one-loop amplitudes}
\label{sec:higherloop}

In the above analysis we have discussed the leading behaviour of boundary and bulk loop amplitudes that contribute, in the limit $s,t,u\to 0$,  to low order terms in the low-energy expansion.  The expansions of these expressions to higher orders in $s,t,u$  is straightforward since the field theory box diagram has a simple expansion.  At least at low orders in this expansion the amplitude can be separated into an analytic part that and the part that contains non-analytic threshold behaviour.  The first of these thresholds gives contributions of order $t_8 \hat F^4 s\log s$ or $t_8 t_8 \curv^4 s\log s$ in the ten-dimensional theory, although the nature of the singularity changes when compactified.  For example, in nine dimensions the gauge theory amplitude has a threshold term of the form $t_8 \hat F^4 s^\half$. 
As stressed earlier, only the first few terms in the low-energy expansion are likely to be protected by supersymmetry against receiving  corrections beyond those  exhibited by the Feynman diagrams we are considering.

\sm 

In addition to the higher order terms obtained by expanding the loop diagrams we have already considered in powers of $s$, $t$ and $u$, there are other one-loop Feynman diagrams that contribute to Yang--Mills amplitudes in supergravity in the Ho\u rava--Witten background.  The low-energy limit of these diagrams starts with higher powers of the Mandelstam invariants than those we have considered so they do not affect the terms that we expect to be protected by supersymmetry, but the systematics of their contributions may nevertheless be of some interest.  
 
\subsection{Yang--Mills one-loop amplitude with one gravity propagator: $s\, t_8\tr_i F_i^4$}
\label{dtwoffour}

 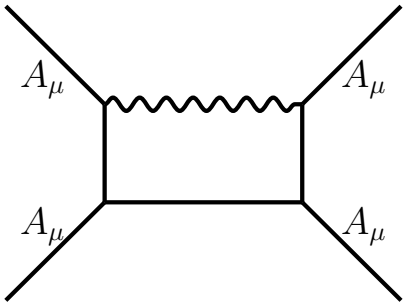
\begin{figure}[h]
\begin{center}
\begin{tikzpicture}[scale=.65]
\draw [sugra , ultra thick] (-2,0) --(2,0);
\draw [suym, ultra thick] (-2,-2) --(2,-2);
\draw [suym, ultra thick] (-2,0) --
(-2,-2);
\draw [suym, ultra thick] (2,0) -- 
(2,-2);
\draw [suym, ultra thick](-4,2) -- (-2,0);
\draw [suym, ultra thick](-4,-4) -- (-2,-2);
\draw [suym, ultra thick](4,2) -- (2,0);
\draw [suym, ultra thick](4,-4) -- (2,-2);
\draw (3.25,-2.5) node{\Large $A_{\mu}$};
\draw (3.25,.5) node{\Large $A_{\mu}$};
\draw (-3.25,-2.5) node{\Large $A_{\mu}$};
\draw (-3.25,.5) node{\Large $A_{\mu }$};

\end{tikzpicture}
\end{center}
\caption{A Yang--Mills four gauge boson loop amplitude (localised on one boundary) with an internal gravity propagator}
\label{onegrav}
\end{figure}
The first of these diagrams is illustrated in figure~\ref{onegrav}.  In this contribution to the four gauge particle amplitude one of the propagators in figure \ref{fig:oneloopgauge} is replaced by a gravitational propagator.  Two vertices now have an extra power of momentum and the  low-energy limit contributes to the $d^2 \tr F^4$ interaction in the low-energy expansion of the type I four gauge particle  disk  amplitude.  

\sm 

Before discussing the details of the diagram, we note that S-duality must act in a non-trivial manner - much as  we saw with the $t_8t_8\curv^4$ interaction, it cannot act term by term.   This can be seen from the fact that there is a disk amplitude in the type I theory that gives a contribution of the form $\zeta(3)\, d^2 \tr F^4/\gib$.   Transforming to the Einstein frame produces no powers of the dilaton and therefore, applying the type I/heterotic S-duality transformation, $\gib\to  \gho^{-1}$, results in a HO interaction of order $\gho$.  This does not make sense, which means that the coefficient of this interaction must be a non-trivial duality invariant function of the coupling constant, which gives a HO tree contribution of the same form as the type I disk contribution.  Another feature to note for  this interaction is that it  vanishes when two $F$s belong to each $E_8$ subgroup of $E_8 \times E_8$. This follows since $d^2\, \tr F^4$ is then a total derivative  (equivalently, $s+t+u=0$). 

\sm 

As before, we will consider the amplitude compactified on $S^1$ to nine dimensions in the presence of the Wilson line that breaks the symmetry to $SO(16)\times SO(16)$.  In this case only adjoint $SO(16)$ gauge particles can propagate in the loop in figure \ref{onegrav}  since the external states are in the adjoint representation and the graviton is a $SO(16)$ singlet. 
A straightforward extension of earlier arguments leads to the expression for the low-energy limit of this amplitude of the form
\begin{eqnarray} 
\frac{48}{(2\pi)^{10}}
C(\raten, \raele)\,  d^2 \tr F^4\,.
\label{d2f4}
\end{eqnarray} 
Here $C(\raten, \raele)$ is the $s,t,u \to 0$ limit of a ten-dimensional scalar  box diagram compactified on the $x^{10}$ circle and with the unusual feature that one propagator is of the form \eqref{Iprop8}, which involves a sum over the Kaluza--Klein momentum in the $x^{11}$ direction, which is $m/(\lepl\raele)$ in the following expressions (whereas the Kaluza--Klein momentum in the $x^{10}$  is $n/(\lepl \raten)$.
Including the volume factor $2 \pi \lepl \raten$ and ignoring an overall normalisation constant we have
\begin{eqnarray} 
C(\raten, \raele) =\frac{\lepl^2}{\raele}\sum_{m,n}\int_{\mathcal{M}^9} d^9p \frac{1}{\left(p^2+\frac{n^2}{\lepl^2 \raten^2} \right)^3}\times 
 \frac{1}{p^2+\frac{n^2}{\lepl^2\raten^2} +\frac{m^2}{\lepl^2 \raele^2}}
\label{Ical4.1}
\end{eqnarray}
which can be expressed as 
\begin{eqnarray}
\frac{\lepl^2}{2\raele}\sum_{m,n}\int_{\mathcal{M}^9} d^9p 
\int_0^\infty d\sigma_1 d\sigma_3\ \sigma_3^2\ 
\exp\left(-(\sigma_1+\sigma_3)
\left(p^2+\frac{n^2}{\lepl^2 \raten^2}\right) 
-\sigma_1
 \frac{m^2}{\lepl^2 \raele^2}
\right)
\label{Ical4.2}
\end{eqnarray}
Performing the $p$ integrations gives
\begin{eqnarray}
C(\raten, \raele)=
\frac{\pi^{9/2}\lepl ^2}{2\raele }\sum_{m,n}
	\int_0^\infty \frac{d\sigma_1 d\sigma_3\ \sigma_3^2}{(\sigma_1+\sigma_3)^{9/2}}
	\exp\left(-(\sigma_1+\sigma_3)
\frac{n^2}{\lepl^2 \raten^2} 
-\sigma_1
 \frac{m^2}{\lepl^2 \raele^2}
\right) \,.
\label{Ical4.3}
\end{eqnarray}
The sum over Kaluza--Klein charges $m,n$ is converted to  sum over winding numbers $(\hat m,\hat n)$ by Poisson summations.  In the first step the Poisson sum over $n$ gives
\begin{eqnarray}
C(\raten, \raele)=	\frac{\pi^5 }{2}  \frac{\lepl^3 \raten }{\raele }\sum_{m,\hat n}
	\int \frac{d\sigma_1 d\sigma_3\ \sigma_3^2}{(\sigma_1+\sigma_3)^{5}}
	\exp\left(-
\frac{\pi^2  \hat n^2 \lepl^2 \raten^2}{\sigma_1+\sigma_3} 
-\sigma_1
 \frac{ m^2}{\lepl^2  \raele^2}
\right) \, ,
\label{Ical4.4}
\end{eqnarray}
and the subsequent summation over $m$ gives
\begin{eqnarray}
C(\raten, \raele)= \frac{\pi^{11/2}}{2}  \lepl^4 \raten
\sum_{\hat m, \hat n}
	\int_0^\infty \frac{d\sigma_1 d\sigma_3\ \sigma_3^2}{(\sigma_1+\sigma_3)^{5}\sqrt{\sigma_1}}
	\exp\left(-
\frac{\pi^2 \hat n^2 \lepl^2 \raten^2}{\sigma_1+\sigma_3} 
-\frac{\pi^2 \hat m^2\lepl^2  \raele^2}{\sigma_1}
\right) \,.
\label{Ical4.5}
\end{eqnarray} .
We may now analyze the expansion of this expression in the perturbative HO limit in which $\gho=\raele/\raten  \to 0$ or the perturbative type I limit in which  $\gib=\raten/\raele  \to 0$. 

\sm 

As in the previous examples, the ultraviolet divergence of this Feynman diagram is contained in the zero winding, $\hat m=\hat  n=0$ term.    A high momentum cut-off at a momentum scale $~ \lepl^{-1}$, which regularises this divergence, translates into a cut-off at the lower endpoint of the $\sigma_1$ and $\sigma_3$ integrations. Substituting in \eqref{Ical4.5}, the renormalised value of this contribution is
\begin{eqnarray} 
C(\raten, \raele)\bigl |_{UV\, divergence} = \tilde C \,  \lepl \raten= \tilde C \lesth \rahe\,,
\label{uvc}
\end{eqnarray}  
where $\tilde C$ is a dimensionless constant.   
Its value is arbitrary, but since we know that there is no $t_8\tr F^4$ interaction in the HE theory (where the trace is in the fundamental representation of a $SO(16)$ subgroup of $E_8$), the only consistent value is $\tilde C=0$.

\sm 

The remaining non-zero winding terms in \eqref{Ical4.5} are finite and can be interpreted in the string parameterisation in the following manner.

\subsubsection*{Tree  coefficient in heterotic string theory}

The tree-level term arises as the most singular contribution in the small-$\ghe$  (or small-$\raele/\raten$) limit, which comes by setting $\hat n=0$ in \eqref{Ical4.5} 
\begin{eqnarray}
	C(\raten, \raele)\bigl |_{HO\ tree}&=&\frac{\pi^{11/2}}{2} \lepl^4 \raten\sum_{\hat m}
	\int \frac{d\sigma_1 d\sigma_3\ \sigma_3^2}{(\sigma_1+\sigma_3)^{5}\sqrt{\sigma_1}}
	\exp\left(-\frac{\pi^2 \hat m^2\lepl^2\raele^2}{\sigma_1}
\right)
\nonumber\\
&=& \frac{\pi^3 \lepl  \raten}{24 \raele^3}\, \zeta(3)
=  \lesth \raho  \frac{1}{\gho^2} \frac{\pi^3  \zeta(3)}{24} \,.
\label{proptree}
\end{eqnarray}
This has the right form, including the presence of the $\zeta(3)$ factor,  to correspond to the tree-level terms compactified to nine dimensions in both the heterotic theories (although we have not kept track of rational prefactors and powers of $\pi$ in the overall coefficient).  

\subsubsection*{Loop contributions to heterotic string theory}
In order to extract the loop contributions to $C(\raten, \raele)$ we need to consider the intermediate summation given in  \eqref{Ical4.4}.

 Firstly consider the $m=0$ term.  This gives
\begin{eqnarray}
	C(\raten, \raele)\bigl |_{HO\ loop} &=&\frac{\pi^5}{2}   \frac{\lepl^3 \raten} { \raele }\sum_{\hat n\ne 0}
	\int \frac{d\sigma_1 d\sigma_3\ \sigma_3^2}{(\sigma_1+\sigma_3)^{5}}
	\exp\left(-
\frac{\pi^2 \hat n^2\lepl^2 \raten^2}{\sigma_1+\sigma_3} 
\right)\nonumber\\
&=& \frac{\pi^3}{3}  \frac{\lepl}{\raten\raele} \zeta(2) = \lesth \raho  \frac {\pi^3 \zeta(2)}{3}  \,.
\label{oneloo}
\end{eqnarray}

\sm

Further perturbative terms arise from $m\ne 0$ terms in \eqref{Ical4.4}.   The integral can be evaluated explicitly for $m, \hat n \ne 0$.  Converting to the HO parameterisation, these terms give
\begin{eqnarray} 
C(\raten,\raele)\bigl |_{m,\hat n \ne0} &=&\pi^3\lesth \raho \left(  \frac{2 \zeta(2)^2}{ \pi^2}  g_{ho}^2
- \frac{ 8\zeta(4)^2}  { \pi^4 }  g_{ho}^4+ \frac{24\zeta(6)^2} {\pi^6}g_{ho}^6     \right. \nonumber\\
&& \left. \qquad\qquad - \frac{8  }{\pi^2} g_{ho}^2 \sum_{k >0} \frac{1}{k^2} \sigma_{-2} (k)\, K_4\left( 2\pi k/ g_{ho}\right) \right)  \,,
\label{instpert}
\end{eqnarray} 
where $k = m \hat n$ and $\sigma_{-2}(k) = \sum_{d|k} 1/d^2$.  In the weakly coupled HO limit the terms in the first line of this equation are contributions of the form expected for two, three and four loop HO string contributions.  The last term containing the Bessel function gives rise to instantonic contributions, as we will see shortly.

\sm

The complete contribution of perturbative terms to the HO amplitude arising from the diagram in figure~\ref{onegrav}  therefore has the form
\begin{eqnarray}
	C(\raten, \raele)\bigl |_{ho\ pert}\, d^2 \tr F^4 
&=&	\lesth\raho  \pi^3\left(\frac{\zeta(3)}{24\gho^2}+ \frac{\zeta(2)}{3}+ \frac{2 \zeta(2)^2}{ \pi^2}  g_{ho}^2
	\right. \nonumber\\
&&\left. \qquad\qquad - \frac{ 8\zeta(4)^2}  { \pi^4 }  g_{ho}^4+ \frac{24\zeta(6)^2} {\pi^6}g_{ho}^6    	\right)  d^2 \tr F^4
	\nonumber\\
&=&	\lesth\rahe \pi^3\left(\frac{\zeta(3)}{24\gcou_{\textrm{he}}^2}+\frac{\zeta(2)}{3\rahe^2}  + \frac{2 \zeta(2)^2}{ \pi^2} \frac{ g_{het}^2}{\rahe^4}	\right. \nonumber\\
&&\left. \qquad\qquad    
- \frac{ 8\zeta(4)^2}  { \pi^4 }  \frac{g_{ho}^4}{\rahe^6}+ \frac{24\zeta(6)^2} {\pi^6} \frac{g_{ho}^6}{\rahe^6}      \right)d^2 \tr F^4 \,,
\label{pertcomp}
\end{eqnarray}
where we have used T-duality to relate the HO and HE amplitudes in the last step.   

\subsubsection*{Instanton contribution in the HO theory}
The last line of  \eqref{instpert}  gives rise, in the $g_{ho} \to 0$ limit,  to  an infinite set of instanton contributions to the interaction \eqref{d2f4} of the form
 \begin{eqnarray} 
C(\raten,\raele)\bigl |_{HO\ inst} &=& -8\pi \lesth \raho g_{ho}^2 \sum_{k >0} \frac{1}{k^2} \sigma_{-2} (k)\, K_4\left( 2\pi k/ g_{ho}\right) \nonumber\\
&=& -4\pi \lesth \raho g_{ho}^{5/2} \sum_{k >0}   \frac{1}{k^{5/2}} \sigma_{-2} (k)\,    e^{-\frac{2\pi k}{\gho}} (1+O(\gho)) \,,
\label{instcon}
\end{eqnarray} 
where the instanton number is $k= |\hat n m|$.  So, we have an indication that, ss in the case of the $t_8t_8\curv^4$ interaction discussed in section \ref {subsec:gravloop1}, the  ten-dimensional HO amplitude contains the contribution of an infinite sequence of D-instantons.  As commented earlier, we do not have an explanation of the origin of such instantons witihn the HO string theory.   

\subsubsection*{Tree coefficient  in type I string theory}
We may now consider the weakly coupled type I limit, in which $\raele >> \raten$.  In that case the tree coefficient is obtained by setting  $\hat m=0$ in  \eqref{Ical4.5}  giving
\begin{eqnarray}
C(\raten, \raele)\bigl |_{I\ tree}&=&
\pi^{11/2} \lepl^4 \raten
\sum_{\hat n\ne 0}
	\int \frac{d\sigma_1 d\sigma_3\ \sigma_3^2}{(\sigma_1+\sigma_3)^{5}\sqrt{\sigma_1}}
	\exp\left(-
\frac{\pi^2 \hat n^2\lepl^2 \raten^2}{\sigma_1+\sigma_3} 
\right)
\nonumber\\
&=&
\frac{\pi^{3}\lepl}{\raten^2} 
\frac{16 }{15}\zeta(3)   = \pi^{3}\raib \lesti\frac{1}{\gib} \frac{16}{15}  \zeta(3) \,.
\label{itree}
\end{eqnarray}
The presence of the $\zeta(3)$  is again in qualitative agreement with the expression obtained by expanding the type I tree-level amplitude.  However, the ratio of the heterotic tree-level coefficient in \eqref{proptree} to the type I coefficient in \eqref{itree} does not correspond to the result obtained by explicit calculation in string perturbation theory, which is not surprising since this is not expected to be a protected process.

\subsubsection*{Two-loop  (and absence of one-loop) coefficient  in type I string theory}
Higher order perturbative terms in the type I theory can be obtained by  performing a Poisson summation over the integer $m$ in \eqref{Ical4.3} instead of over $n$.  In that case we are led to the expression 
\begin{eqnarray}
	C(\raten, \raele) =\pi^5\lepl^3 \sum_{n}\sum_{\hat m \ne 0}
	\int \frac{d\sigma_1 d\sigma_3\ \sigma_3^2}{(\sigma_1+\sigma_3)^{9/2}\sqrt{\sigma_1}}\ 
	\exp\left(-(\sigma_1+\sigma_3)
\frac{n^2}{\lepl^2\raten^2} 
- \frac{\pi^2 \hat m^2 \lepl^2\raele^2}{\sigma_1}
 \right)  \, ,
 \nonumber\\
 \label{typeiloopaa}
\end{eqnarray}
which is analogous to \eqref{Ical4.4}, but with sums over $\hat m$ and $n$ so the r\^oles of Kaluza--Klein momentum and winding number reversed.  The term in this expression that corresponds to the lowest order perturbative loop term in type I string theory is obtained by  setting $n=0$ and is a {\it two}-loop contribution.  In this case, changing integration variables to $\tilde \sigma_i=(\pi^2\raele^2)^{-1}\sigma _i$, gives
\begin{eqnarray}
	C(\raten, \raele)  \Bigl|_{I\ 2-loop}&=&\frac{\pi^3 \lepl}{\raele^2}\sum_{\hat m}
	\int \frac{d\tilde \sigma_1 d\tilde \sigma_3\ \tilde  \sigma_3^2}{(\tilde \sigma_1+\tilde \sigma_3)^{9/2}\sqrt{\tilde \sigma_1}}\
	\exp\left(-\frac{\hat m^2 }{\tilde \sigma_1}
 \right)
 \nonumber\\
&=&\frac{\pi^3 \lepl}{ \raele^2} \frac{16}{105} \zeta(2)= \pi^3 r_I \lesti  \gib  \frac{16}{105} \zeta(2) \,.
\label{twoloopamp}
\end{eqnarray}
There are undoubtedly higher order perturbative terms arising from terms in the integral \eqref{typeiloopaa} with $n\ne 0$, but since  we do not have a useful closed form expression for the integral we have not extracted them.  The perturbative contributions in the weakly coupled type I and IA limits limit  that we have extracted are summarised in the type I and IA theories by
\begin{eqnarray}
C(\raten,\raele)\bigl |_{I\ pert}	 d^2 \tr F^4&=& \frac{16}{15} \pi^3 r_I\lesti \left(\frac{\zeta(3)}{\gib}+\frac{\zeta(2)}{7} \gib + \dots
	\right) \, d^2 \tr F^4
\nonumber\\
&=&\frac{16}{15} \pi^3 \lesti \left(\frac{\zeta(3)}{\gia}+\frac{\zeta(2)}{7} \frac{\gia}{\raia^2} + \dots
	\right) \, d^2 \tr F^4  \,.
\label{perttwo}
\end{eqnarray}

\subsubsection*{Instanton contribution in type I string theory}

The  D-instanton contribution in the type I theory also arises from the terms in \eqref{typeiloopaa}  with $n\ne 0$.   Although there appears to be no closed-form expression in terms of Bessel functions for this integral, it is easy to make use of a saddle point analysis  to find the terms that are exponentially damped when  $\raten/\raele  = \gib << 1$. They give a series of the form
\begin{eqnarray} 
\label{insti}
C[\raten, \raele] \bigl |_{inst} &=& \frac{\lepl}{\raten \raele}
\sum_{k>0} c_k\,\exp\left(-2\pi k\frac{ \raele}{\raten}\right)  \left(1+ O(\raten/\raele) \right) \nonumber\\
&=&  \ell_{I} r_I
\sum_{k>0} c_k\, \exp\left(- \frac{2\pi k}{\gib}\right)\left(1+ O(\gib) \right) \,,
\end{eqnarray} 
where $k = \hat m n$ and $c_k$ is a constant that can be determined by standard saddle point methods.

\subsubsection*{Comments and summary of features of the  $s\, t_8 \tr F^4$ calculation}

We do not expect that the loop diagram in figure \ref{onegrav} should generate the exact coefficient of the $d^2 \tr F^4$ interaction, but we have included it because it does generate a coefficient that demonstrates several of the expected features of string theory. A summary of these is as follows.

\begin{itemize}

\item  The expression for the coefficient of the $s\, \tr F^4$ interaction contains perturbative terms up to  four loops in the HO theory.  Strikingly, the one-loop term is absent  in the weakly coupled type I limit - this agrees with the explicit string calculation \cite{Rudra:2016xx}.   Since we expect that there are further contributions to this interaction from other sources, it is  quite possible that there is an infinite number of contributions in the full perturbation expansion.

\item The origin of the type IA two-loop term in string perturbation theory can be traced to the world-sheet diagram shown in  figure \ref{twolooptypeI}, which is a torus with a boundary  localised on a stack of eight-branes coincident with one of the orientifold planes. The four gauge particles are attached to these eight-branes. This contribution is of the same order as the disk world-sheets with two holes or cross-caps inserted, which contribute to  $\gia\, t_8 (\tr F^2)^2$ as we discussed earlier.

\begin{figure}[h]
\begin{center}
\begin{tikzpicture}[scale=.5]
\begin{scope}[shift={(0,0)}]
\begin{scope}[rotate=0]      
\filldraw [gray!50] (0,0) ellipse (4 and 3);
\draw [ultra thick] (0,0) ellipse (4 and 3);

\filldraw [white, ultra thick]     
(-1.3,-.4) .. 
controls (-.5,.6) and (.5,.6) ..
(1.3,-.4)
(-1.3,-.4) .. 
controls (-.5,-.9) and (.5,-.9) ..
(1.3,-.4)
; 
\draw [white, ultra thick] (-1.3,-.4)  -- (1.3,-.4);
\draw [ultra thick]     
(-1.3,-.4) .. 
controls (-.5,.6) and (.5,.6) ..
(1.3,-.4)
(-1.3,-.4) .. 
controls (-.5,-.9) and (.5,-.9) ..
(1.3,-.4)
;    
\draw [ultra thick]
(-2,0) .. 
controls (-.5,-1) and (.5,-1) ..
(2,0)      
; 
\end{scope}
\end{scope}

\begin{scope}[shift={(3.4,0)}] 
\filldraw [white] (0,0) ellipse (.7  and 1.6); 
\draw [ultra thick] (0,0) ellipse (.8 and 1.6);
\draw [ultra thick, suym](2,.8) -- (.8,.5);
\draw [ultra thick, suym](2,-.8) -- (.8,-.5);    
\draw [ultra thick, suym](1,1.8) -- (-.8,.5);
\draw [ultra thick, suym](1,-1.8) -- (-.8,-.5); 
\end{scope}

\end{tikzpicture}
\end{center}
\caption{A contribution of order  $\gib  \,s \, t_8 \tr F^4$ in Type I superstring theory.}
\label{twolooptypeI}
\end{figure}
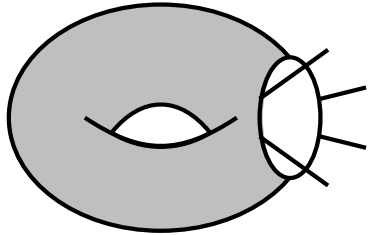

\item  Just as we saw for the $t_8 t_8 \curv^4$ interaction that was obtained from the bulk supergravity loop, the coefficient function $C(\raten,\raele)$ contains an infinite sequence of non-perturbative instantonic terms. These make exponentially small contributions of order $e^{-2\pi k/g_{ho}}$ in perturbative HO theory and $e^{-2\pi k/g_{I}}$ in perturbative type I theory.  

\end{itemize}

\section{Discussion}
\label{sec:result}
 
The arguments of this paper, based on perturbative supergravity in the Ho\u rava--Witten background, $\mathbb{M}^9 \times S^1 \times S^1/\mathbb{Z}_2 $,  lead to an interpretation of a number of features of the low-energy expansion of scattering amplitudes for gauge particles and gravitons in $\mathcal{N}=1$ string theories in $D=9$ and $D=10$ dimensions (although we did not discuss mixed gauge/gravity amplitudes).  Several of these are known features of superstring perturbation theory that would seem mysterious without such an interpretation and some of them are indications of non-perturbative features, such as non-renormalisation theorems and the contributions of instantons.  The interpretation of supergravity Feynman amplitudes in terms of string theory has some unusual features, the most striking of which are summarised here.

\sm 
\begin{itemize}

\item 
The gravitational propagator in the Ho\u rava--Witten bulk depends of the interval length, $\lepl\raele$.  As a result, the gauge boson tree diagrams in section \ref{sec:YMtrees} and  graviton tree diagrams in section \ref{sec:gravitrees}, depend on the string coupling constants induced by the orbifold geometry.  One consequence is that the low-energy expansion of these tree amplitudes contains a power series in $s\, \raele^2 \lepl^2 = \ghe^2 s\,\lesth^2= s\, \lesti^2 /\raib^2 $ multiplying $(\tr F^2)^2$.   This infinite series of terms is therefore interpreted in terms of  contributions to loop amplitudes in HE string theory to all orders in the coupling constant $\ghe^2$.  In the type I description, this sum of this series reproduces the factor \eqref{typeiloop} in the one-loop cylinder amplitude  (figure~\ref{fig:nonplanar})   that comes from the sum of  the ground states of closed string winding modes.  These modes, which correspond to the Kaluza--Klein modes in the Ho\u rava--Witten interval, are unstable.  However, the agreement of the supergravity and string amplitudes up to order $s \ghe^2 t_8(\tr F^2)^2$  is in accord with expectations based on supersymmetry and suggests that the effects of this instability enter at higher order in the low-energy expansion.
 
\sm

\item
The tree amplitudes were generalised in section \ref{subsec:itetree} to tree contributions with ``iterated'' propagators induced by the $\curv^2$ and $(\partial H)^2$ interactions localised on the boundaries. This generated another infinite series of powers of $s \,\lepl^2/\raele = s\, \lesth^2 =\gib s\, \lesti^2 $.
This corresponds an infinite sequence of tree-level contributions in the HE and HO theories that reproduces the explicit factor in the heterotic tree amplitude displayed in  \eqref{grosssloan}.   In the type I theory interpretation this low-energy expansion of a tree level HO factor is  interpreted as an infinite series of higher order terms corresponding to world-sheets with arbitrary numbers of boundaries.

\sm

\item

In section \ref{sec:loopamp} we gave a novel analysis of the ten-dimensional $E_8$ gauge theory loop amplitude compactified on $S^1$, which   is relevant to a loop that is localised in either of the two Ho\u rava--Witten boundaries.  
Before compactifying on $S^1$ the states circulating in the loop are in the  adjoint representation (the ${\bf 248}$) of $E_8$ and the amplitude is an ill-defined ultraviolet divergent integral multiplying $t_8(\Tr_{\bf 248} F^2)^2$.  After compactification, with the gauge group broken by Wilson lines  to $SO(16)$, the states circulating in the loop are the massless $SO(16)$ adjoint  gauge states together with their Kaluza--Klein tower, as well as  the Kaluza--Klein tower of massive $SO(16)$ spinor states.  These states complete the adjoint representation of $E_8$ in the large-$\raten$  limit.  

The sum over Kaluza--Klein modes circulating in the loop was transformed into a sum over windings of the loop around the $x^{10}$ circle by means of a Poisson summation and the ultraviolet divergence was thereby isolated in the zero winding term.  We argued that, after renormalisation,  this zero-winding  term gives a contribution that has no sensible string theory interpretation and so its renormalised value should be taken to vanish.  The non-zero winding terms give a sum of finite contributions proportional to $1/\rahe = \raho$ that were interpreted as contributions to the $S^1$ compactification of the  HO theory.  Making use of the conspiracy between adjoint and spinor traces in  \eqref{diffcas}, which is a special feature of $SO(16)$ these non-zero winding terms contribute to $t_8\,\tr_i F_i^4$ ($i=1,2$), where the traces are in the {\it fundamental} representation of  either $SO(16)$, which agrees with the result in HO perturbation theory.   Strikingly, this would {\it not}  be the structure of the one-loop amplitude in conventional $SO(16) \times SO(16)$ gauge theory, where the traces would be in the adjoint representation of  either $SO(16)$ (and the loop would be ultraviolet divergent).  The trace has to be in the fundamental representation in order to agree  with HO/type I duality, since the type I amplitude arises from a disk diagram with a Chan--Paton factor, which obviously gives single trace in the fundamental representation.

\sm

\item 
The arguments concerning the contributions of one-loop four-graviton amplitudes considered  in section~\ref{sec:gravloop} provided suggestive illustrations of HO/type I duality in the $\mathbb{M}^9 \times S^1 \times S^1/\mathbb{Z}_2 $ background.  There were  two kinds of loops.  The first was one in which the external gravitons coupled to a loop of gauge particles localised on either boundary, which generalised the gauge theory loop amplitude summarised in the previous item and led to a combination of $t_8 \tr \curv^4$ and $t_8 (\tr \curv^2)^2$ interactions. 

\sm 

The second contribution came from a loop in which the external gravitons coupled to a loop of gravitons propagating in the bulk.  In this case the amplitude involved a sum over the Kaluza-Klein momentum, $p_{11}$, in the eleventh dimension.   We argued that the $p_{11}=0$ contribution to this loop integral added a term to the gauge loop contribution, leading to a total one-loop effective action in the HO theory proportional to  $t_8 \, \ygs(\curv)$.  This is the parity conserving partner of the anomaly cancelling ten-form $\epsilon_{10} B \,\ygs(\curv)$ and is expected to be one-loop exact.

\sm

The $p_{11}\ne 0$ terms in the loop integral generated a $t_8t_8 \curv^4$ interaction, analogous to that of the type II theories, but with a coupling constant dependence described by  $(\gho^{-\half}\, E_\threeh(i/\gho) - 2\zeta(2))$, where $E_\threeh(i/\gho)$   is an Eisenstein series that has a weak coupling expansion containing just two perturbative terms (tree-level and one-loop) and an infinite series of $\mathbb{Z}_2$ D-instanton contributions  The presence of $- 2\zeta(2)$  subtracts the one-loop term of order $\gho^0$ from the Eisenstein series, which accounts for the absence of the $p_{11}=0$ term in this loop contribution.  It follows that the  $t_8t_8 \curv^4$ interaction in the HO theory only has a perturbative tree-level contribution together with D-instanton contributions.  In addition there is a one-loop contribution proportional to $t_8\, \ygs(\curv,F)$, which is the parity-conserving partner of the anomaly-cancelling ten-form  (the curvature dependent piece was obtained in \eqref{efftrr4}).    After applying HO/type I duality we conclude that the type I theory has contributions to $t_8 t_8 \curv^4$  of order  $1/\gib^2$ and $\gib^0$, corresponding to spherical and toroidal world-sheets. 

\sm

This analysis  hints at the non-renormalisation of the $\curv^4$ interactions beyond one  loop in any of the $\mathcal{N}=1$ string theories -- a feature that ties in with expectations based on perturbative supergravity and string theory calculations  \cite{Bern:2012cd,Tourkine:2012ip}. 

\sm 

 \item 
 The presence of D-instanton terms in the expansion of the function $E_\threeh(i/\gho)$ that multiplies the $t_8t_8 \curv^4$ interaction in \eqref{hotend}  is  crucial in ensuring its invariance under HO/type I duality, $\gho \to \gib=1/\gho$.    Although the necessity of type I D-instanton contributions is well documented \cite{Witten:1998cd}, there is much less evidence that there should be well-defined  contributions of HO D-instantons.
After all, these objects originate from world-lines of {\it unstable} D0-branes in the HE theory wound around the $x^{10}$ direction.   Such unstable heterotic D-branes are generally not expected to play a preferred r\^ole, so while the coupling constant dependence in \eqref{hotend} is interesting, it is by no means proven to be exact.

\sm 

\item 

Similar considerations determine the structure of the parity-violating gauge and gravitational effective  interactions that are necessary to ensure the absence of chiral gauge, gravitational and mixed anomalies. These are contained in the ten-form $B\wedge \xgs $ and were discussed in the context of the HE theory in \cite{Horava:1996ma}.  The expression for $\xgs $ is given  in \eqref{hegsterm} as a sum of the bulk ``Vafa--Witten'' term, $\xvw $, and $ (\tr_i F_i^2 - \tr \curv^2/2)^2$ terms localised on the two boundaries.   In this paper we have seen how the parity-conserving partners of these interactions arise as the sum of the $t_8t_8 \curv^4$ interaction induced by a gravitational loop and the combination $t_8 (\tr_i F_i^2- \tr \curv^2/2)^2$ induced by a sum of  tree amplitudes with vertices localised in each boundary.   

\sm
In the HO case the expression for $\xgs= \epsilon \ygs $ is given by \eqref{hogsterm} (and $\ygs$ is given in \eqref{gsfull}).   We found that the corresponding parity-conserving  terms of the form $t_8\, \ygs$ arise in a rather different fashion since they originate from a loop of  gauge particles localised in either compactified boundary, coupling to external gauge particles (as in section \ref{sec:loopamp}), or  gravitons (as in section \ref{sec:gravloop}), or a mixture of both gauge particles and gravitons (which we have not explicitly considered).  The  anomaly-cancelling terms in the HO theory defined by \eqref{hogsterm} are clearly determined by analogous  parity-violating  loop amplitudes with an external $B$-field coupling to four gauge or gravitational particles, although we have not explicitly evaluated these amplitudes in this paper. As in the parity-conserving case, such an analysis would explain the occurrence of the fundamental trace in the $\tr F^4$ term in \eqref{hegsterm} and \eqref{gsfull}.
   
\end{itemize}

More generally, it is obvious that the Feynman diagram approximation is not adequate for understanding the physics of M-theory beyond the low-energy approximation.  Even in the context of the low-energy expansion in terms of Feynman diagrams, there are many other sources of contributions to higher derivative interactions of the form $d^{2k} F^4$,  $d^{2k} \curv^4$ and mixed gauge/gravity amplitudes, that are not generally expected to be protected against receiving higher order corrections.  These include contributions from higher order terms in the low-energy expansion of tree-level and one-loop Feynman amplitudes considered in this paper, as well as  from higher-loop amplitudes that we have not considered.  Clearly, understanding the dynamics of M-theory beyond the first few terms in the low-energy expansion requires a deeper understanding of intrinsically stringy effects that are not probed by  the supergravity approximation.  It is nevertheless of interest to probe the extent to which low order terms are determined by supersymmetry.

\paragraph{Acknowledgements}

We thank Malcolm Perry, Ashoke Sen, David Skinner, Piotr Tourkine, Pierre Vanhove, Edward Witten for useful discussions. The research leading to these results has received funding from the European Research Council under the European Community's Seventh Framework Programme (FP7/2007-2013) / ERC grant agreement no. [247252]. A.R. would like to thank TIFR, Mumbai, where a preliminary version of this work was presented. A.R. would also like to thank HRI, Allahabad, where a final version of this work was presented.  We are also grateful to Michael Haack for pointing out the error in \eqref{hoeffstring} in the first archive version of this paper.

\appendix  

\section{The M-theory/string theory dictionary in the Ho\u rava--Witten background}
\label{sec:dictionary}

\subsection{Dualities relating $\mathcal{N}=1$ superstring theories}
\label{subsec:stduality}

There are four  distinct $\mathcal{N}=1$ superstring  perturbation ``theories''  in ten-dimensional Minkowski space (each with sixteen supercharges), namely, heterotic  $\textrm{E}_8 \times  \textrm{E}_8$, heterotic $\textrm{Spin}(32)/\mathbb{Z}_2$, type I superstring theory and type IA superstring theory. These theories have different perturbative expansions but they are related to each other by stringy dualities.  These dualities relate the moduli associated with each of these theories in the following manner.

 \sm 

\begin{enumerate}
\item {\bf Heterotic T-duality}

T-duality in the direction of the $x^{10}$ circle maps the heterotic string vacuum with unbroken $\textrm{Spin}(32)/\mathbb{Z}_2$ on $\mathbb{M}^9\times S^1$ to itself, and maps the heterotic string vacuum with unbroken $\textrm{E}_8 \times \textrm{E}_8$ on $\mathbb{M}^9\times S^1$ to itself.  However, T-duality is more interesting when the gauge group is broken by Wilson lines in the compactified theory.  The HE theory with  $\textrm{E}_8 \times  \textrm{E}_8$ broken to $\textrm{SO}(16) \times  \textrm{SO}(16)$ is related by T-duality to the HO theory with $\textrm{Spin}(32)/\mathbb{Z}_2$ broken to $\textrm{SO}(16) \times  \textrm{SO}(16)$ \cite{Narain:1985jj, Narain:1986am, Ginsparg:1986bx}. The parameters of two heterotic string theories are related to each other by  
\begin{eqnarray}
\rahe= \frac{1}{\raho}\,, \hspace{30pt} \ghe=\frac{\gho}{\raho}\,.
\label{hettdual1}
\end{eqnarray}

\sm

\item{\bf Type I theory from type IIB Orientifold}

The type I theory is a theory of unoriented  open and closed strings with $\textrm{SO}(32)$ gauge group that is equivalent to type IIB in the presence of an orientifold nine-plane \cite{Horava:1989vt} and sixteen D9-branes, which are needed to neutralise the total R-R charge \cite{Polchinski:1995mt}. This explains the origin of the $\textrm{SO}(32)$ gauge group in terms of D-branes.  The coupling constants of the type II and type I theories are related by
\begin{eqnarray}
	\giib =\sqrt 2\ \gib \,,
\end{eqnarray}
which fits in with the  understanding that the world volume coupling constant of D-branes in the type I theory is twice that in the type IIB theory.

 \sm

\item  {\bf Relationship of type IA and type IIA theories}

The type IA theory has two orientifold eight-planes located at the fixed points of the orbifold of the $x^{11}$ circle, There are sixteen D8-branes positioned at points on the $x^{11}$ axis  between the fixed planes, together with their images.  In the $SO(16) \times SO(16)$ case considered here, there eight $D8$-branes coincide with each orientifold plane, also coinciding with their images. Type IA theory can be equivalently thought as type IIA theory in an orbifold $\mathbb{M}^9 \times S^1/\mathbb{Z}_2 $.  The coupling constants of the IA and IIA theories are equal and given by 
\begin{eqnarray}
\giia =\gia \,.
\label{iaiia}
\end{eqnarray}

\sm

\item  {\bf T-duality of type I and type IA theories}

When compactified to nine dimensions on a circle of radius $\raib$ (in string units)  the type I theory is T-dual to the type IA theory compactified on a circle of radius $\raia$.  This is the image of the transformation that relates the type IIA and type IIB closed string theories compactified on a circle.  The radii and the coupling constants are related by
\begin{eqnarray}
	\raib= \frac{1}{\raia} \hspace{30pt}\frac{\raib }{\gib}=\frac{1}{\gia} \,.
\label{itdual1}
\end{eqnarray}
The details of this duality are again particularly simple in the situation in which the symmetry group is $SO(16)\times SO(16)$, which is the case in which the dilaton charge of the orientifold planes is locally screened.  
 
\sm

\item {\bf Heterotic $Spin(32)/\mathbb \mathbb{Z}_2$ /Type I strong coupling duality} 

 Heterotic $\textrm{Spin}(32)/\mathbb{Z}_2$ theory and type I $\textrm{SO}(32)$ theory are conjecturd to be related by S-duality  \cite{Witten:1995ex, Polchinski:1995df}  with the following relationships between the patrameters of the theories
\begin{eqnarray}
\gho=\frac{1}{\gib} \hspace{30pt} \raho= \frac{\raib}{\gib^{1/2}}\,\hspace{30pt} \lesti  =\lesth   (\gho)^{1/2} \,.
\label{Idua5}
\end{eqnarray}
According to S-duality the D-string of the type I theory can be identified with fundamental heterotic string \cite{Polchinski:1995df}.  It was noted in \cite{Witten:1998cd} that in order for this duality to be satisfied there have to be non-perturbative $\mathbb{Z}_2$ instanton effects in the type I theory that are associated with the breaking of $O(32)$ to $SO(32)$.  We will comment on these in an explicit calculation later in this paper.  

\sm 

{\it An important comment:}\hfill\break
\noindent
A term of fixed loop number in the HO theory has a low-energy expansion in powers of $\lesth^2\, s$.  Since these powers translate into powers of $\gib \lesti^2   \,s$ in the type I theory  this expansion may be reinterpreted as a sum of terms of higher order in type I perturbation theory, with the number of world-sheet boundaries increasing as the power of $s$ increases.  However, this  identification of individual terms in the heterotic and type I/IA  low-energy effective actions only applies to special terms.  More generally the strong/weak coupling duality does not allow the identification of specific terms in the expansion of the HO theory with specific terms in the type I theory, as is seen explicitly in the body of the paper.
\end{enumerate}

\subsection{Relationships between $\mathcal{N}=1$ and type II string theories.}

We will here review the relations between the parameters in maximally supersymmetric ($\mathcal{N}=2$)  string theories (type II theories) compactified on a circle and the $\mathcal{N}=1$ theories compactified on $S^1$.  The objective is to clarify certain factors of $\sqrt 2$ that arise in passing from the type II theories to the heterotic and type I theories.

\begin{enumerate}
	\item   
	{\bf Relations between the parameters.}
	
	The parameters of the type IIA and IIB theories compactified on $S^1$ to nine dimensions are related by 
\begin{eqnarray}
\frac{2\pi \lestii\, \raiia}{\giia^2}	
=\frac{2\pi \lestii\, \raiib}{\giib^2}\,,
\label{factortwo1}
\end{eqnarray}
where $\raiia=1/\raiib$ and $\lestii$ is the type IIA or IIB string length scale. 

The type IA  theory is obtained from type IIA  by compactifying on $S^1/\mathbb{Z}_2$ together with a world-sheet orientation reversing operator, $\Omega$.  The  type I theory is simply obtained by acting on the type IIB theory with $\Omega$ and is reduced to nine dimensions   by compactifying on $S^1$. So the relation \eqref{factortwo1} between type II theories becomes a relation between type IA and IB if  the  $S^1$ is replaced by the orbifold $S^1/\mathbb{Z}_2$ on the left-hand side of the equation, which  halves the volume of the compact direction but  the right-hand side is unaltered.\footnote{See the comment below eqn 13.3.30 in page 151 of \cite{Polchinski:1998rr}}. In that case the above relation is replaced by  
\begin{eqnarray}
\frac{\pi \lestii  \raiia}{\giia^2}	
=\frac{2\pi \lestii  \raiib}{\giib^2}\,.
\label{factortwo2}
\end{eqnarray}

The  relation between the type I and type IA theories takes the standard form if  we define 
\begin{subequations}
\begin{eqnarray} 
&&\giia=\gia \qquad\qquad \giib=\sqrt{2}\ \gib	
\label{factortwo3.1}
\\ 
&&\raia=\raiia \qquad\qquad \raib=\raiib
\label{factortwo3.2}
\end{eqnarray}
and equate the type I  and type II string lengths so that 
\begin{eqnarray}
\lestii=\lesti\,.
\label{factortwo3.3}
\end{eqnarray}	
\end{subequations}
Equation \eqref{factortwo2} can then be written as 
\begin{eqnarray}
\frac{2\pi \lesti \raia}{\gia^2}	
&=&\frac{2\pi \lesti  \raib}{\gib^2}
\qquad,\qquad
\raia=\frac{1}{\raib}	\,,
\label{factortwo4}
\end{eqnarray}
or
\begin{eqnarray} 
\frac{\raib}{\gib}=\frac{1}{\gia}\,.
\label{types}
\end{eqnarray}

\item 
{\bf Dp-brane tension in type II and type I theories.}

The tension of a Dp brane in the type II theory is 
\begin{eqnarray}
\frac{T_p}{\gii}	\qquad\qquad T_p=\frac{2\pi}{(2\pi \lestii )^{p+1}} \,.
\label{factortwo5}
\end{eqnarray}
The D-brane tension in  the unoriented theory is smaller by a factor of $1/\sqrt 2$ \footnote{See footnote 8 in page 21 of \cite{Polchinski:1996fm} } and hence the tension in type I theory is
\begin{eqnarray}
\frac{T_p}{\sqrt 2\, \giib}\,.	
\label{factortwo6}
\end{eqnarray} 
From the relation between the type IIB and type I coupling constants in \eqref{factortwo3.1},  it follows that the tension in the type I theory is 
\begin{eqnarray}
\frac{T_p}{ 2 \gib}	\, ,
\label{factortwo7}
\end{eqnarray}
which is consistent with the claim on page 151 of \cite{Polchinski:1998rr}.

\sm

\item  {\bf Relation between orientable and unorientable closed string loop amplitudes.} 

An important point of relevance to the interpretation of the graviton loop calculations in section \ref{sec:gravloop}  involves the relationship of the type II one-loop amplitude to that of the type I theory.  The type II loop amplitudes are defined on orientable world-sheets  while $\hat \Omega$ is the orientation reversing operator. Hence, the orientable part of the $n$-loop diagram in type I theory comes with a factor of 
\begin{eqnarray}
\frac{(\giib)^{2(n-1)}}{2^n} \,,
\label{factortwo8}
\end{eqnarray}
using the relation between the type IIB coupling and type I coupling \eqref{factortwo3.1} this is simply a factor of 
\begin{eqnarray}
\half \,(\gib)^{2(n-1)} \,.
\label{factortwo9}
\end{eqnarray}

This is relevant to the one-loop calculation in section~\ref{subsec:gravloop1}, where we suggest that the ratio of tree-level to one-loop amplitudes arising from $I^{bulk}$ (that enters in \eqref{Idef}) in the HO theory.is the same as in the type I theory.      The factor of $\half$ in \eqref{factortwo9}, together with the fact that the closed-string tree-level amplitude (given by four vertex operators attached to a spherical world-sheet)  is proportional to $1/\giib^2= 1/(2\gib^2)$ from \eqref{factortwo3.1}, is in accord with this suggestion.

\sm

\item {\bf Type I T-duality and D-particles in type IA theory} 

Recall that a D0-brane of the type IA theory moving in the bulk (i.e., in the fundamental domain, $0 <x^{11} < \pi \lepl  \raele$)  is identical to the D0-brane of the type IIA theory and has a mass
 \begin{eqnarray}
\frac{T_0}{\gia} =\frac{T_0}{\giia} \,.		
\label{factortwo12}
\end{eqnarray}
This is the description in the downstairs formalism in which $x^{11}$ is restricted to an interval with two boundaries.  
In the upstairs formalism, where $x^{11}$ spans the circle, $0\le x^{11} \le 2\pi \lepl  \raele $,  the fields are subject to the $\mathbb{Z}_2$ orbifold condition (so that for a scalar field, $\Phi(x^{11}) = \Phi(- x^{11})$). In this formalism bulk D0-branes always come in pairs comprising a D0-brane at $0 <x^{11} < \pi \lepl  \raele $ with mass $T_0/(2 \gia)$ and its mirror image at $- x^{11}$ with the same mass. The mass of the pair agrees with the mass of the bulk D0-brane in the downstairs description.

A  type IA D0-brane can be its own mirror image if it is ``stuck'' to either fixed point of the orbifold $(x^{11} = 0$ or $x^{11} =\pi \lepl \raele $) and cannot move in the bulk.  The mass of such a stuck type IA D0-brane is 
\begin{eqnarray}
	\frac{T_{0}}{2\gia}\,.
\label{factortwo11}
\end{eqnarray}
 T-duality along $x^{11}$ identifies such a  D0-brane with a type I D1-brane wrapping once around  the circle of radius $\lesti  \raib$.  The D1-brane of the type I theory has a tension
\begin{eqnarray}
	\frac{T_{1}}{2\gib}\, ,
\label{factortwo10}
\end{eqnarray} 
and from  \eqref{factortwo5} it is easy to see that  its mass agrees with that of the stuck type IA D0-brane in \eqref{factortwo11}.

\end{enumerate}

\subsection{Relationship between string theory parameters and M-theory parameters}
\label{subsec:mthpara}

We will here give a brief summary of the relationships between the parameters $\raele$ and $\raten$ of the Ho\u rava--Witten  geometry on a circle ($\mathbb{M}^{9}\times  S^1 \times S^1/\mathbb{Z}_2 $) and the  parameters of the various $\mathcal{N}=1$ string theories compactified to nine dimensions.  For each of these theories these parameters consist of the coupling constant and radius of the tenth dimension: ($\ghe,\, \rahe$), ($\gho,\, \raho$), ($\gia,\, \raia$), ($\gib,\, \raib$), for the HE, HO, type IA and type I theories, respectively.
Detailed arguments for these relationships can be found in \cite{Horava:1995qa,Horava:1996ma}.

 The Ho\u{r}ava-Witten geometry is obtained from eleven-dimensional Minkowski space by compactifying on the orbifold of a circle in the eleventh direction, where the generator of the orbifold group acts on the eleventh dimension $x^{11}$ by reflection $x^{11}\rightarrow -x^{11}$, as well as acting on the three-form field of eleven-dimensional supergravity, $C\rightarrow-C$, i.e, the three form field $C$ is odd under parity reflection
  Only the components $C_{11 \mu \nu }$ are even under reflection and hence survive the $\mathbb{Z}_{2}$ projection. Similarly the $h_{11\mu}$ components of the graviton are odd under reflection and hence projected out. 
The gauge fields are vector fields that propagate on the boundary of the space-time.

The  radius of the orbifold circle and the other circle, measured in  $11$-dimensional Planck units, are  $\raele$ and  $\raten$, respectively\footnote{This  is the convention used in  \cite{Horava:1996ma}, but differs from that of \cite{Horava:1995qa} .}.  The physical length of the interval in the $11^{\textrm{th}}$ direction  is then given by 
\begin{eqnarray}
\ciele= \pi \,\raele \lepl  \,.
\label{Idua0}
 \end{eqnarray}
 The metric on the cylinder is given by 
\begin{eqnarray}
	G^{(2)}_{ij} = 
\lepl^2	\frac{\mathcal{V}}{\Omega}
\left(
\begin{matrix}
 \Omega^2 & 0 \\
  0 & 1
 \end{matrix}
 \right)
=2 \pi^2\lepl^2 \left(
\begin{matrix}
 \raten^2 & 0 \\
  0 & \raele^2
 \end{matrix}
 \right) \,,
 \hspace{30pt}
\label{Idua2.3}
\end{eqnarray}
\begin{eqnarray} 
\mathcal{V} =  2\pi^2 \raten \raele\,,\qquad \Omega  = \frac{\raten}{\raele} \,.
\label{Idual2.4.a}
\end{eqnarray}
The following summarises the relationships between the M-theory pararmeters, $\raten$ and $\raele$ and the string theory parameters for each of the $\mathcal{N}=1$ string theories.   It also summarises the relations between the eleven-dimensional Planck length, $\lepl$  and the string length in each of the nine-dimensional string theories.
  
\begin{itemize}
\item
{\bf Heterotic $ E_8 \times E_8$ theory}

\noindent
The ten-dimensional heterotic $E_8 \times E_8$ theory has a coupling constant that is expressed as 
\begin{eqnarray}
\ghe=\raele^{3/2}\, .
\label{Idua1}
\end{eqnarray}
while the radius of the spatial circular dimension is given in string units by 
\begin{eqnarray}
\rahe= \raten\sqrt{\raele}\,.
 \label{Idua2}
\end{eqnarray} 
The heterotic string length $\lesth$ is related to the eleven-dimensional Planck length $\lepl$ by
\begin{eqnarray}
\lesth  =\frac{\lepl}{\sqrt{\raele}} \,.
\label{Idua3}
\end{eqnarray}

\item {\bf Heterotic   $Spin(32)/\mathbb{Z}_2$ theory} 

Using \eqref{hettdual1} and  \eqref{Idua1} - \eqref{Idua3}, we find the relations between the $Spin(32)/\mathbb{Z}_2$ heterotic string theory parameters and $\raele $ and  $\raten $ 
\begin{eqnarray}
	\gho=\frac{\raele}{\raten}
\hspace{30pt} 
\raho= \frac{1}{\raten\sqrt{\raele}}
\hspace{30pt} 
\lesth  =\frac{\lepl}{\sqrt{\raele}} \,.
\label{Idua2ho}
\end{eqnarray}
 
\sm

\item
{\bf Type IA theory }

Upon compactification of the Ho\u rava--Witten geometry on the circle of radius $\raten$ the theory may be interpreted in terms of type IA string theory.   The relation between type IA parameters and M-Theory parameters 
\begin{eqnarray}
	\gia=\raten^{3/2}
\hspace{30pt} 
\raia= \raele\sqrt{\raten}
\hspace{30pt} 
\lesti  =\frac{\lepl}{\sqrt{\raten}} \,.
\label{Idua3a}
\end{eqnarray}
 
\sm

\item{\bf Type I theory }

T-duality  of type IA  string theory along the orbifolded direction (with radius $\raele$) results in the type I description, with parameters that are related to those of M-theory by\footnote{This corrects a typographical error in equation (3.2) in \cite{Horava:1995qa}.}
\begin{eqnarray}
	\gib=\frac{\raten}{\raele}
\hspace{30pt} 
\raib= \frac{1}{\raele\sqrt{\raten }}
\hspace{30pt} 
\lesti  =\frac{\lepl}{\sqrt{\raten }} \,.
\label{Idua4}
\end{eqnarray}
\end{itemize}

\subsection{Particle states in nine dimensions. }
\label{sec:particlestates}

We will here briefly summarise the spectrum of particle states that arise in the M-theory orbifold $\mathbb{M}^9 \times S^1 \times S^1/\mathbb{Z}_2 $ and the corresponding string theories, compactified on $S^1$ to nine dimensions, as described in \cite{Horava:1996ma}.    Although the M2-brane states do not enter into the amplitude calculations in the body of this paper,  the particle states that arise from wrapping it on $S^1/\mathbb{Z}_2 \times S^1$ enter into a discussion of the multiplets of states in nine-dimensions. 

These particle states may be obtained starting from the BPS states in the maximally supersymmetric theory obtained from M-theory on $\mathbb{M}^9\times S^1\times S^1$.    Following \cite{Horava:1996ma} (with a slight change of notation) the masses  of these states are given by
\begin{eqnarray}
\frac{|m|}{\lepl  \raele }\,, \quad\qquad \frac{|n|}{\lepl  \raten}\,,  \qquad\quad \frac{|w|\, \raten \raele }{\lepl } \,,
\label{n2mass}
\end{eqnarray}
where $m,n,w \in \mathbb{Z}$.
These are the  Kaluza--Klein modes in the $x^{11}$ and $x^{10}$ directions, with charges $m$ and $n$, respectively, together with the wrapped M2-brane states with wrapping number $w$. 

The states of interest to us are those that arise in the $\mathbb{M}^9\times S^1/\mathbb{Z}_2\times S^1$ compactification of M-theory, which must be invariant under the action of the $\mathbb{Z}_2$ orbifold group.  This identifies $x^{11}$ with $-x^{11}$ and so its action on the states is  $|m,n,\ell\rangle \to \pm |-m,n,\ell\rangle$.  The Kaluza--Klein charge $m$ in the orbifold direction is therefore not conserved in this background.  The wrapping number $w$ is conserved by virtue of the fact that the orbifold projection acts on both the embedding space-time and the M2-brane world-volume (this is the definition of an orientifold).

The Kaluza--Klein modes in the $x^{10}$ direction,  translate  (using the dictionary in 
appendix~\ref{sec:dictionary}) into stable states with the following masses in the various string theories
\begin{eqnarray}
\frac{|n|}{\lesth  \rahe}\, ,\qquad\quad \frac{|n| \, \raho}{\lesth }\, ,\qquad\quad \frac{|n|}{\lesti  \gia}\, ,\qquad\quad \frac{|n| \raib}{\lesti  \gib}\,.
\label{kk10masses}
\end{eqnarray}
These are Kaluza--Klein modes of the HE theory, winding modes of the HO theory, D0-branes of the type IA theory and winding modes of the D1-brane of the type I theory (which is the heterotic $Spin(32)/\mathbb{Z}_2$ string), respectively.

The unstable Kaluza--Klein modes in the $x^{11}$ direction correspond to unstable states  in the various  nine-dimensional $\mathcal{N} =1$ string theories with masses given by
\begin{eqnarray}
\frac{|m|}{\lesth \, \ghe}\,, \qquad\quad \frac{|m|\, \raho}{\lesth \, \gho}\,, \quad\qquad \frac{|m|}{\raia\lesti }\,,\quad \qquad \frac{|m| \, \raib}{\lesti }\,.
\label{kkmasses}
\end{eqnarray} 
The first of these is the mass of a charge-$m$ $D0$-brane in the HE theory that is inherited from the type IIA theory but its charge is not conserved.  The second is the mass of a ground state of the unstable $D1$-brane in the HO theory with  winding  number $m$ (a  wound type I string), which is  related to the HE theory by T-duality.  The third entry in \eqref{kkmasses} is the mass of the  charge-$m$ non-conserved Kaluza--Klein mode of the type IA closed string, and the fourth entry is a state of the type I closed string with winding number $m$  that is T-dual to the type IA state, and is unstable since the type I string  can break into open strings.  Although the unstable D-branes of the HE and HO theories are motivated by extrapolating from the Ho\u rva--Witten starting point, it is not clear how they can be described directly in the heterotic string theories.\footnote{The argument given by Shenker \cite{Shenker:1990uf} suggested the existence of D-branes and D-instantons based on the divergence of closed string perturbation theory appears to apply not only to the type II theories but also to the heterotic and type I theories.}  The instability of the type I string is well understood.  In the Ho\u rava--Witten description  invariance under the action of the orbifold requires a superposition of type I string states of opposite orientations.

 The stable wrapped M2-brane states with wrapping number $w$ in \eqref{n2mass} translate into states with the fiollowing masses in the string theory descriptions
\begin{eqnarray}
\frac{|w|\rahe}{\lesth }\,, \quad\qquad \frac{|w|}{\lesth  \raho}\,, \quad\qquad 
\frac{|w|\, \raia}{\lesti  }\,, \quad\qquad  \frac{|w|}{\lesti  \raib }\,. 
\label{wrappingmass}
\end{eqnarray}
These are respectively,  winding states of the HE string, Kaluza--Klein states of the  HO string, winding states of the type IA string, and Kaluza--Klein states of the type I theory.
 
\section{Notation and conventions}  
\label{sec:A} 

We will here summarise some well-known features of certain terms in the low-energy effective theory that arise from ten-dimensional  string theories with $\mathcal{N}=1$ space-time supersymmetry.   We will illustrate these in subsection~\ref{sec:duality} by reviewing the low lying terms in the low-energy expansion of the heterotic and type I $SO(32)$ theories that contribute to on-shell three-point amplitudes.  In subsections~\ref{hwsugra} and \ref{sec:gravityreview} we will review some notation relating to parity conserving and parity violating terms that  are related by supersymmetry and enter the effective action with up to five external on-shell particles. 
 
 \sm
 
 \subsection{Duality between effective action of type I and HO theories}
 \label{sec:duality}
 
 We begin by reviewing the low-energy effective actions for the type I and heterotic $SO(32)$ theories, keeping those terms that contribute to on-shell three-point functions.  
 
 \sm
  
The string-frame effective action for the heterotic $\textrm{SO}(32)$ string theory that includes terms contributing to  three-point functions gets contributions entirely from tree-level  interactions since that are not renormalised by loop effects \cite{Ellis:1987yx, D'Hoker:2005jc} .  The  bosonic terms are given by 
\begin{eqnarray}
S^{(3)\, het}&=&\frac{1}{(2\pi)^7\lesth^8  } \int_{\mathcal{M}_{10}} d^{10}x\sqrt{-G}e^{-2\phi^{h}}\Big(
\curv+4\partial_\mu \phi^{h}\partial^\mu \phi^{h}
-\frac{1}{2}| \tilde H_3|^2 
-\frac{\lesth^2 }{2}
\textrm{tr}(|F|^2)
\Big) \nonumber
\\ 
&& + S^{het}_{\curv^2}+ S^{het}_{(\partial \, H)^2}\,.
\label{hoactdef}
\end{eqnarray}
In this expression $\curv $ is the Riemann curvature scalar, $\phi$ is the dilaton,  As in the body of this paper, the symbol $\tr$ indicates a trace of a matrix in the fundamental representation, while $\Tr$ indicates a trace in the adjoint representation.   The three-form field strength for the two-form Neveu--Schwarz/Neveu--Schwarz potential, $B_{2\, \mu\nu}$, includes the modifications due to Yang--Mills and Lorenz Chern-Simons terms.
\begin{eqnarray}
\label{tildh3def}
\tilde H_3 =   dB_2-\frac{\lesth^2}{2}\ \tr\left(A\wedge dA+\frac{2}{3}A\wedge A\wedge A\right) +\frac{\lesth^2}{2}\ \tr\left(\omega\wedge d\omega+\frac{2}{3}\omega\wedge \omega\wedge \omega\right)  \,.
\end{eqnarray}
The two terms in the second line of \eqref{hoactdef} are higher-derivative interactions that contribute to three-point functions and are required by supersymmetry once the Lorentz Chern--Simons term, which is also a higher derivative term,  is included in the action.  These are  manifestations in the HO effective action of the corresponding boundary terms in \eqref{sbound}.
 
 \sm
 
 The equivalent string frame effective action for the bosonic fields of the type I theory is given by 
\begin{eqnarray}
S^{(3)}_{I}&=&\frac{1}{(2\pi)^7(\lesti)^8} \int_{\mathcal{M}_{10}} d^{10}x\sqrt{-G}\Big(
e^{-2\phi^{I}}\Big(\curv+4\partial_\mu \phi^{I}\partial^\mu \phi^{I}\Big)
-\frac{1}{2}|\tilde F_3|^2
-\frac{(\lesti)^2 }{2}e^{-\phi^I}\textrm{tr}(|F|^2)
\Big) 
\Big) \nonumber
\\ 
&&+ S^{I}_{\curv^2}+ S^{I}_{(\partial \, H)^2}
\,.
\label{typeIact}
\end{eqnarray}
 Here the three-form field strength for the  Ramond--Ramond potential, $C_{2\,\mu\nu}$, again includes the presence of the Yang--Mills and Lorenz Chern-Simons terms, and is given by 
\begin{eqnarray}
\tilde F_3 =  dC_2-\frac{\lesti^2}{2 }\, e^{-\phi^I}\,\tr\left(A\wedge dA+\frac{2}{3}A\wedge A\wedge A\right) +\frac{\lesti^2}{2}\, e^{-\phi^I}\, \tr\left(\omega\wedge d\omega+\frac{2}{3}\omega\wedge \omega\wedge \omega\right) \,.
\label{tildfdef}
\end{eqnarray}

\sm

The equivalence of the type I and HO actions is manifest with the identifications 
\begin{eqnarray}
G_{\mu \nu}^{I}=e^{-\phi^{h}}G_{\mu \nu}^{h}\,,
\qquad
\phi^{I}=-\phi^{h}\,,
\qquad
B^{I}_{\mu \nu}=B_{\mu \nu}^{h}\,,
\qquad
A_\mu^I=A_\mu^h\,,
\end{eqnarray}
together with the relation between the string length scales in the two theories in \eqref{Idua1}.

\sm

The corresponding actions for the HE and IA theories have analogous structure.  

\subsection{Supergravity in the  HW background and the Feynman rules}
\label{hwsugra}

In order to express the Feynman rules in a unified manner in the main text we will express the eleven-dimensional gravitational constant and the ten-dimensional gauge coupling  in terms of the eleven-dimensional Planck length, $\lepl$, given in \eqref{gravigauge}.
 
 \sm
 
The Feynman rules that follow from this action have the following general features.  
\begin{itemize}
\item Each vertex coupling three gauge particles contains a single derivative and  contributes a factor of $1/\lepl^6$.  
\item Each vertex coupling a  pair of gauge particles in a boundary to a gravitational particle (the graviton or the antisymmetric three-form, $C$, which couples via the  Lorentz Chern-Simons term)  contains two derivatives and again contributes a factor of $1/\lepl^6$.  
\item
Each  supergravity bulk interaction vertex is quadratic in derivatives and contributes a factor of $1/\lepl^9$. 
\item Gravitational particles also couple via the Lorentz Chern--Simons term, is localised on a boundary  and has four derivatives and contributes a factor of $1/\lepl^6$.  
\item The graviton propagator has a factor of $\lepl^9$ while each gauge propagator contributes a factor of $\lepl^6$.
\item  The expression for an amplitude in Minkowski space-time includes an implicit product of delta functions for momentum conservation in each direction. When a dimension is compactified on a circle of radius $R\lepl$ so the conjugate momentum is quantised in units of the inverse radius, the dleta function  becomes a Kronecker delta conserving the integer Kaluza--Klein charges multiplied by the volume factor $2 \pi \lepl R$.  As discussed in the  paragraph after \eqref{bifunmass} our expressions include this volume factor, but suppress the discrete momentum Kronecker delta.

\end{itemize}

\subsection{Some higher order interactions}
\label{sec:higherorder}

In addition to the terms in the actions \eqref{hoactdef},  \eqref{typeIact}  and the corresponding HE and IA versions, we will encounter a number of interactions that arise in higher-point on-shell gauge particle and graviton amplitudes in the main part of this paper.  Certain  of these are parity-violating terms arise that are crucial for understanding the cancellation of gauge and gravitational anomalies.  These are components of a  $D=10$, $\mathcal{N}=1$ supersymmetry multiplet that also contains analogous parity-conserving terms. These terms arise art one loop in the heterotic theories (so they are independent of the dilaton) and are protected from renormalisation beyond one loop.    It is convenient to introduce a notation that highlights this relationship.  
 
\subsection{The gauge sector}
\label{sec:gaugereview}

The eighth rank tensor $t_8$ is defined by its contractions with the gauge field strength, $F_{\mu\nu}$, which is a matrix in some representation of $E_8 \times E_8$ or $SO(32)$,
\begin{eqnarray}
t_8 \, F^4 & \equiv & t_8^{\mu_1, \nu_1,\dots \mu_4\nu_4}\, F_{\mu_1\nu_1}\, F_{\mu_2\nu_2}\, F_{\mu_3\nu_3}\, F_{\mu_4\nu_4}\nonumber\\
&=& 16 F^{\mu\nu} \, F_{\rho\nu}\, F_{\mu\lambda} \, F^{\rho\lambda} + 8  F^{\mu\nu} \, F_{\rho\nu}\, F^{\rho\lambda} \, F_{\mu\lambda}\nonumber\\
&&- 4  F^{\mu\nu} \, F_{\mu\nu}\, F_{\rho\lambda} \, F^{\rho\lambda} -2  F^{\mu\nu} \, F^{\rho\lambda}\, F_{\mu\nu} \, F_{\rho\lambda}\,.
\label{ymdefs}
\end{eqnarray}
In order to distinguish the single and double traces on the group theory indices we will use the notation $t_8 \Tr F^4$ and $t_8 (\Tr F^2)^2$, where the capital $\Tr$ indicates that trace is in the adjoint representation.  We will use the lower case $\tr$ symbol to indicate a trace in the fundamental representation, where this is appropriate.

\sm 

In a gauge theory scattering amplitude with gauge particle polarisations and momenta labelled by $\epsilon_r$ and $k_r$ ($r=1,2,3,4$) the linearised gauge field  has the form  $\hat F^{A} = T^{ A}\, ( \epsilon_\mu k_\nu - \epsilon_\nu k_\mu)$.  In this case the effective $F^4$ interactions can be represented by  $t_8 \Tr\, F^4= t_8 \hat F^4\, \Tr \, T^4$  or  $t_8 (\Tr\, F^2)^2 = t_8 \hat F^4 \, (\Tr\, T^2)^2$  where 
\begin{eqnarray} t_8 \hat F^4 &=&
- 2 u t (\epsilon^{(1)} \cdot \epsilon^{(2)}) (\epsilon^{(3)} \cdot \epsilon^{(4)})
- 2 s t (\epsilon^{(1)} \cdot \epsilon^{(3)}) (\epsilon^{(2)} \cdot \epsilon^{(4)})
- 2 s u (\epsilon^{(1)} \cdot \epsilon^{(4)}) (\epsilon^{(2)} \cdot \epsilon^{(3)})
\nonumber \\
& & \qquad
+(\epsilon^{(1)} \cdot \epsilon^{(2)})
   \left[ 4 t (\epsilon^{(3)}\cdot k^{(1)}) (\epsilon^{(4)}\cdot k^{(2)})
         +4 u (\epsilon^{(3)}\cdot k^{(2)}) (\epsilon^{(4)}\cdot k^{(1)}) \right]
\nonumber \\
& & \qquad
+(\epsilon^{(3)} \cdot \epsilon^{(4)})
   \left[ 4 t (\epsilon^{(1)}\cdot k^{(3)}) (\epsilon^{(2)}\cdot k^{(4)})
         +4 u (\epsilon^{(1)}\cdot k^{(4)}) (\epsilon^{(2)}\cdot k^{(3)}) \right]
\nonumber \\
& & \qquad
+(\epsilon^{(1)} \cdot \epsilon^{(3)})
   \left[ 4 s (\epsilon^{(2)}\cdot k^{(3)}) (\epsilon^{(4)}\cdot k^{(1)})
         +4 t (\epsilon^{(2)}\cdot k^{(1)}) (\epsilon^{(4)}\cdot k^{(3)}) \right]
\nonumber \\
& & \qquad
+(\epsilon^{(2)} \cdot \epsilon^{(4)})
   \left[ 4 s (\epsilon^{(1)}\cdot k^{(4)}) (\epsilon^{(3)}\cdot k^{(2)})
         +4 t (\epsilon^{(1)}\cdot k^{(2)}) (\epsilon^{(3)}\cdot k^{(4)}) \right]
\nonumber \\
& & \qquad
+(\epsilon^{(1)} \cdot \epsilon^{(4)})
   \left[ 4 s (\epsilon^{(2)}\cdot k^{(4)}) (\epsilon^{(3)}\cdot k^{(1)})
         +4 u (\epsilon^{(2)}\cdot k^{(1)}) (\epsilon^{(3)}\cdot k^{(4)}) \right]
\nonumber \\
& & \qquad
+(\epsilon^{(2)} \cdot \epsilon^{(3)})
   \left[ 4 s (\epsilon^{(1)}\cdot k^{(3)}) (\epsilon^{(4)}\cdot k^{(2)})
         +4 u (\epsilon^{(1)}\cdot k^{(2)}) (\epsilon^{(4)}\cdot k^{(3)}) \right]\,.
\label{t8def}
\end{eqnarray} 
The interactions $t_8 \Tr F^4$ and  $t_8 (\Tr F^2)^2$ are  components of two $D=10$ $\mathcal{N}=1$ superinvariants that also contains the parity-violating $F^4$ terms  that are essential for understanding the absence of chiral gauge anomalies.  The superinvariant that arises at one loop in the HO theory is  the combination
 \begin{eqnarray}
I_1= t_8\, \tr F^4 - \quart \epsilon_{10} B \, \tr F^4\,, 
 \label{sogauge}
 \end{eqnarray}
 where $\tr$ indicates the trace in the fundamental representation of $SO(32)$ and 
 \begin{eqnarray}
\epsilon_{10}\, B F^4 \equiv  \epsilon^{\mu_1, \nu_1,\dots \mu_5,\nu_5}\,  F_{\mu_1\nu_1}\, F_{\mu_2\nu_2}\, F_{\mu_3\nu_3}\, F_{\mu_4\nu_4}\,B_{\mu_5\nu_5}\,.
\label{parityym}
\end{eqnarray}
eqn \ref{sogauge} we have  converted from traces in the adjoint representation of $SO(32)$ to traces in the fundamental representation using 
 \begin{eqnarray}
 \Tr  F^4 =24 \tr  F^4 + 3 (\tr  F^2)^2\,,  \qquad \Tr  F^2 = 30 \tr  F^2 \,.
 \label{adjfun}
 \end{eqnarray} 
There is no independent fourth order Casimir in the $E_8\times E_8$ theory,  and we have $\Tr F^4 = (\Tr F^2)^2/100$ where $\Tr$ denotes the trace in the $248 \times 248$-dimensional adjoint representation.   The terms that arise in the $E_8 \times E_8$ one-loop effective action for the HE theory form the combination 
\begin{eqnarray}
I_2= t_8\, (\tr_i  F_i^2)^2 - \quart \epsilon_{10} B \, (\tr_i F_i^2)^2 \,.
\label{e8gauge}
\end{eqnarray}

\subsection{The gravitational sector} 
\label{sec:gravityreview}
 We will again introduce a notation that emphasises the relationship between terms in the ten-dimensional effective action that are integrals of ten-forms  with analogous scalar expressions.   
 For example, the Vafa-Witten term in the type IIA theory  will be denoted 
\begin{eqnarray}
\epsilon\,B\,  \yvw (\curv)\equiv B\wedge \xvw (\curv)=  B\wedge \left(\tr(\curv\wedge \curv\wedge \curv\wedge \curv )- \quart \tr(\curv\wedge \curv)\, \tr(\curv\wedge \curv) \right)\,,
\label{vwterm}
\end{eqnarray}
where $\xvw(\curv)$ is an eight-form (that is inherited from a ten-dimensional characteristic class).  
The eleven-dimensional version of this term \cite{Duff:1995wd} is the eleven-form, $C \wedge \xvw (\curv)$.
We have introduced the notation $\xvw(\curv) = \epsilon \yvw(\curv)$ since the Vafa--Witten term  is related by $\mathcal{N}=2$ supersymmetry to a scalar term formed from four curvatures, which can be written as
\begin{eqnarray}
t_8 \yvw (\curv)&\equiv& \frac{1}{24}\,  t_8 t_8 \curv^4 \nonumber\\
 &=&    t_8^{\mu_1, \nu_1,\dots \mu_4\nu_4}t_{8\, \mu'_1, \nu'_1,\dots \mu'_4\nu'_4}\, \curv^{\mu'_1\nu'_1}_{\mu_1\nu_1} \dots  \curv^{\mu'_4\nu'_4}_{\mu_4\nu_4} 
 \nonumber\\
&=& t_8 \left(\tr \curv^4-\quart (\tr \curv^2)^2 \right) \, .
\label{vwdef}
\end{eqnarray}
The third line  follows from the earlier  definition of $t_8$ and uses the notation in which the curvature is viewed as a matrix in the  fundamental representation of the tangent space Lorentz group, $SO(9,1)$.
As with the corresponding term in the gauge sector, the parity-violating term is replaced by the parity-conserving term simply by exchanging a factor of $\epsilon B$ for a factor of $t_8$.

\sm

The  same combination of four powers of the  curvature, $\yvw(\curv)$, enters the $\mathcal{N}=1$ theories as the $\mathcal{N}=2$ theories.  In addition, a different combination involving the fourth power of curvatures and Yang--Mills field strengths arises in the $\mathcal{N}=1$ $D=10$ theories, which is a key ingredient necessary for the absence of anomalies.  The parity-violating piece is the Green--Schwarz ten-form given by
\begin{eqnarray}
\epsilon B\, \ygs (\curv,F)\equiv B\wedge  \xgs (\curv,F) \, ,
\label{gsterma}
\end{eqnarray}
where $B$ is either the Neveu--Schwarz/Neveu--Schwarz  two-form in the heterotic theories or the Ramond--Ramond two form in the type I theory and\footnote{The anomaly cancelling term in the action has the symbolic form $-1/(2^{17} \pi^{5}3) \int d^{10}x \epsilon B\, \ygs$ (this value differs from the normalisation in \cite{Horava:1996ma} because of the different definition of the anomaly cancelling term, as stressed in footnote \ref{fn:6}). }
 \begin{eqnarray}
\ygs (\curv, F) &=&\left( 8\, \tr F^4+ \tr \curv^4 + \quart\, (\tr \curv^2)^2   -  \tr F^2\, \tr \curv^2 \right)\, .
\label{gsfull}
\end{eqnarray}
Again the low-energy expansion involves a parity-conserving partner of this ten-form, which can be written as $t_8 \ygs (\curv,F)$, which includes the $t_8 \, \tr F^4$ term that follows from \eqref{ymdefs}. 

\sm 

It is notable that the ratio of the coefficient of the $\tr \curv^4$ term to that of the $(\tr \curv^2)^2 $ term in \eqref{vwdef} has  the opposite sign to the ratio of these coefficients in \eqref{gsfull}, so that
\bea
 \ygs (\curv,0)= \yvw (\curv) + \half (\tr \curv^2)^2\,.
\label{diffgs}
\eea
The $(\tr \curv^2)^2/2$ term is a boundary contribution that was explained in the context of M-theory in the  Ho\u rava--Witten  background in \cite{Horava:1996ma}.

\subsubsection*{$\curv^4$ superinvariants }
\label{sec:superinvariants}

It is useful  for the discussion in sections \ref{sec:gravitrees} and \ref{sec:gravloop} to identify which combinations of the above $\curv^4$ terms are bosonic components of superinvariants.  We will here summarise the discussion of Tseytlin in \cite{Tseytlin:2000sf}, adapted to our present  conventions. We will also make use of the detailed analysis in \cite{deRoo:1992sm, deRoo:1992zp}.

 \subsubsection*{Type II invariants}
 
 The type II  theories have $\mathcal{N}=2$ supersymmetry. The following combination of bosonic terms involving the Riemann curvature is a superinvariant that enters  into the tree-level  effective action of both the IIA and IIB theories.
 \begin{eqnarray}
 J_0 = t_8t_8 \curv^4 -\frac{1}{8} \epsilon_{10}\epsilon_{10} \curv^4  \,.
 \label{j0def}
 \end{eqnarray}
The two ten-dimensional epsilon tensors contract into the sixteen indices of $\curv^4$ leaving two pairs to contract into each other.  The eight-dimensional analogue of the $\epsilon_{10}\epsilon_{10} \curv^4$ term is proportional to the Euler invariant.

\sm 

At one loop the type IIA effective action receives an extra contribution.  This is proportional to the  superinvariant, $\mathcal{I}_2$ defined by
\begin{eqnarray}
 \mathcal{I}_2 = - \frac{1}{8}\,\epsilon_{10}\epsilon_{10} \curv^4 + 6 \, \epsilon_{10} B \yvw  (\curv)  \, .
 \label{supdef}
\end{eqnarray}
 Recall that $\epsilon_{10} B \yvw(\curv)  $ is the odd-parity Vafa--Witten term, which is here seen to be related by supersymmetry to $\epsilon_{10}\epsilon_{10} \curv^4 $.
The $\curv^4$ terms in the  type IIA effective action only arise at tree-level and one loop and can be summarised by an effective action proportional to
\begin{eqnarray}
\label{oneloop2a}
S_{\curv^4}^{IIA} = \frac{1}{\lestii^2}\int_{\mathcal{M}_{10}} d^{10}x \sqrt{-G} \left(\frac{2 \zeta(3)}{\giia^2} J_0 - \frac{2\pi^2}{3}\, (J_0 - 2\mathcal{I}_2) \right) \,,
\end{eqnarray}
where $\lestii$ is the type IIA or IIB string length scale.
 
\sm 

It is notable that in the type IIA theory the combination of $t_8 t_8 \curv^4$ and $\epsilon_{10}\epsilon_{10}\curv^4$ 
 arises at one loop with the opposite relative sign to the tree-level combination.  The one-loop odd parity Vafa--Witten term in the type IIA theory is protected against renormalisation at higher loops. It follows that  $ \epsilon_{10}\epsilon_{10} \curv^4$  is also protected against higher loop corrections.    In the type IIA theory these tree-level and one-loop contributions are the only contributions to $\epsilon_{10}\epsilon_{10} \curv^4$.
 
 \sm 

In the type IIB theory there is no Vafa--Witten term and the only invariant containing $\curv^4$ is  $J_0$.   In this case the relative signs of the $t_8 t_8 \curv^4$ and $\epsilon_{10}\epsilon_{10}\curv^4$ terms are the same in both the tree-level and one-loop terms.  There is no reason to expect $J_0$ to be protected against getting higher loop or D-instanton corrections, and in fact the dilaton dependence of the coefficient of $J_0$  enters as a  modular invariant function of the complex scalar, $\Omega$.
The type IIB theory effective $\curv^4$ action  is proportional to 
\begin{eqnarray}
\label{2beffect}
S_{\curv^4}^{IIB}  = \frac{1}{\lestii^2}\int_{\mathcal{M}_{10}} d^{10}x \sqrt{-G} \,\giib^{-\half}\, E_{\threeh}(\Omega) \, J_0 \,,
\end{eqnarray}
where  $E_s(\Omega)$ is a non-holomorphic Eisenstein series of weight $s$ that is modular function of the complex scalar $\Omega= C^{(0)} + i/\giib$  and is discussed in section \ref{sec:graviamps}.   The  coupling constant dependence of $E_\threeh(\Omega)$ shows that that this $t_8t_8\curv^4$ interaction has contributions from tree level and one loop in string perturbation theory and from an infinite set of D-instantons.

  \subsubsection*{Heterotic and type I invariants}
  \label{sec:n1inv}
  
  The heterotic or type I effective actions contain $\mathcal{N}=1$ superinvariants.  These can be chosen to be $J_0$ (the $\mathcal{N}=2$ invariant defined in \eqref{j0def}, and $X_1$ and $X_2$, defined as follows
 \begin{eqnarray}
 X_1 &= t_8 \tr \curv^4 - \quart \epsilon_{10} B\, \tr \curv^4\,, \qquad\quad X_2 = t_8 (\tr \curv^2)^2 - \quart \epsilon_{10} B \, (\tr \curv^2)^2  \,.
 \label{x1x2def}
 \end{eqnarray}
 The combination that contains the $\mathcal{N} =1$ odd-parity anomaly-cancelling term is the $\mathcal{N}=1$ invariant 
 \begin{eqnarray}
 J_2= X_1 + \quart X_2 = (t_8 - \quart \epsilon_{10} B)\, \ygs  (\curv,0) \,,
  \label{gsterm}
\end{eqnarray}
which should not to be confused with  the  $\mathcal{N}=2$ invariant $\mathcal{I}_2$ defined earlier, which is the combination
\begin{eqnarray}
\mathcal{I}_2 = J_0 -24(X_1-\quart X_2) =   \frac{1}{8} \epsilon_{10}\epsilon_{10} \curv^4 +6 \, \epsilon_{10} B\, \yvw(\curv) \,.
\label{i2comb}
\end{eqnarray}

\section{Relations between traces} 
\label{sec:traces}

In the body of the paper we make use of a number of well-known  identities between traces of  matrices in various representations of $SO(N)$ and $E_8$, which we summarise in this appendix.

\subsection{$SO(N)$ traces}
The relations between traces of products of up to six  matrices in the adjoint and fundamental representations of $SO(N)$ are given by 
\begin{subequations}
\begin{eqnarray}
	\Tr_{ad_{N}} F^2&=& (N-2) \tr_N F^2\,,
\label{e8trace51a}\\
\Tr_{ad_N} F^4&=& (N-8) \tr_N F^4 +3 (\tr_N F^2)^2 \,,
\label{e8trace51b}	\\
\Tr_{ad_N} F^6 &=& (N-32)\tr_N F^6+ 15 \tr_N F^4 \tr_N F^2	 \,,
\label{e8trace51c}	
\end{eqnarray}
\end{subequations}
where the symbol $\Tr_{ad_N}$ indicates the trace in the adjoint representation of $SO(N)$ while  $\tr_N$ denotes the trace in the $N$-dimensional representation..

\sm 

The traces of products of up to six matrices in the  Weyl spinor representation of $SO(2M)$  (indicated by $\trs $), which is $2^{M-1}$ dimensional, are (see, for example, pages 274-276 of \cite{Bastianelli:2006rx})
\begin{subequations}
\begin{eqnarray}
	\trs F^2&=& 2^{M-4}\tr_{2M} F^2 \,,
\label{e8trace52a}	\\
\trs F^4&=& -2^{M-5} \tr_{2M} F^4 +3\cdot 2^{M-7} (\tr_{2M}F^2)^2 \,,
\label{e8trace52b}	\\
\trs F^6 &=& 2^{M-4}\tr_{2M} F^6-5\cdot 3\cdot 2^{M-8} \tr_{2M}F^4 \tr F^2	+5\cdot 3\cdot 2^{M-10} (\tr_{2M} F^2)^3	\, .
\label{e8trace52c}	
\end{eqnarray}
\end{subequations}
Note that the ratio of coefficients is independent of $M$, while the overall normalization  is proportional to $2^M$. The same relations also hold for the spinor representation of $SO(2M-1)$.  

\subsection{$E_8$ traces}
The adjoint representation of $E_8$ is ${\bf 248} $  dimensional and it is given by the sum of the adjoint and spinor representations of $SO(16)$
\begin{eqnarray}
\bf 
248 = 120+ 128 \,.
\label{e8trace53}	
\end{eqnarray}
The quadratic Casimir of $E_8$ in the adjoint representation can be expressed in terms of $SO(16)$ traces by 
\begin{eqnarray}
\Tr_{ E_8}F^2 &=&\tr_{\bf 120}F^2+\tr_{\bf 128}F^2	
\nonumber\\
&=& 14\,
\tr_{\bf 16} F^2+16\, \tr_{\bf 16} F^2=30\, \tr_{\bf 16} F^2\,,
\label{e8trace54}	
\end{eqnarray}
where we have denoted the adjoint trace by $\Tr_{ E_8}F^2\equiv\Tr_{\bf 248}F^2$.
Similarly, the quartic Casimir  of $E_8$ is given by 
\begin{eqnarray}
\Tr_{E_8}F^4&=&\tr_{\bf 120}F^4+\tr_{\bf 128}F^4	
\nonumber\\
&=& 9\,
(\tr_{\bf 16} F^2)^2=\frac{1}{100} \,(\Tr_{E_8} F^2)^2\,.
\label{e8trace55}	
\end{eqnarray}
The sixth order Casimir in the adjoint representation of $E_8$ is given by 
\begin{eqnarray}
\Tr_{E_8}F^6\equiv\tr_{\bf 248}F^6&=&\tr_{\bf 120}F^6+\tr_{\bf 128}F^6	
\nonumber\\
&=& \frac{15}{4}\,
(\tr_{\bf 16} F^2)^3=\frac{1}{7200}\, (\Tr_{E_8} F^2)^3\, .
\label{e8trace555}
\end{eqnarray}
This implies that  $E_8$ has no independent fourth and sixth order Casimirs.
Using \eqref{e8trace55} we can write \eqref{e8trace555}  as 
\begin{eqnarray}
\Tr_{E_8} F^6 &=& \left[ \frac{1}{48} \Tr_{E_8} F^4 -\frac{1}{14400} (\Tr_{E_8} F^2)^2\right]\Tr_{E_8} F^2\,.
\label{e8trace2}	
\end{eqnarray}
 $E_8$ does not possess a fundamental representation but it has become conventional to {\it define} a quantity $\tr_{E_8}$ by\begin{eqnarray}
	\tr_{E_8} F^2&\equiv & \frac{1}{30} \Tr_{E_8} F^2\,.
\label{e8trace3}
\end{eqnarray}
With this definition,  several normalisations of coefficients in the $E_8\times E_8$ heterotic theory coincide with those of the $SO(32)$ theory if $\tr_{E_8}$ is interchanged with $\tr_{SO(32)}$.
With the above definition substituted in \eqref{e8trace55}  and \eqref{e8trace555}  we get
\begin{subequations}
\begin{eqnarray}
\Tr_{E_8}F^4&=& 9\, (\tr_{E_8} F^2)^2 \,,
\label{e8trace4a}	
\\
\Tr_{E_8}F^6&=&	\frac{15}{4}\, (\tr_{E_8} F^2)^3  \,.
\label{e8trace4b}	
\end{eqnarray}
\end{subequations}

\section{Loop amplitude in compactified $SO(32)$ gauge theory}
\label{sec:holoopamp}

We will here consider the one-loop amplitude in the $SO(32)$ gauge theory with gauge group broken to  $SO(16) \times SO(16)$ by the Wilson line in \eqref{cartanho}.  This is analogous to the loop amplitude in the $E_8$ gauge theory considered in section \ref{sec:heloopamp}. However, as emphasised earlier, this loop contribution does not arise in supergravity in the Ho\u rava--Witten background so it is not directly relevant to the bulk of this paper, but the structure of the following argument complements that of the $E_8$  case.   
 
We will again consider the loop where all the external states are massless in nine dimension and are in the adjoint representation of the $SO(16)\times SO(16)$ gauge group. The states circulating in the loop can either be $SO(16)\times SO(16)$ adjoint states with masses given by \eqref{cartane8}, or the  $SO(16)\times SO(16)$ bi-fundamental states with masses given by \eqref{bifunmass}.

The amplitude is the sum of  contributions of the circulating massless gauge states and their massive Kaluza--Klein recurrences in the adjoint of $SO(16)\times SO(16)$ and of the tower of massive  bi-fundamental states propagating around the loop, so that 
\begin{eqnarray}
A^{1-loop} = A_{adj} + A_{bifun}\,.
\label{hoamp}
\end{eqnarray}
The piece containing circulating adjoint states is given by
\begin{eqnarray}
A_{adj} &=&\frac{2}{3(2\pi)^{10}}t_8\hat F^4\mathcal{C}_{\textrm{adj}} I_{adj}(s,t,u; \raho ) 	\,,
 \label{gluoncal11}
\end{eqnarray}
where  the colour factor is the sum of two copies of \eqref{gluoncal2}
\begin{eqnarray}
 \mathcal{C}_{\textrm{adj}}
&=&
\sum_{i=1}^{2}
\left[(N-8) \tr_i(T^{a_1}T^{a_2}T^{a_3}T^{a_4}) +( \tr_i(T^{a_1}T^{a_2})\tr_i(T^{a_3}T^{a_4})+\textrm{perms})\right] \,,
 \label{gluoncal12}
\end{eqnarray}
where we will later set $N=16$, and  the dynamical factor $I_{adj}(s,t,u; \raho )$ (which we have chosen to express in terms of the HO parameters, using $\lesth^2 \raho^2  = \lepl^2(\raten \raele )^{-2} $) is again given by a ten-dimensional scalar box diagram.  Repeating the Poisson summation argument that led from \eqref{gluoncal3} to \eqref{gluoncal3a} gives  
\begin{eqnarray}
I_{adj}(0,0,0; \raho )&=&
2\pi^{11/2}
\int_0^\infty 
\frac{d\tau}{\tau^{3/2}}
\sum_{m\in \mathbb{Z}
} e^{-\tau\left(\frac{m}{\lesth\raho}\right)^2}
\nonumber\\
&=&
2\pi 
\left[\hat C_1\frac{\raho}{\gho^{2/3}\lesth}
+
\frac{\pi^3}{(\raho \lesth)}
\zeta(2)\right] \,.
 \label{gluoncal13}
\end{eqnarray}
Here we have again regularised the divergent zero winding number term and assigned it an arbitrary renormalised value  $\hat C_1 R_{10}/\ell_{11}$ in $\ell_{11}$ units.  Whereas a cut-off  in $\ell_{11}$ units was natural for the $E_8$ theory in the context of supergravity in the Ho\u rava--Witten background, it is not so clear that the M-theory Planck scale provides a natural cut-off in the context of the $SO(32)$ theory under consideration.  Our treatment of this renormalised term, which is proportional to $\raho$, may  therefore be questionable.  However, with this choice of renormalisation the power of the string coupling in the first term in  \eqref{gluoncal13} makes no sense, so we need to set $\hat C_1=0$.

The other piece of the four gauge boson  loop amplitude, in which $SO(16)\times SO(16)$ bi-fundamental states are circulating, has the form
\begin{eqnarray}
A_{bifun}&=&\frac{2}{3(2\pi)^{10}}t_8\hat F^4\, \mathcal{C}_{bifun} I_{bifun}(s,t,u; \raho ) \,,
 \label{gluoncal14}
\end{eqnarray}
where $ \mathcal{C}_{bifun}$ is the colour factor for the loop of  bi-fundamental states and is given by
\begin{eqnarray}
\mathcal{C}_{bifun}&=&	\left[N\sum_{i=1}^2\tr_i(T^{a_1}T^{a_2}T^{a_3}T^{a_4})+ \tr_1(T^{a_1}T^{a_2})\tr_2(T^{a_3}T^{a_4})\right] \, .
 \label{gluoncal15}
\end{eqnarray}
The low-energy limit is obtained by setting $s,t,u =0$ in the dynamical factor $I_{bifun}$, which leads to 
\begin{eqnarray}
I_{bifun}(0,0,0; \raho )&=&
2\pi^{11/2}
\int_0^\infty 
\frac{d\tau}{\tau^{3/2}}
\sum_{m\in \mathbb{Z}
} e^{-\tau\left(
\frac{m-1/2}{\lesth\raho}\right)^2}
\nonumber\\
&=&
2\pi 
\left[\hat C_2\frac{\raho}{\gho^{2/3}\lesth} - \frac{1}{2}\frac{\pi^3}{\raho \lesth}\zeta(2)\right]\,,
 \label{gluoncal16}
\end{eqnarray}   
where, as before we have used a cut-off in $\ell_{11}$ units, and we again need to set the arbitrary renormalised coefficient of the zero winding number term to zero ($\hat C_2=0$) since it multiplies an unphysical power of the string coupling.

The total amplitude is given by adding  \eqref{gluoncal13} and \eqref{gluoncal16}. We see that  setting  $N=16$ in the effective action gives
\begin{eqnarray}
 A^{1-loop} &=& \left( A_{adj} + A_{bifun}\right) \nonumber\\
&=&t_8\left[\sum_{i=1}^2(\tr_i F_i^2)^2 -(\tr_1 F_1^2)(\tr_2 F_2^2)\right]  
	\frac{1}{\lesth\raho }\zeta(2) 
\nonumber\\
&=&t_8\left[\sum_{i=1}^2(\tr_i F_i^2)^2 -(\tr_1 F_2^2)(\tr_2 F_2^2)\right] 
\frac{\rahe}{\lesth }\zeta(2)\,,
\label{gluoncal17}
\end{eqnarray}  
and so the double-trace  amplitude survives in the ten-dimensional limit of the HE theory ($\rahe \to \infty$).
The result is identical to the expression we obtained earlier by expanding tree amplitudes in \eqref{oneloophe}, and  agrees with the direct evaluation of the low-energy limit of the HE loop amplitude in string perturbation theory.

As noted above,  the renormalisation procedure used to obtain the values of the zero winding terms in \eqref{gluoncal13}  and \eqref{gluoncal16}, which are proportional to $\raho$, is sensitive to our choice of renormalisation procedure.  This is not determined by present considerations since the $SO(32)$ gauge theory does not originate from a local action analogous to that of  supergravity in the Ho\u rava--Witten background.  With the choice of cut-off in M-theory Planck units used in  \eqref{gluoncal13}  and \eqref{gluoncal16}  we are led to the expression  \eqref{gluoncal17}, which vanishes in the limit $\raho\to \infty$.  Therefore the ten-dimensional HO expression  is not apparent from this perspective, although it was determined from the $E_8$ gauge theory loop  as the $\raho\to \infty$ limit of  \eqref{heterf4} in section \ref{sec:heloopamp}.

\end{document}